\documentclass[11pt]{article}
\usepackage{graphicx}
\usepackage[margin=1.25in]{geometry}
\usepackage[usenames,dvipsnames]{color}
\usepackage{url}
\usepackage[colorlinks = true,
            linkcolor = blue,
            urlcolor  = blue,
            citecolor = blue,
            anchorcolor = blue]{hyperref}

\usepackage{lineno,hyperref,color}
\modulolinenumbers[5]
\usepackage{todonotes}
\usepackage{ulem}
\usepackage{booktabs}
\usepackage{float}
\usepackage{wrapfig}

\usepackage{appendix}
\usepackage{pdfpages}
\usepackage{hyperref}
\usepackage{geometry}                
\usepackage[parfill]{parskip}    
\usepackage{graphicx, subfigure}
\usepackage{amssymb}
\usepackage{amsmath}
\usepackage{epstopdf}
\usepackage{array}
\usepackage{lineno}
\usepackage{scrextend}

\newcommand\pubnumber{arXiv }
\newcommand\pubdate{\today}

\def\Title#1{\begin{center} {\LARGE #1 } \end{center}}
\def\Author#1{\begin{center}{ \sc #1} \end{center}}

\newcommand\pubblock{\rightline{\begin{tabular}{l} \pubnumber\\
         \pubdate \end{tabular}}}
\newenvironment{Abstract}{\begin{quotation} \begin{center}
                       ABSTRACT
     \end{center}\bigskip  }{\end{quotation}}

\def\ie{{\it i.e.}}
\def\eg{{\it e.g.}}
\def\etc{{\it etc.}}

\def\beq{\begin{equation}}
\def\eeq#1{\label{#1}\end{equation}}
\def\eeqn{\end{equation}}

\newenvironment{Eqnarray}%
   {\arraycolsep 0.14em\begin{eqnarray}}{\end{eqnarray}}
\def\beqa{\begin{Eqnarray}}
\def\eeqa#1{\label{#1}\end{Eqnarray}}
\def\eeqan{\end{Eqnarray}}

\let\bar=\overbar

\def\lsim{\mathrel{\raise.3ex\hbox{$<$\kern-.75em\lower1ex\hbox{$\sim$}}}}
\def\gsim{\mathrel{\raise.3ex\hbox{$>$\kern-.75em\lower1ex\hbox{$\sim$}}}}

\def\W{{\cal W}}

\def\del{\partial}
\def\Dslash{\not{\hbox{\kern-4pt $D$}}}
\def\dslash{\not{\hbox{\kern-2pt $\del$}}}
\def\pslash{\not{\hbox{\kern-2pt $p$}}}
\def\ETmiss{\not{\hbox{\kern-4pt $E$}}_T}

\def\Dlr{\mathrel{\raise1.5ex\hbox{$\leftrightarrow$\kern-1em\lower1.5ex\hbox{$D$}}}}

\def\ee{\mathrm{e^+e^-}}
\newcommand{\alphas}{\alpha_{S}}
\newcommand{\alphasmZ}{\alphas(m^2_\mathrm{Z})}
\def\MSB{{\bar{M \kern -2pt S}}}
\def\msb{{\bar{\scriptsize M \kern -1pt S}}}

\def\drb{{\bar{\scriptsize D \kern -1pt R}}}

\def\keV{{\rm keV}}
\def\MeV{{\rm MeV}}
\def\GeV{{\rm GeV}}
\def\TeV{{\rm TeV}}

\input{defs}

\newcommand\snowmass{\begin{center}\rule[-0.2in]{\hsize}{0.01in}\\\rule{\hsize}{0.01in}\\
\vskip 0.1in Submitted to the  Proceedings of the US Community Study\\ 
on the Future of Particle Physics (Snowmass 2021)\\ 
\rule{\hsize}{0.01in}\\\rule[+0.2in]{\hsize}{0.01in} \end{center}}

\begin{document}

\pubblock

\Title{The Future Circular Collider: a Summary for the US 2021 Snowmass Process}

\bigskip 
 {\large \bf Section Editors and Readers}

\Author{APC, Paris, France}{G.~Bernardi}
\Author{Brookhaven National Laboratory, New York, USA}{E.~Brost, D.~Denisov}
\Author{Brown University, Providence, USA}{G.~Landsberg}
\Author{CERN}{M.~Aleksa, D.~d'Enterria, P.~Janot, M.L. Mangano, M. Selvaggi, F.~Zimmermann}
\Author{CIEMAT, Madrid, Spain}{J.~Alcaraz Maestre}
\Author{DESY, Hamburg and Humboldt-Universit\"at zu Berlin, Germany}{C.~Grojean}
\Author{Fermilab, Illinois, USA}{R.M.~Harris}
\Author{IFIC, University of Valencia - CSIC, Valencia, Spain}{A.~Pich, M.~Vos}
\Author{IFT (UAM/CSIC), Madrid, Spain}{S.~Heinemeyer}
\Author{INFN, Bologna, Italy}{P.~Giacomelli}
\Author{INFN, Padova, Italy}{P.~Azzi}
\Author{INFN, Pisa, Italy}{F. Bedeschi}
\Author{KIT, Karlsruhe, Germany}{M.~Klute}
\Author{LPNHE, Paris, France}{A.~Blondel}
\Author{Massachusetts Institute of Technology, Cambridge, MA, USA}{C.~Paus}
\Author{Max-Planck-Institute for Physics, Munich, Germany}{F.~Simon}
\Author{Niels Bohr Institute, Copenhagen University, Denmark}{M.~Dam}
\Author{Northeastern University, Boston, USA}{E.~Barberis, L.~Skinnari}
\Author{Stanford University, California, USA}{T.~Raubenheimer}
\Author{University of Basel, Basel, Switzerland}{S.~Antusch}
\Author{University of California -- Santa Cruz, USA}{W. Altmannshofer}
\Author{University of Chicago, Illinois, USA}{L.-T.~Wang}
\Author{University of Granada, Granada, Spain}{J.~de Blas}
\Author{University of Maryland, Maryland, USA}{S.~Eno, Yihui~Lai}
\Author{University of Massachusetts, Amherst, USA}{S.~Willocq
\Author{University of Michigan, Ann Arbor, USA}{J.~Qian, J.~Zhu}
}\Author{University of New Mexico, Albuquerque, USA}{R.~Novotný, S.~Seidel}
\Author{University of Notre Dame, Indiana, USA}{M.D.~Hildreth}
\Author{University of Pennsylvania, Philadelphia, USA}{E.J.~Thomson}
\Author{University of Rochester, New York, USA}{R.~Demina}
\Author{University of Silesia, Katowice, Poland}{J.~Gluza}
\Author{University of Zurich, Switzerland}{G.~Isidori}
\Author{Uppsala University, Uppsala, Sweden}{R.~Gonzalez Suarez}

~~
~~

{\huge See the appendix for the list of supporters of U.S. involvement in a future FCC program}

\hfill \break
\hfill \break
\hfill \break

\medskip

 \begin{Abstract}
\noindent In this white paper for the 2021 Snowmass process, we give a description of the proposed Future Circular Collider (FCC) project and its physics program.  The paper summarizes and updates the discussion submitted to the European Strategy on Particle Physics. After construction of an $\approx$ 90\,km tunnel, an electron-positron collider based on established technologies allows world-record instantaneous luminosities at center-of-mass energies from the Z resonance through the ZH and WW and up to $\ttbar$ thresholds, enabling a very rich set of fundamental measurements including Higgs couplings determinations at the subpercent level, precision tests of the weak and strong forces, and searches for new particles, including dark matter, both directly and via virtual corrections or mixing. Among other possibilities, the FCC-ee will be able to (i) indirectly discover new particles coupling to the Higgs and/or electroweak bosons up to scales $\Lambda \approx 7$ and 50~TeV, respectively; (ii) perform competitive SUSY tests at the loop level in regions not accessible at the LHC; (iii) study heavy-flavor and tau physics in  ultra-rare decays beyond the LHC reach, and (iv) achieve the best potential in direct collider searches for dark matter, sterile neutrinos, and axion-like particles with masses up to $\approx90$~GeV. The tunnel can then be reused for a proton-proton collider, establishing record center-of-mass collision energy, allowing unprecedented reach for direct searches for new particles up to the $\approx 50$~TeV scale, and a diverse program of measurements of the Standard Model and Higgs boson, including a precision measurement of the Higgs self-coupling, and conclusively testing weakly-interacting massive particle scenarios of thermal relic dark matter. The FCC-ee and FCC-hh physics and accelerator programs are remarkably synergistic and complementary.  The program builds on the stable funding provided by the CERN member states and the existing, long-standing worldwide partnerships built via the LHC, but requires substantial contributions both intellectual and financial from the US and other non-CERN-members to become a reality.

\end{Abstract}

\snowmass

\def\thefootnote{\fnsymbol{footnote}}
\setcounter{footnote}{0}

\tableofcontents


\section{Introduction}
\label{sec:introduction}
\textcolor{red}{Editors: A. Blondel, D. Denisov, S. Eno, and P.~Janot}

The LEP Electroweak Working Group, using results from precision measurements at LEP  (CERN) and SLC (SLAC), was able to predict~\cite{ALEPH:2005ab} the top quark mass and the W mass within the context of the Standard Model (SM) to be $m_\mathrm{t} = 173^{+13}_{-10}$~\GeV~and $m_\mathrm{W} = 80.362^{+0.032}_{-0.031}$~\GeV~respectively, assuming that no particle but the (yet unobserved) Higgs boson would impact the quantum corrections to these measurements. These predictions were found to be in striking agreement with their direct measurements at the Tevatron and the LHC ($m_\mathrm{t} = 172.76 \pm 0.30$~\GeV~and $m_\mathrm{W} = 80.379 \pm 0.012$~\GeV)~\cite{pdg}. The combination of the LEP, SLC, Tevatron, and LHC precision measurements were then used to constrain the quantum effects of the Higgs boson and to predict its mass to be $m_\mathrm{H} = (90^{+21}_{-18})~\GeV$ and less than 135~\GeV at the 95\% confidence level (CL)~\cite{Haller:2018nnx} within the SM. In 2012, data from the LHC confirmed that prediction
by observing a new particle with properties consistent with those of the Higgs boson and a mass near 125\unit{\GeV}~\cite{higgscms,higgsatlas}. 
All of these results came from collaborations with strong participation
from the United States.  This adventure can continue with U.S. participation in the FCC accelerator program, which starts with a new circular electron-positron collider (FCC-ee)~\cite{FCC-ee-accelerator} operating at center-of-mass energies from the Z pole to the top-pair production thresholds, followed by proton-proton collisions at 100\,TeV (FCC-hh)~\cite{Benedikt:2018csr}, and potentially electron-ion, electron-proton, and heavy ion collisions as well.

 The  discovery of the Higgs boson was a formidable success of the Standard Model; however it leaves a number of questions unanswered, and in fact marks the opening of a new and exciting era of exploration.
 The Higgs boson  is the only known elementary spin zero particle and, in spite of remarkable work at the LHC,  many of its predicted SM properties will still require testing 
 at the end of the upgraded LHC exploitation. The Higgs coupling to all charged fermions is predicted  to be equal to their masses -- can we verify this, even for the electron and the light quarks that compose the ordinary matter? Is the Higgs boson also at the origin of neutrino masses? Is there any sign that there are other similar Higgs bosons, as in Supersymmetry, or that there exist other sources of mass generation? The Higgs boson generates the breaking of electroweak symmetry because it couples to itself: can this be  verified and the form of the associated potential established? The FCC program will provide a thorough investigation of these questions. The FCC-ee, with its high luminosity and two (possibly four) interaction regions, will produce more than one million (possibly two millions) Higgs bosons. By precise measurement of the fully inclusive cross section $\epem \to \mathrm{ZH}$ at the production maximum, FCC-ee will provide the $g_\mathrm{ZH}$ coupling with per mil precision.
 The capability of FCC-ee to measure the ZH and $\rm WW \to H$ cross sections at two different centre-of-mass energies (240 and 365\,GeV) offers a 2--4$\sigma$ sensitivity (depending on the number of interaction points) to the Higgs self-coupling, via a loop diagram. Many other measurements of the Higgs boson, that will lay the foundation of highly  precise measurements and investigations  both at FCC-ee and later at FCC-hh, will be made possible for the first time. 
 
The FCC physics motivation does not end with the Higgs boson. We still do not know what dark matter (DM) is, and if it has any interactions other than gravitational with the particles of the Standard Model. 
We know neutrinos are massive but very little about the value of neutrino masses and whether neutrinos are Dirac or Majorana particles.   
We do not know how the Universe came to be composed of matter, while all antimatter seems to have disappeared.
We do not know why the Higgs mass is near 125\unit{GeV} when loop corrections, especially from the top quark, would naturally push it towards the Planck scale.  These experimental facts as well as many other theoretical problems call for new  particles  and/or phenomena beyond those which constitute the minimal Standard Model. Will these questions and other theoretical problems find their answers in a  more fundamental theory at a high energy scale of which the Standard Model would be a low energy approximation? Or will some answers come from a ``dark sector" comprising  one or several particles that lie at or below the weak scale, and which, because of extremely weak couplings to the known particles, may have so far escaped detection? 

Because of the unknown properties and huge possible ranges of masses of the required new particles, this quest calls for a broad exploration, with much improved measurement precision and sensitivity to rare processes, and ultimately more energy. We will see that the FCC programme (FCC-ee then hh) will address this challenge  very
effectively. 
The agreement between the observed top quark and Higgs boson masses with the values predicted by the minimal SM is highly non trivial. It strongly constrains the properties of possible new physics, whose quantum effects must be smaller than the current experimental uncertainties. One of the main tools of exploration of the next $\epem$ collider would therefore be to reduce  these uncertainties to a level that may offer sensitivity to these smaller effects.
In the very clean environment of an electron-positron collider, this multifaceted program of precision measurements and searches for new particles and rare or forbidden phenomena will give  access to high energy scales via loop diagrams or to tiny couplings via mixing or forbidden processes. Searches for dark matter, heavy neutrinos and other feebly coupled particles provide real opportunities for direct discovery.  The reduced uncertainties will also improve our knowledge of the fundamental parameters of the SM. The strong, weak and electromagnetic coupling constants at the weak scale will be measured with uncertainties reduced by up to two orders of magnitude. 

 Ultimately, more energy is required. The hadron collider would increase the kinematic reach for direct production of new particles by a factor of five with respect to HL-LHC.  A 100~TeV\, collider would provide over 10 billions of Higgs bosons and a precise measurement of the Higgs self-coupling. 
The unprecedented samples of e.g., bottom quarks, top  quarks, and W and Z bosons would allow further ``intensity frontier" probes of new physics. 

The synergy and complementarity between the FCC-ee and FCC-hh programs are stunning. Just considering the Higgs sudies: (i) FCC-ee measures in an absolute manner the HZZ coupling and the Higgs boson width, which in turn transform into absolute determinations the coupling ratios measured previously at LHC, and then at FCC-hh; (ii) FCC-ee measures the top electroweak coupling ttZ, which is needed to turn the ttH/ttZ measurement at FCC-hh to an absolute ttH measurement;  (iii) the  ttH measurement is essential for the determination of the Higgs self-coupling; finally, (iv) the FCC-ee will measure precisely the large couplings that are difficult to access at the hadron colliders ($\rm b\bar{b}$, $\rm c\bar{c}$, $gg $), while the FCC-hh, will profit from the huge data sets collected to precisely measure the $\mu\mu$, $\gamma\gamma$, and ${\rm Z} \gamma$ rare decays. 



\begin{figure}[hbtp]
\centering
\resizebox{0.9\textwidth}{!}{\includegraphics{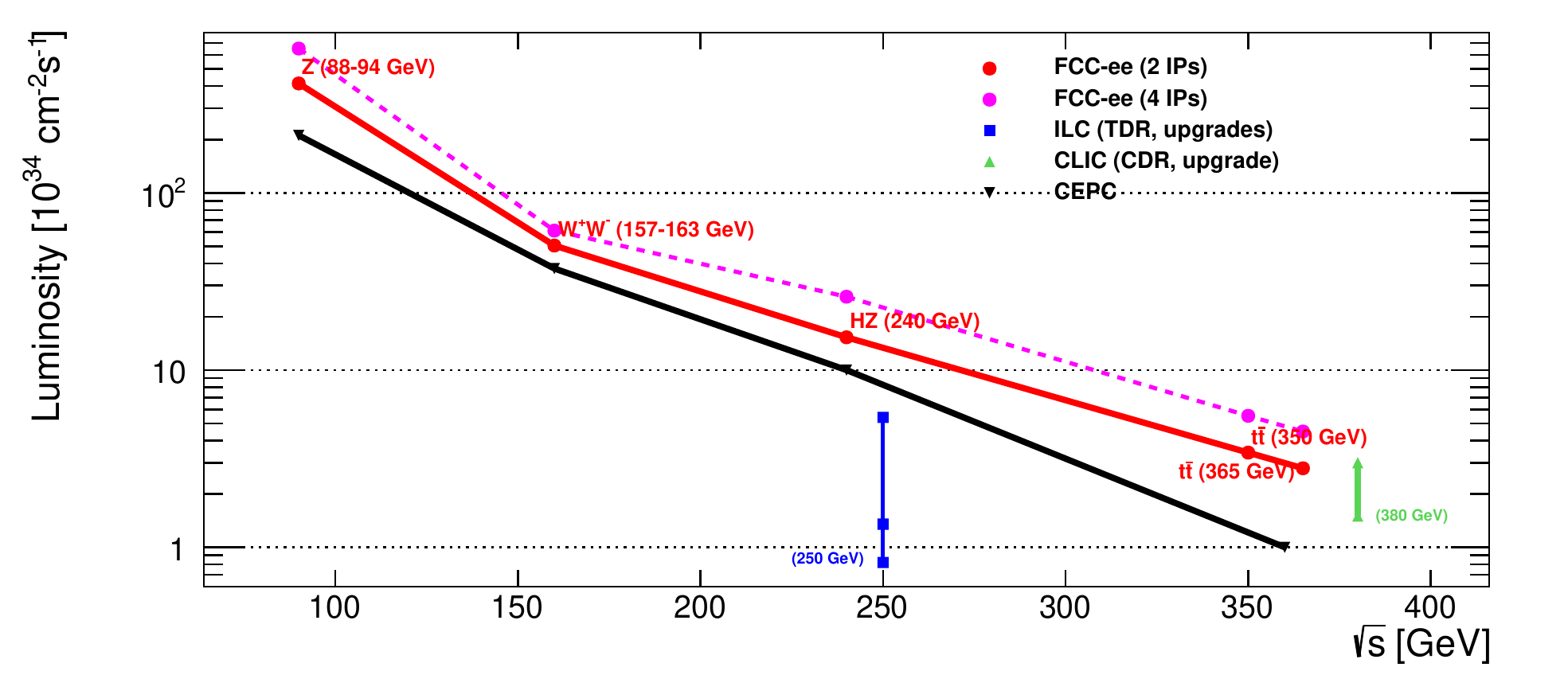}}
\caption{
Potential instantaneous luminosity versus center-of-mass energy for FCC-ee.
}\label{fig:runandx}
\end{figure}

\begin{figure}[hbtp]
\centering
\resizebox{0.6\textwidth}{!}{\includegraphics{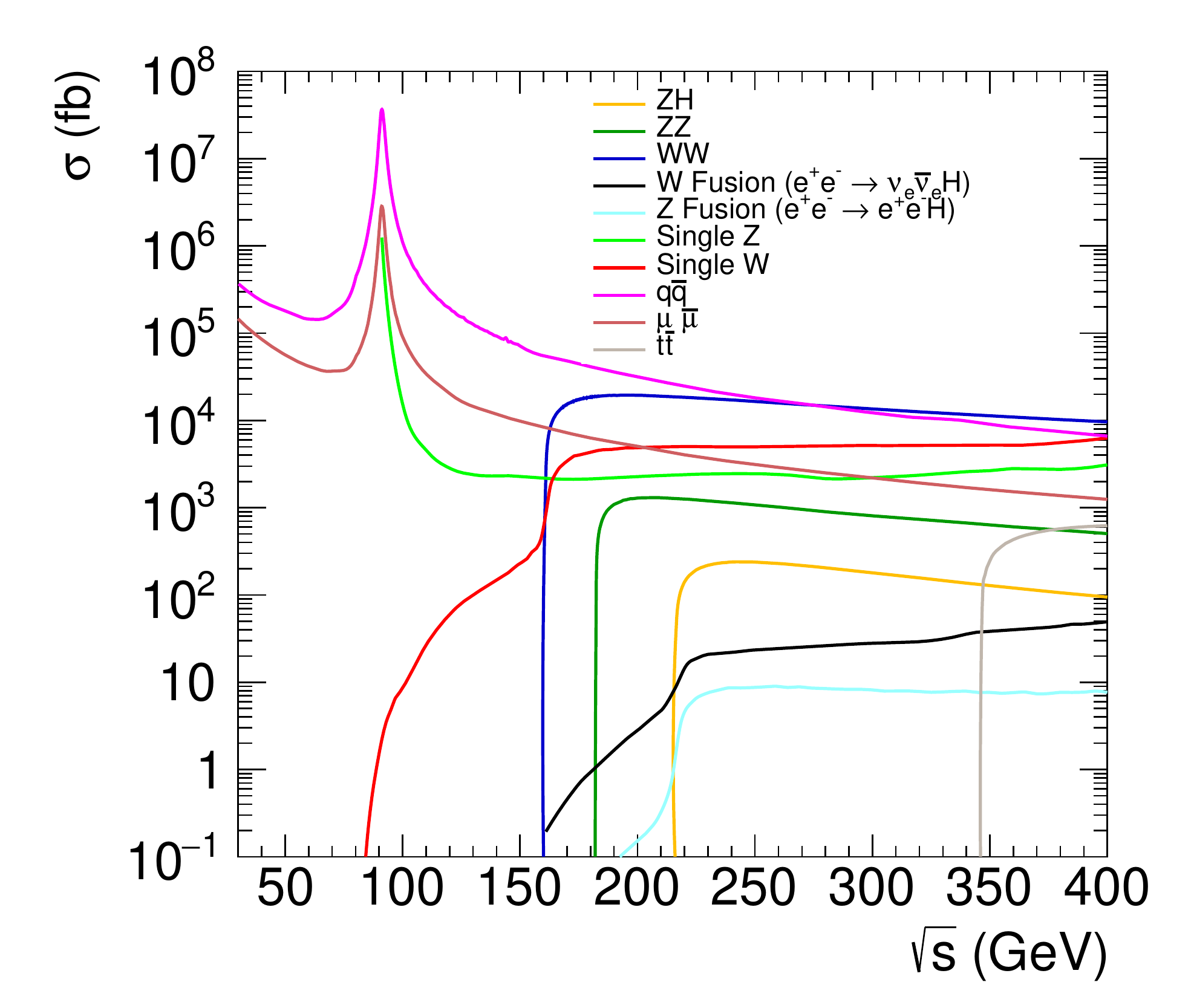}}
\caption{
Cross sections for various processes in $\ee$ collisions versus center-of-mass energy.
}\label{fig:runandx2}
\end{figure}

It is compelling to consider the sheer sizes of the samples of W and Z bosons, bottom and charm quarks, light quarks and gluons, tau leptons, top quarks and other ``intensity frontier'' samples, that the huge luminosity at low energies the first stage, FCC-ee, will provide. Predicted instantaneous luminosities such as those in Fig.~\ref{fig:runandx}, together with the production cross sections in Fig.~\ref{fig:runandx2}, lead to event samples similar to those listed in Table~\ref{tab:fcceesamplesize}. 
The collider is designed from the early stage to provide a remarkable 100 keV precision on the center-of-mass determination  by  resonant depolarization~\cite{Blondel:2019jmp}, which is central to the program of high precision measurements. 
In addition, these large samples can be easily multiplied by a factor of nearly two by implementing four instead of two interactions points (IPs), thus allowing a more complete exploitation of the multiple physics opportunities with varied detector designs. 
%

The US now has the opportunity to continue to take a leading role in the search for answers to these fundamental questions by enthusiastically collaborating with our international partners in the FCC program.
The FCC program builds on the very productive collaborations developed in the LHC era among the world's high energy physics communities and on CERN's accumulated experience and existing accelerator infrastructure, in a successful 70-years-old international organization.  A strength is the stable funding provided by the CERN member states.
A set of CDRs~\cite{FCC-ee-accelerator,Abada:2019lih,Benedikt:2018csr} exist with a well developed accelerator design, physics program, and straw-man detectors capable operating in the FCC experimental environment. The basic vision of the FCC  matches the vision endorsed by the European Strategy Group~\cite{ESPP}.
 The quality and timely realization of this exciting physics program will be strongly enhanced by substantial scientific and financial participation from the US and other global partners of CERN.

In this white paper, we summarize the extensive work elaborating the exciting physics potential of such a circular electron-positron collider.  We start in Chapter~\ref{sec:iaccelerator} with
a discussion of the R\&D needed for the accelerator infrastructure and the role the US could play, and summarize the existing work on the accelerator design.  Chapter~\ref{sec:higgs}  summarizes studies of its sensitivity to the properties of the Higgs boson.
Chapter~\ref{sec:ewk} discusses the unprecedented precision that the enormous integrated luminosity collected at the Z resonance and W pair production threshold will give to our knowledge of these bosons. In Chapter~\ref{sec:top}, the improvement beyond existing measurements of the properties of the top quark is detailed, and in Chapter~\ref{sec:bsm}, the sensitivity to direct searches for new particles, including dark matter and Heavy Neutral Leptons, are summarized.  Chapter~\ref{sec:qcd} discusses the improved understanding of the strong force enabled by FCC-ee.  Chapter~\ref{sec:flavor} discusses the probes of very high energy scales accessible to study exploiting very large samples of heavy quarks (excluding top) and leptons.  Chapter~\ref{sec:fcchh}  outlines the proposals for a long-term future physics program with p-p collisions at the energy frontier enabled by the resulting infrastructure.  Chapter~\ref{sec:instrumentation}  describes the R\&D on detectors needed to realize systematic uncertainties smaller than the statistical and theoretical uncertainties and the role the US could play in their development.  The paper ends with a summary and a vision for the path forward.

\begin{table}[ht!]
\centering
\caption{\small First five columns: Baseline FCC-ee operation model, listing the center-of-mass energies, instantaneous luminosities for each interaction point (IP), integrated luminosity per year summed over the 2 IPs corresponding to 185 days of physics per year and 75\% efficiency. As a conservative measure, the yearly integrated luminosity is further reduced by 10\% in this table and in all physics projections. The total luminosity distribution is set by the physics goals which in turn set the run time at each energy. The luminosity is assumed to be half the design value for commissioning new hardware during the first two years at the Z pole and in the first year at the ${\rm t\bar t}$ threshold. The sixth and last column is not part of the baseline FCC-ee operation model, but indicates possible numbers for an additional run at the H resonance, to investigate the electron Yukawa coupling. \vspace{0.25cm} 
\label{tab:fcceesamplesize}}
\renewcommand\arraystretch{1.3}
\resizebox{\textwidth}{!}{%
\begin{tabular}{l|c|c|c|c|c|c|c}
\hline
Working point & Z \small {years~1-2} & Z, {\small later} & WW & HZ & \multicolumn{2}{|c|}{${\rm t\bar t}$  } & ($s$-channel H)  \\ \hline
$\sqrt{s}$ {\footnotesize (GeV)} & \multicolumn{2}{|c|}{88, 91, 94} & 157, 163 & 240 & {\small 340--350} & 365 & $\rm m_H $\\ \hline
{\small Lumi/IP {\footnotesize ($10^{34}\,{\rm cm}^{-2}{\rm s}^{-1}$)}} & 115 & 230 & 28 & 8.5 & 0.95 & 1.55 & (30) \\ \hline
{\small Lumi/year {\footnotesize (${\rm ab}^{-1}$, 2 IP)}} & 24 & 48 & 6 & 1.7 & 0.2 & 0.34 & (7) \\ \hline
Physics goal {\footnotesize (${\rm ab}^{-1}$)} & \multicolumn{2}{|c|}{150} & 10 & 5 & 0.2 & 1.5& (20) \\ \hline
Run time {\footnotesize (year)} & 2 & 2 & 2 & 3 & 1 & 4 & (3) \\ \hline
 & \multicolumn{2}{|c|}{} & & $10^6$ HZ + & 
   \multicolumn{2}{|l|}{$10^6$ ${\rm t\bar t}$} & \\
Number of events &  
\multicolumn{2}{|c|}{$5\times 10^{12}$ Z} & $10^8$ WW & 25k WW $\to$ H & \multicolumn{2}{|l|}{$+200$k HZ} & (6000) \\
 & \multicolumn{2}{|c|}{} & &  &  \multicolumn{2}{|l|}{$+50$k\,${\rm WW}\to {\rm H}$} &  \\
\hline
\end{tabular}
}
\end{table}

\section{Accelerator}
\label{sec:iaccelerator}
\textcolor{red}{Editors: T.~Raubenheimer and F.~Zimmermann }

The Future Circular Collider is a set of high energy colliders to provide a century of physics.  The critical infrastructure is a $>$90 km tunnel located close to CERN and Geneva, Switzerland that would support a high luminosity and high precision lepton collider to study the Higgs and Electroweak physics, a high energy hadron collider that would probe the highest energies with a possible lepton-hadron collider that could run concomitantly.   Following proposals in 2010--2012, a formal study of the FCC hadron and lepton colliders was launched in 2014, and in 2018 conceptual designs were developed for the main collider options  \cite{FCC-ee-accelerator,Benedikt:2018csr,Abada:2019lih}.  

The conceptual designs were provided as input to the 2019/2020 Update of the European Strategy on Particle Physics \cite{ESPP}, which endorsed a program of FCC-ee followed by FCC-hh with the statements: ``An electron-positron Higgs factory is the highest-priority next collider. For the longer term, the European particle physics community has the ambition to operate a proton-proton collider at the highest achievable energy." and “Europe, together with its international partners, should investigate the technical and ﬁnancial feasibility of a future hadron collider at CERN with a center-of-mass energy of at least 100~TeV and with an electron-positron Higgs and electroweak factory as a possible ﬁrst stage. Such a feasibility study of the colliders and related infrastructure should be established as a global endeavor and be completed on the timescale of the next Strategy update.''  To address this need, in summer 2021, the Future Circular Collider Feasibility Study was launched \cite{FCCFS1, FCCFS2} to really examine the challenges of building and operating FCC-ee.  

The FCC-ee builds on 60 years of operating colliding beam storage rings. It takes advantage of advanced concepts demonstrated at VEPP-4M, DA$\Phi$NE, the SLAC and KEK B-factories,
and the present SuperKEKB. 
All critical parameters have been demonstrated in operating accelerators.  The design is robust and will provide high luminosity over the desired center-of-mass energy range from 90 to 365~GeV. The FCC-ee is also the most sustainable of all Higgs and electroweak factory proposals, in that it implies the lowest energy consumption for a given value of total integrated luminosity \cite{RevModPhys.93.015006, naturefcc}.  These features of FCC-ee along with those of the FCC-hh hadron collider (discussed further below) have attracted a large number of collaborators from around the world. 

The FCC-ee collider has separate rings for electrons and positrons with a full-energy top-up booster ring.  All three rings, the e$^+$ and e$^-$ collider rings and the full-energy booster, will be located in the same tunnel.  The design is optimized for a maximum of 50 MW of synchrotron radiation power per collider ring across the operating energy range.  An asymmetric interaction region (IR) layout limits the photon energies of the synchrotron radiation (SR)  emitted by the incoming beams over the last 500 m upstream of the IP towards detectors, and, at the same time, generates the large crossing angle of 30 mrad \cite{koide} at the IP’s.  Finally, the crab-waist technique is used to optimize the luminosity.  The crab-waist collision scheme was developed at DA$\Phi$NE \cite{Raimondi:1182932,PhysRevLett.104.174801}, and the specific ``virtual'' crab waist optics first developed for the FCC-ee
\cite{koide}  
is successfully being used at the SuperKEKB B-factory since 2020.

\begin{figure}[htbp]
\centering
\resizebox{0.7\textwidth}{!}{%
  \includegraphics{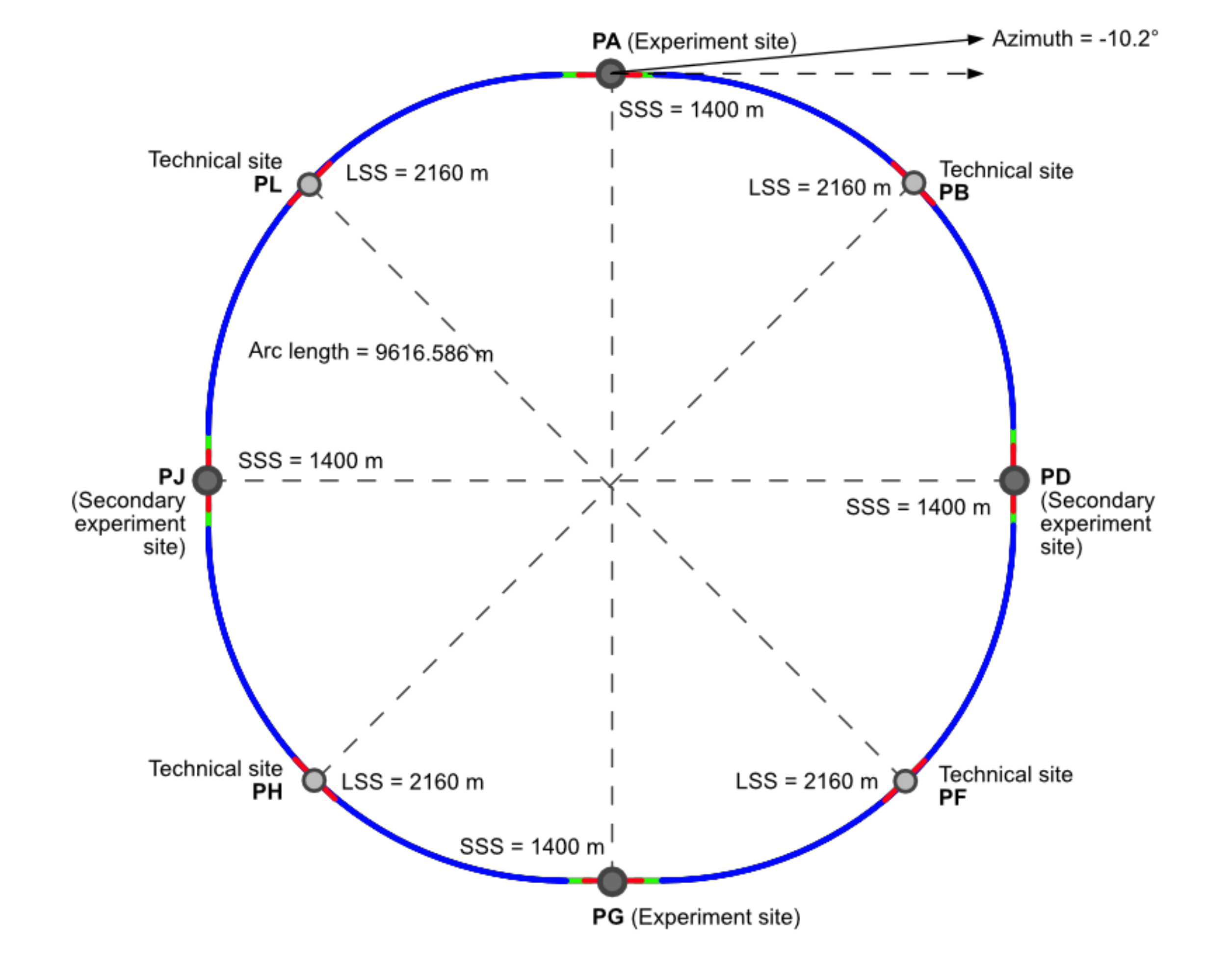}
}
\caption{The new FCC layout referred to as PA31-1.0; four possible experiments could be located at PA, PD, PG, and PJ while RF stations would be located PH and PL and injection/extraction and collimation could be located in PB and PF straights.}
\label{fig:newlayout}       
\end{figure}

Since the CDR, detailed site evaluations have led to a modified configuration of the FCC layout with a slightly smaller circumference of 91 km rather than 98 km and 8 access shafts instead of 12, and with a four-fold superperiodicity \cite{agapov_ilya_2021_5643134,ecfa-NL-2021}. The FCC-ee design is now being developed for either 2 or 4 symmetric IP’s located at four of the access points and with RF, collimation, and injection/extraction occupying the other 4 straight sections (Fig. \ref{fig:newlayout}).  The baseline configuration is based on 400 MHz RF systems
with Nb/Cu cavities, similar to those 
used at LEP and the LHC but capable of sustaining higher gradient, which are augmented by an 800 MHz RF system with bulk-Nb cavities 
to reach \ttbar operation.
The layout of the FCC-ee pre-injector complex, consisting
of a warm copper linac and a damping ring, 
has been re-optimized,
now allowing for positron production with a primary electron energy of at least 6~GeV. 
The rate of positrons required, e.g., for top-up injection in the FCC-ee collider, is comparable to the production rates achieved at the SLC and at SuperKEKB.


The operation model has the FCC-ee collider first operating at 91~GeV to study the Z boson with very high luminosity, and the lowest number of single-cell 400 MHz RF cavities. For this early operation up to at least 160~GeV there is a single RF section, to ensure that the centre of mass energies are identical at all interaction points. Beam energies are calibrated with sub-ppm precision using resonant depolarization~\cite{Blondel:2019jmp}. The high beam current at the Z will allow for a rapid beam conditioning of the vacuum system.  
Then an upgrade of the baseline 400 MHz RF system through the installation of multi-cell 400 MHz RF cavities would allow the beam energy to increase to 160~GeV to study the W$^\pm$.  An additional RF upgrade with a minor reconfiguration of the beam lines in the RF region will allow a further energy increase to 240~GeV to study the Higgs.  Finally, the addition of an 800 MHz RF system in the second RF region would allow the collider to reach $>$350~GeV for ${\rm t}\bar{\rm t}$ production. An example of a 15 year run plan is illustrated in Table\,\ref{tab:fcceesamplesize}.
Different operation models are conceivable, e.g., starting with Higgs production at 240~GeV, which would require a subsequent removal and later re-installation of a large number of RF cryomodules, and may imply  additional downtime periods for disassembly and reassembly,
but which, on the other hand, might facilitate initial operation at lower beam current and lower luminosity than foreseen on the Z pole.

At this time, in the middle of the Feasibility Study, there is not a formal update of the FCC-ee collider parameters although, as described, progress has been made in identifying a possible site and modifying the design developed in the CDR to the site.  The present interim parameters are listed in Table\,\ref{tab:param}.  The circumference of the rings changes slightly (10 cm) as the energy is increased from 160~GeV to $>$200~GeV due to the different path length of the separated versus shared rf cavities.  Detailed optimization is still in progress, and it is expected that the parameters such as the RF frequencies, tunes, the charge per bunch, and the bunch spacing will all evolve as the design is developed.

\begin{table}[htbp]

\centering
\caption{FCC-ee machine parameters for each beam energy with PA31-1.0 and 4 IPs \cite{KoideEPJ}, where ``(SR)'' indicates beam parameters 
determined by regular synchrotron radiation without collisions, 
and ``(BS)'' includes the additional effect of 
beamstrahlung due to the beam-beam interaction.\vspace{0.25cm}}
\label{tab:param}  
\resizebox{\textwidth}{!}{%
\begin{tabular}{l|c|c|c|c|c}
\toprule
Beam energy&[GeV]& 45.6 & 80 & 120 & 182.5 \\
\midrule  
Layout & & \multicolumn{4}{c}{PA31-1.0}\\
\# of IPs & & \multicolumn{4}{c}{4}\\ 
Bending radius of arc dipole &[km]& \multicolumn{4}{c}{9.937} \\  
SR power / beam &[MW]&\multicolumn{4}{c}{50}\\ 
Circumference & [km] & \multicolumn{2}{c|}{91.174117}&\multicolumn{2}{c}{91.174107}\\
Energy loss / turn &[GeV]&  0.0391 & 0.370 & 1.869 & 10.0 \\
Beam current&[mA]& 1280 & 135 & 26.7 & 5.00 \\ 
Bunches / beam & & 9600 & 880 & 248 & 36 \\ 
Bunch population &[$10^{11}$]&  2.53 & 2.91 & 2.04 & 2.64 \\
Horizontal emittance $\varepsilon_x$ &[nm]& 0.71 & 2.16 & 0.64 & 1.49\\ 
Vertical emittance $\varepsilon_y$ &[pm]&  1.42 & 4.32 & 1.29 & 2.98 \\
Arc cell & & \multicolumn{2}{c|}{Long 90/90} & \multicolumn{2}{c}{Short 90/90} \\ 
Momentum compaction $\alpha_p$ &[$10^{-6}$]& \multicolumn{2}{c|}{28.5} & \multicolumn{2}{c}{7.33}\\
Arc sextupole families  & & \multicolumn{2}{c|}{75}&\multicolumn{2}{c}{146} \\ 
$\beta^*_{x/y}$ &[mm] & 150 / 0.8 & 200 / 1.0 & 300 / 1.0&1000 / 1.6\\ 
Transverse tunes/IP $Q_{x/y}$ & & \multicolumn{2}{c|}{53.563 / 53.600}& \multicolumn{2}{c}{100.565 / 98.595}\\ 
Energy spread (SR/BS) $\sigma_\delta$ &[\%]&  0.039 / 0.130 & 0.069 / 0.154 & 0.103 / 0.185 & 0.157 / 0.229 \\
Bunch length (SR/BS) $\sigma_z$ &[mm]& 4.37 / 14.5 & 3.55 / 8.01 & 3.34 / 6.00 & 2.02 / 2.95 \\  
RF voltage 400/800 MHz & [GV] & 0.120 / 0 & 1.0 / 0 & 2.08 / 0 & 4.0 / 7.25\\
Harmonic number (400 MHz) & & \multicolumn{4}{c}{121648} \\
RF frequency (400 MHz) & MHz & \multicolumn{2}{c|}{399.994581} & \multicolumn{2}{c}{399.994627}\\
Synchrotron tune $4Q_s$ & & 0.0370 & 0.0801 & 0.0328 & 0.0826\\ 
Long. damping time &[turns] & 1168 & 217 & 64.5 & 18.5\\
RF acceptance &[\%]& 1.6 & 3.4 & 1.9 & 3.1 \\ 
Energy acceptance (DA) &[\%]& $\pm1.3$ & $\pm1.3$ & $\pm1.7$ &-2.8 +2.5\\ 
Beam-beam $\xi_x/\xi_y$\footnote{including hourglass.} & & 0.0040 / 0.152 & 0.011 / 0.125 & 0.014 / 0.131 & 0.096 / 0.151\\ 
Luminosity / IP&[$10^{34}$/cm$^2$s] & 189 & 19.4 & 7.26 &1.33\\
Lifetime (q + BS) & [sec] &  \multicolumn{2}{c|}{--} & 1065 & 2405\\ 
Lifetime (lum) & [sec] &  1089 & 1070 & 596 & 701\\
\bottomrule
\end{tabular}
}
\end{table}

The layout of the hadron collider FCC-hh has been adapted to be consistent with the optimized lowest-risk placement.
The experimental caverns for the two primary and two secondary hadron collision points coincide with those of the two or four IP's of the FCC-ee lepton collider, respectively. An optics solution has been developed which would bring the hadron collider collision exactly, i.e.~also transversely, on top of the e$^+$e$^-$ collision point. Equal IP locations minimize the size of the experimental caverns and 
might allow for a modular evolutionary detector design, where parts of the magnet system of the FCC-ee experiments could be reused for the FCC-hh detectors \cite{mannelliALPHA}.
The total proton-proton luminosity production of FCC-hh over 25 years of operation
is expected to exceed 30 ab$^{-1}$. 

The key technological challenge for FCC-hh is the 
design optimization, feasibility demonstration and cost-effective production of 
the high-field accelerator magnets. 
In addition to the use of Nb$_{3}$Sn superconductor, 
which is brittle 
and limited to a maximum field of about 16 T, 
the time scale and proposed staging of the FCC 
allow the parallel pursuit of alternative options based on 
high-temperature superconductor (HTS), which might enable higher fields,
operation at elevated temperature (with 
reduced power consumption for cryogenics), or reduced cost. 

Over the past 7 years, US institutes have made numerous  contributions to the FCC design effort. 
For example, JLAB has built a prototype five-cell 800 MHz bulk Nb cavity, required for the FCC-ee ${\rm t}\bar{\rm t}$ operation, whose actual  
performance exceeded the FCC design 
specifications \cite{Marhauser:2018jxn}.  
Accelerator specialists from SLAC have 
contributed, among others, to the development of the FCC-ee machine detector interface, e.g.~\cite{Boscolo:2021dxi}, to impedance assessments, to the optics design of lepton and hadron colliders, and to the FCC-ee study coordination, while experts at FNAL, Cornell, and the University of New Mexico contribute to essential studies of FCC-ee beam selfpolarization, 
e.g.~\cite{PhysRevAccelBeams.19.101005,Blondel:2019jmp,Blondel:2021zix}. 
Extremely impactful is the US effort on high-field magnets. The US D.O.E.'s Magnet Development Program \cite{usmdp2016} aimed (or aims) at demonstrating the feasibility of 16 T magnets based on Nb$_3$Sn conductor, 
as considered in the FCC-hh conceptual design \cite{Benedikt:2018csr}, 
and at assessing the feasibility of accelerator magnets based on HTS materials, that might enable  
potentially higher fields.
As a result of this activity, 
at FNAL in 2019 a 15 T cos$\theta$ Nb$_3$Sn dipole short-model demonstrator \cite{zlobin-napac19} reached a field of 14.1 T at 4.5 K \cite{zlobin2020test14T}, which in 2020 could be further increased to 
14.5 T  at 1.9 K.  
US magnet experts also optimized magnet design concepts, e.g.~\cite{caspi2013canted}. 
Higher field is facilitated by a higher-quality conductor. Advanced US Nb$_3$Sn wires with Artificial Pinning Centers (APCs) produced by two different teams (FNAL, Hyper Tech Research Inc., and Ohio State; and NHMFL, FAMU/FSU) reached the target critical current density for FCC of 1500 A/mm$^2$ at 16 T \cite{uswire1, uswire2}, which is 50\% higher than for the HL-LHC wires. The APCs allow for better performance; they decrease magnetization heat during field ramps, improve the magnet field quality at injection, and reduce the probability of flux jumps \cite{xu2014refinement}. 

The FCC collaboration engaged in the FCC Feasibility Study spans the world and has engaged $>$145 institutions, including at least ten universities in the US. The US government and the US D.O.E.~joined the FCC effort through an agreement in December 2020 which also covers some participation by the US National Laboratories.  To support the Feasibility Study, CERN is investing roughly 100 M CHF for technology and site development.  Additional support comes from the European Union and the FCC collaborators.  The technology development is focused on both value engineering studies such as developing a complete 1/2 cell of the arc lattice and vacuum system as well as advanced technology development such as high efficiency RF sources that would reduce the FCC-ee baseline energy consumption. 
In parallel, a CERN high-field magnet development program, funded separately and at a comparable level, aims at  developing the magnet technology required for the FCC-hh.

\section{Higgs physics}
\label{sec:higgs}
\textcolor{red}{Editors: G. Bernardi, E. Brost,  D. d'Enterria, C. Grojean,  P. Janot, M. Klute, C.~Paus}

%

With the discovery of the Higgs boson at the LHC in 2012~\cite{higgscms,higgsatlas} its mass (around $125~\GeV$) was known. It is therefore possible to design a Higgs factory to study its properties in detail quite similar to what was done for the W and Z bosons in the LEP project after these two gauge bosons were discovered in 1983 at the $\mathrm{Sp\bar{p}S}$. Studies of the various Higgs boson decay channels are sensitive to tree- or quantum-level corrections to the corresponding couplings. While the LHC will eventually reach a precision of order 5\% to the heaviest particle couplings (albeit with some model assumptions), FCC-ee will start a new era of precision, removing any model dependence, reaching subpercent level precision for most couplings, and adding access to several lighter Higgs Yukawa couplings. Typically, new particles at the $1~\TeV$ mass scale lead to deviations in the couplings at the 5\% or smaller level, namely any deviation $\delta g_{_{\rm HXX}}$ relative to the SM value $g_{_{\rm HXX}}^{_{\rm SM}}$ can be approximately translated into BSM scale limits:
$\rm \small \Lambda\gtrsim (1\,TeV)/\sqrt{(\delta g_{_{\rm HXX}}/g_{_{\rm HXX}}^{_{\rm SM}})/5\%}$. 
The expected 0.15\% uncertainty for the most precise coupling $g_{\rm HZZ}$ will thus set the most stringent bounds, $\Lambda\gtrsim~7$~TeV, on new physics coupled to the scalar sector of the SM. 

\begin{figure}[ht]
\centering
\includegraphics[width=\textwidth]{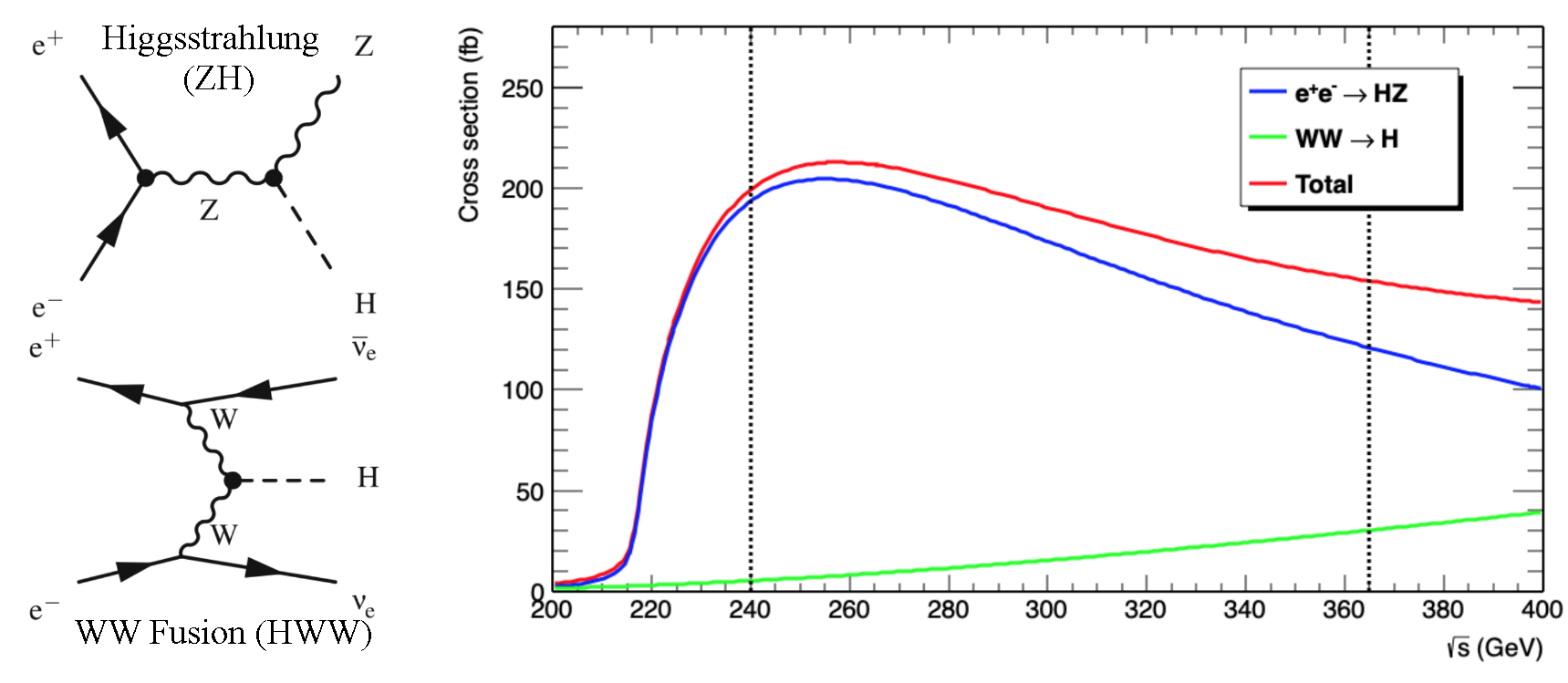}
\caption{\label{fig:higgs-production}
 Lowest order Feynman diagrams for WW fusion and Higgsstrahlung (left) and the corresponding cross sections versus the center-of-mass energy  per production process along with their sum (right). The default running scenarios at 240~GeV and 365~GeV are indicated with dashed lines. Figure from~\cite{Azzurri:2021nmy}.}.
\label{fig:couplings}
\end{figure}

The FCC-ee program  offers a unique opportunity to measure the Higgs couplings. The two most important Higgs production processes are Higgsstrahlung (ZH), $\epem \to \mathrm{ZH}$, and WW fusion to a Higgs boson (WWH), $\epem \to \mathrm{H}\nu_\text{e}\bar{\nu}_\text{e}$. The lowest order Feynman diagrams for these two production mechanisms are displayed in Figure~\ref{fig:higgs-production} (left) together with their corresponding cross sections versus the center-of-mass energy (right). The predictions include initial state radiation~\cite{ISR} using the HZHA program~\cite{Altarelli:1996gh} and the small interference term present in the WWH final state diagrams. Given the cross sections and the planned FCC-ee running scenario, and with two interaction points~\cite{Blondel:2021ema}, over a million ZH events and almost one hundred thousand WWH events will be collected at various center-of-mass energies. These numbers drive the statistical uncertainties for the following studies.


Our goal for this report is not to lay out the details of all studies possible with these large data samples, but to pick out the studies that demonstrate the key capabilities of FCC-ee in terms of Higgs boson physics that have been documented. In the following we will present the sensitivity for the  Higgs cross section, mass, and width, and then summarize the status of the projected precision on the various coupling constants of the Higgs bosons to bosons and fermions, including the Higgs boson self-coupling.

\subsection{Production cross Sections, mass, and width}

The FCC-ee running scenario at $\sqrt{s}=240~\GeV$ was optimized as a tradeoff between ZH production rate and luminosity. A feature unique to lepton colliders is the measurement of the Higgs boson properties using the recoil system in the ZH production mode. The well-determined four-momenta of the initial state leptons and the fully reconstructed Z boson (recoil system) in the final state allow clean recovery of the  Higgs boson kinematics independent of the Higgs boson decay mode. This cannot be accomplished at a hadron collider (or in the WWH production mode) and gives access to the total ZH cross section. Assuming perfect efficiency and no background, the statistical precision is as small as one per mill.

The various cross section measurements to exclusive Higgs final states are used to determine the corresponding branching fractions. The expected uncertainties are based on studies in Ref.~\cite{FCC-ee-accelerator} and summarized in Table~\ref{tab:higgs-branching}. These fundamental measurements will serve in Section~\ref{subsec:higgs-couplings} as an input to the determination of the Higgs boson couplings.

The primary tool used in the measurement of the Higgs boson mass is the already introduced recoil system in the ZH production mode of the Higgs boson. Without initial state radiation and neglecting the beam energy spread, the Higgs boson mass is given by the mass of the difermion recoil system: $ m_\text{recoil} = s + m_{f\overline{f}} - 2\sqrt{s} (E_{f} + E_{\overline{f}}) $. Initial ZH studies utilized the Z boson decays to dimuons because it has excellent energy resolution and very low background. Studies based on DELPHES simulations~\cite{FCC-ee-accelerator, Azzurri:2021nmy} show that rather conservatively a statistical precision of $6-9~\MeV$ will be reached on the Higgs boson mass, depending on the choice of the tracking detectors. The energy calibration of the collider, expected to reach a precision below $100~\keV$~\cite{Blondel:2019jmp}, as well as the muon momentum measurement, are the essential experimental uncertainties to be controlled. Given the clean environment and the large background from single and double vector boson production, it is plausible that the experimental uncertainties can be controlled well within the required precision.

Adding the dielectron decays of the Z boson into the analysis could decrease the uncertainty on the Higgs boson mass to  $4~\MeV$~\cite{FCC-ee-accelerator}. Finally -- based on estimates derived from similar measurements with the CMS detector in a much harsher environment -- the addition of all exclusive final states decays using energy constraints or the inclusion of the direct measurement of the Higgs boson reconstructed mass should allow an $m_\mathrm{H}$ measurement with a precision of $2.5~\MeV$, and maybe even $2~\MeV$, if the magnetic field is increased from $2~\text{T}$ to $3~\text{T}$~\cite{FCC-ee-accelerator}.

The use of the cross-section lineshape for ZH production at threshold -- similar to the WW threshold scan at LEP -- to measure the Higgs boson mass and width have been proposed but are not included in the default running plan for the FCC-ee. Options like this underline the fact that there is further room for improvement should that be needed.

Additional studies will be necessary, many of which have not yet been performed at the most realistic level and are clearly outside of the scope of this document.  However the studies documented so far~\cite{FCC-ee-accelerator} support that the Higgs boson mass can be measured at the few-$\MeV$ level, and its width at the percent level.

\begin{table}{}
\centering
\caption{\label{tab:higgs-branching} Expected uncertainties in percent (\%) for the Higgs branching ratios times the total cross section, as measured in the listed ZH and WWH production modes~\cite{FCC-ee-accelerator}.\vspace{0.25cm}}
\tabcolsep=4.5mm
\begin{tabular}{l|c|c|c|c}
\hline
$\sqrt{s}$ & \multicolumn{2}{|c|}{$240~\GeV$} & \multicolumn{2}{|c}{$365~\GeV$} \\
\hline
Int. Luminosity & \multicolumn{2}{|c|}{$5~\text{ab}^{-1}$} & \multicolumn{2}{|c}{$1.5~\text{ab}^{-1}$} \\
\hline
Channel & ZH & WWH & ZH & WWH \\
\hline
H$\to$ any & $\pm 0.5$ & & $\pm 0.9$ & \\
H$\to\text{b}\overline{\text{b}}$ & $\pm 0.3$ &  $\pm 3.1$ & $\pm 0.5$ &  $\pm 0.9$ \\
H$\to\text{c}\overline{\text{c}}$ & $\pm 2.2$ & & $\pm 6.5$ &  $\pm 10$ \\
H$\to\text{gg}$ & $\pm 1.9$ & & $\pm 3.5$ &  $\pm 4.5$ \\
H$\to\text{W}^+\text{W}^-$ & $\pm 1.2$ & & $\pm 2.6$ &  $\pm 3.0$ \\
H$\to\text{Z}\text{Z}$ & $\pm 4.4$ & & $\pm 12$ &  $\pm 10$ \\
H$\to\tau^+\tau^-$ & $\pm 0.9$ & & $\pm 1.8$ &  $\pm 8$ \\
H$\to\gamma\gamma$ & $\pm 9.0$ & & $\pm 18$ &  $\pm 22$ \\
H$\to\mu^+\mu^-$ & $\pm 19$ & & $\pm 40$ & \\
H$\to$ invisible & $< 0.3$ & & $< 0.6$ & \\
\hline
\end{tabular}
\end{table}

\subsection{Higgs Boson couplings to Standard Model particles}
\label{subsec:higgs-couplings}

Once the ZH and WWH production cross sections, and the branching fractions for each Higgs boson decay $\mathrm{H} \to X\bar{X}$ have been measured they are used to determine all other couplings using the following basic relationships:

\begin{equation}
    \sigma_{\mathrm{ZH}}\times \mathcal{B}(\mathrm{H} \to X\bar{X}) \propto \frac{g^2_\mathrm{HZZ} \times g^2_\mathrm{HX\bar{X}}}{\Gamma_{\mathrm{H}}}
\end{equation}
and 
\begin{equation}
    \sigma_{\mathrm{\mathrm{H}\nu_{e}\bar{\nu}_{e}}}\times \mathcal{B}(\mathrm{H} \to X\bar{X}) \propto \frac{g^2_\mathrm{HWW} \times g^2_\mathrm{HX\bar{X}}}{\Gamma_{\mathrm{H}}}\,.  
\end{equation}

In particular, the measurement of the ZH cross section (described in previous section) gives a model-independent measurement of $g_\mathrm{HZZ}$, which is used to normalize the measurements of other Higgs boson couplings, including those at current or future hadron colliders. This is a unique feature of low-energy $\epem$ colliders.

The expected precision on measurements of Higgs couplings in the kappa framework~\cite{LHCHiggsCrossSectionWorkingGroup:2012nn,LHCHiggsCrossSectionWorkingGroup:2013rie} are presented in  Table~\ref{tab:h-couplings}. The five columns summarize best estimates for various combinations of data from HL-LHC,  FCC-ee, and the entire FCC program labeled ``FCC-INT'' (FCC-ee, FCC-hh, and FCC-eh). The results were produced by the \textit{Higgs@FutureColliders} working group in preparation for the European Strategy Update in 2020. The HL-LHC fit \cite{Cepeda:2019klc} is performed while fixing $g_\text{Hcc}$ and $\Gamma_\text{H}$ to their Standard Model values. This table illustrates the complementarity of the Higgs physics program at lepton and hadron colliders. In the FCC-ee and HL-LHC combination, the HL-LHC makes important contributions in channels with very low production rates at lepton colliders ($\gamma \gamma$, $\mu \mu$, and Z$\gamma$) and in the determination of the top Yukawa coupling, which is out of reach of the low-energy $\epem$ program. On the other hand, FCC-ee dominates the measurement of the Higgs couplings to vector bosons and charm quarks, the latter of which is difficult at the HL-LHC. Not considered in this study are the electron Yukawa coupling, a rare process that is uniquely accessible at low-energy $\epem$ colliders such as the FCC-ee, and the Higgs self-coupling, which is accessible through loop corrections at the FCC-ee. Both are further discussed in the following subsections.

\begin{table}{}
\centering
\caption{\label{tab:h-couplings} Expected sensitivity to measurements of various Higgs boson couplings at the HL-LHC, the FCC-ee with two IPs, and the full FCC program (FCC-ee, FCC-hh, and FCC-eh), in the kappa framework. Combined sensitivities of the FCC-ee and the HL-LHC, and the full FCC and the HL-LHC programs are also presented. The HL-LHC sensitivities are evaluated \cite{Cepeda:2019klc} while fixing the charm Yukawa coupling and the total Higgs boson width to their Standard Model values, and assuming no BSM Higgs decays. All numbers are in \% and indicate the 68\% confidence level intervals. Results from the \textit{Higgs@FutureColliders} group \cite{deBlas:2019rxi}, Table adapted from \cite{Blondel:2021ema}. The self-coupling projection for the FCC-hh is updated in \cite{Mangano:2020sao}. \vspace{0.25cm}}
\tabcolsep=2.5mm
\begin{tabular}{l|c|c|c|c|c}
\hline
Collider    & HL-LHC & FCC-ee$_{240 \to 365}$  & FCC-ee   & FCC-INT & FCC-INT \\
            &        &                         & + HL-LHC &         & + HL-LHC \\

\hline
Int. Lumi (ab$^{-1}$) &   3    &    5 + 0.2 + 1.5 &    -- &  30 & -- \\
\hline
Years       &   10     & 3 + 1 + 4 &  -- &    25 & --    \\
\hline   
$g_\text{HZZ}$           (\%) & 1.5 & 0.18 & 0.17 & 0.17 & 0.16 \\
$g_\text{HWW}$           (\%) & 1.7 & 0.44 & 0.41 & 0.20 & 0.19 \\
$g_\text{Hbb}$           (\%) & 5.1 & 0.69 & 0.64 & 0.48 & 0.48 \\
$g_\text{Hcc}$           (\%) & SM & 1.3 & 1.3 & 0.96 & 0.96 \\
$g_\text{Hgg}$           (\%) & 2.5 & 1.0 & 0.89 & 0.52 & 0.5 \\
$g_{\text{H}\tau\tau}$   (\%) & 1.9 & 0.74 & 0.66 & 0.49 & 0.46 \\
$g_{\text{H}\mu\mu}$     (\%) & 4.4 & 8.9 & 3.9 & 0.43 & 0.43 \\
$g_{\text{H}\gamma\gamma}$ (\%) & 1.8 & 3.9 & 1.3 & 0.32 & 0.32 \\
$g_{\text{HZ}\gamma}$    (\%) & 11. & -- & 10. & 0.71 & 0.7 \\
$g_\text{Htt}$           (\%) & 3.4 & -- & 3.1 & 1.0 & 0.95 \\
\hline
$g_\text{HHH}$           (\%) & 50. & 44. & 33. & 3--4 & 3--4\\
\hline
$\Gamma_\text{H}$        (\%) & SM & 1.1 & 1.1 & 0.91 & 0.91 \\
\hline
\end{tabular}
\end{table}

\subsubsection{Electron Yukawa coupling}

The smallest Yukawa coupling in the SM (with zero-mass neutrinos) is that of the electron: $y_\mathrm{e} = \sqrt{2} m_\mathrm{e}/v = 2.9 \times 10^{-6}$ for $m_\mathrm{e} = 0.511 \times 10^{-3}$\,GeV. Measuring the Higgs coupling to the electron is impossible at hadron colliders because of its tiny branching fraction $\mathcal{B}(\mathrm{H}\to\epem) = 5.22 \times 10^{-9}$: the signal is overwhelmed by the Drell--Yan $\epem$ continuum, which has a cross section that is many orders of magnitude larger. Measurements at the LHC lead today to an upper bound on the branching fraction of $\mathcal{B}(\mathrm{H}\to\epem) < 3.6\times10^{-4}$ at 95\% CL, corresponding to an upper limit on the Yukawa coupling $y_\mathrm{e}\propto \mathcal{B}(\mathrm{H}\to\epem)^{1/2}$ of 260 times the SM value~\cite{Khachatryan:2014aep,ATLAS:2019old}. This result translates into a lower bound on the energy scale of any BSM physics affecting $y_\mathrm{e}$, of $\Lambda_{\textsc{bsm}} \approx v^{3/2} (\sqrt{2}m_\mathrm{e} (y_\mathrm{e}/y^\mathrm{\textsc{sm}}_\mathrm{e}))^{-1/2} \gtrsim 8.8$~TeV~\cite{Altmannshofer:2015qra}. Assuming that the sensitivity to the $\mathrm{H} \to \epem$ decay scales with the square root of the integrated luminosity, the upper bound will be $y_\mathrm{e}\lesssim 120 y^\mathrm{\textsc{sm}}_\mathrm{e}$ (\ie\ $\Lambda_{\textsc{bsm}}\gtrsim 13$~TeV) at the end of HL-LHC running.

The FCC-ee offers a unique opportunity to measure $y_\mathrm{e}$ via resonant $s$-channel Higgs production, $\epem \to \mathrm{H}$, in a dedicated run at $\sqrt{s} = m_\mathrm{H}$~\cite{dEnterria:2017dac,dEnterria:2021xij}. The signature for direct Higgs production is a small rise in the cross sections for final states consistent with Higgs decays over the expectations from SM background processes involving $\mathrm{Z}^*$, $\gamma^*$, or $t$-channel exchanges alone. Performing such a measurement is remarkably challenging for four main reasons. First, the low value of the e$^\pm$ mass leads to a tiny $y_\mathrm{e}$ coupling and correspondingly small cross section: $\sigma_\mathrm{ee\to H} \propto m_\mathrm{e}^2 = 0.57$~fb accounting for ISR $\gamma$ emission. 
Second, the $\epem$ beams must be monochromatized such that the spread of their center-of-mass (\com) energy, $\delta_{\sqrts}$, is commensurate with the narrow width of the SM Higgs boson, $\Gamma_\mathrm{H} = 4.1$\,MeV, while keeping large beam luminosities.
\begin{figure}[htpb!]
\centering
\includegraphics[width=0.47\columnwidth,height=5.5cm]{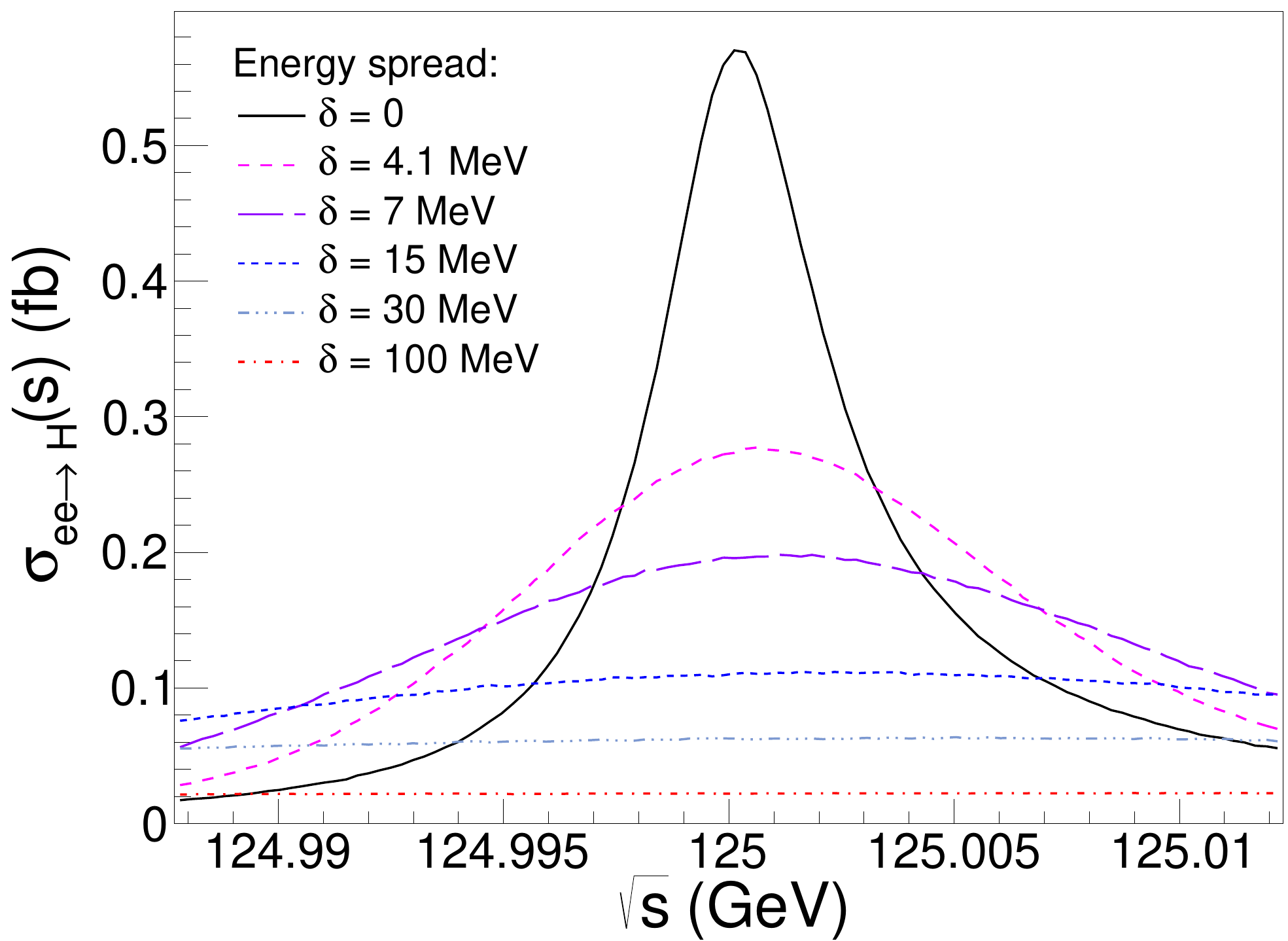}
\includegraphics[width=0.51\columnwidth]{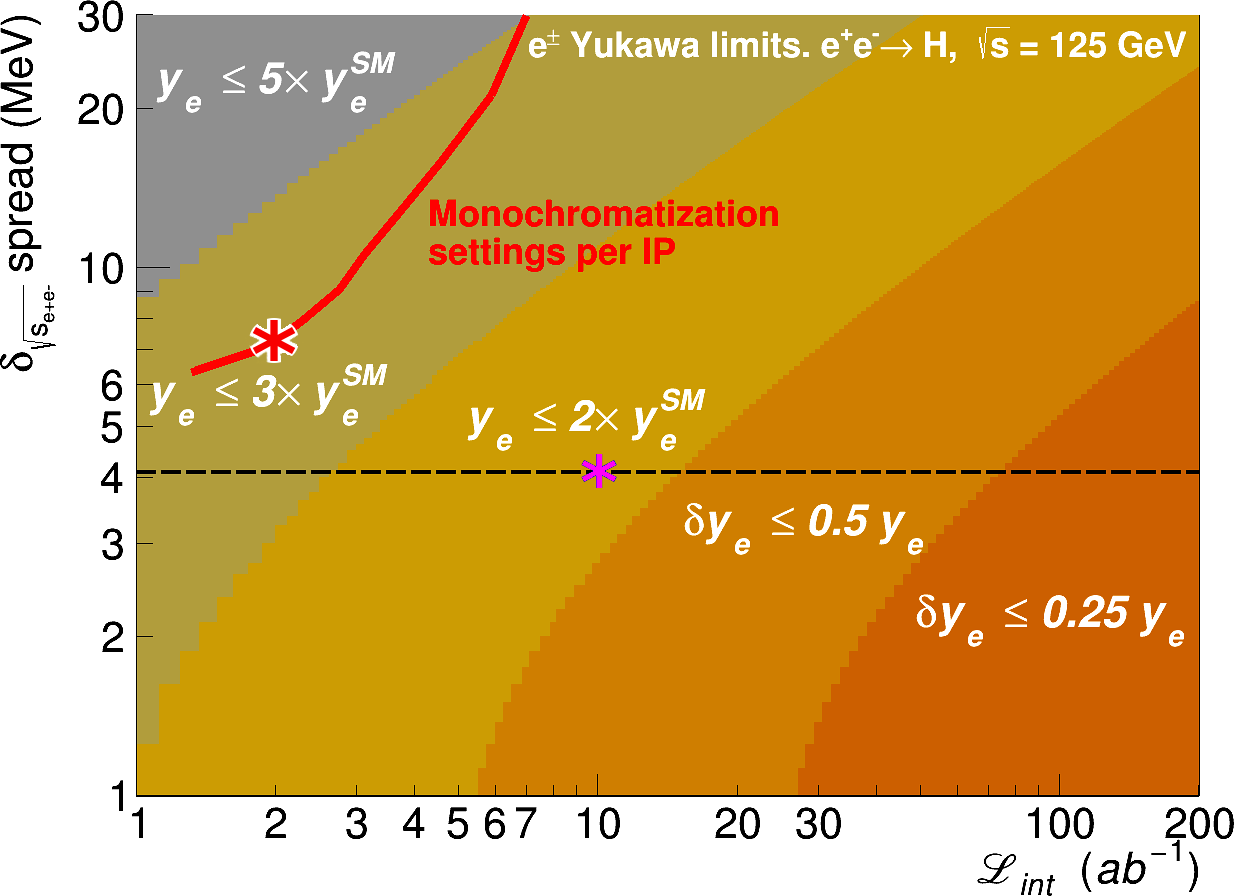}
\caption{Left: Resonant Higgs production cross section, including ISR effects, for several values of the $\epem$ \com\ energy spread~\cite{Jadach:2015cwa}.
Right: Upper limits contours (95\% CL) on $y_\mathrm{e}$. The red curve shows the range of parameters presently reached in FCC-ee monochromatization studies~\protect\cite{Zimmermann:2017tjv,ValdiviaGarcia:2019ezi}. The red star indicates the best signal strength monochromatization point in the plane and the pink star
indicates the ideal baseline point assumed in our default analysis. All results are given per IP and per year. Figures from~\cite{dEnterria:2021xij}.
\label{fig:eYukawa}}
\end{figure}
Figure~\ref{fig:eYukawa} (right) shows the Higgs lineshape for various $\delta_{\sqrts}$ values. The combination of ISR plus $\delta_{\sqrts} = \Gamma_\mathrm{H} = 4.1$\,MeV reduces the peak Higgs cross section by an overall factor of about six, down to $\sigma_\mathrm{ee\to H} = 0.28$~fb. Third, the Higgs boson mass must be known beforehand with a few-MeV accuracy in order to operate the collider at the resonance peak, which is possible but challenging for FCC-ee~\cite{Azzurri:2021nmy}. Last but not least, the cross sections of the background processes are many orders of magnitude larger than those of the Higgs decay signals. A preliminary generator-level study of eleven Higgs decay channels
identifies two final states as the most promising ones in terms of statistical significance: $\mathrm{H}\to gg$ and $\mathrm{H}\to \mathrm{W}\mathrm{W}^*\!\to \ell\nu$\,+\,2\,jets. For a benchmark monochromatization with  $\delta_{\sqrt{s}} = 4.1~\MeV$ 
and 10\,ab$^{-1}$ of integrated luminosity, a $1.3$ standard deviation signal significance can be reached, corresponding to an upper limit on the e$^\pm$ Yukawa coupling at 1.6 times the SM value: $|y_\mathrm{e}|<1.6|y^\mathrm{\textsc{sm}}_\mathrm{e}|$ at 95\% confidence level, per IP and per year~\cite{dEnterria:2021xij}.

The expected final significance of the $\sigma_{\epem\to\mathrm{H}}$ measurement, and associated 95\% CL limits on $|y_\mathrm{e}|$, derived for a benchmark $\delta_{\sqrts} = 4.1$\,MeV collision-energy spread and $\LumiInt = 10\,\mathrm{ab}^{-1}$, can be translated into any other combination of $(\delta_{\sqrts},\LumiInt)$ values achievable through beam monochromatization. Figure~\ref{fig:eYukawa} (right) shows the bidimensional maps for the
95\% CL upper limits on the electron Yukawa, as a function of both parameters. The signal significance and associated upper limits improve with the square-root of the integrated luminosity, and diminish for larger values $\delta_{\sqrts}$ following the relativistic Voigtian dependence of the signal yield on the energy spread shown in Fig.~\ref{fig:eYukawa} (left). The red curve shows the current expectations for the range of $(\delta_{\sqrts},\LumiInt)$ values achievable at FCC-ee with the investigated monochromatization schemes~\cite{Zimmermann:2017tjv,ValdiviaGarcia:2019ezi}. Without monochromatization, the FCC-ee natural collision-energy spread at $\sqrts = 125$\,GeV is $\delta_{\sqrts} \approx 50$\,MeV due to synchrotron radiation. Its reduction to the few-MeV level desired for the $s$-channel Higgs run can be accomplished by means of monochromatization, \eg\ by introducing nonzero horizontal dispersions at the IP ($D_x^*$) of \textit{opposite sign} for the two beams in collisions without a crossing angle. Compared to the baseline working point (pink star in Fig.~\ref{fig:eYukawa}, right), the signal significance for the currently best monochromatization settings, $(\delta_{\sqrts},\LumiInt) \approx (7\,\mathrm{MeV}, 2\,\mathrm{ab}^{-1})$, drops to $\mathcal{S}\approx 0.4$ standard deviations per IP and per year, and the corresponding upper bound becomes $y_\mathrm{e}\lesssim 2.5 y^\mathrm{\textsc{sm}}_\mathrm{e}$ (95\% CL) per IP and per year. Assuming two years of FCC-ee operation at the Higgs pole and combining four IPs, this would translate into a $1.2$ standard deviations significance and a $y_\mathrm{e}\lesssim 1.6 y^\mathrm{\textsc{sm}}_\mathrm{e}$ limit. Such a result, although short of evidence for $s$-channel Higgs production, is still about 100 (30) times better~\cite{Blondel:2019yqr} than that reachable at HL-LHC (FCC-hh~\cite{Benedikt:2018csr}), and implies setting a unique constraint on new physics affecting the electron-Higgs coupling above $\Lambda_{\textsc{bsm}}\gtrsim 110$~TeV.


\subsubsection{Higgs boson self-coupling}

The determination of the Higgs boson self-coupling is a major goal of the HEP community due to its relationship with the shape of the Higgs boson potential. The current HL-LHC projections estimate a precision on the self-coupling of 50\%, based on measurements of Higgs boson pair production. The  self-coupling can also be constrained at low-energy $\epem$ colliders \cite{McCullough:2013rea, Maltoni:2018ttu, DiVita:2017vrr, Blondel:2018aan, DiMicco:2019ngk} via loop corrections to single Higgs boson processes. Through the comparison of the Higgs boson production cross sections at 240 and 365~GeV, the self-coupling can be determined with greater precision than that of the HL-LHC. The combination of the HL-LHC and FCC-ee results brings that precision to 33\%. Few-percent level determination of the Higgs self-coupling requires the FCC-hh. More discussion of the synergies between the FCC-ee and FCC-hh are given in Section~\ref{sec:fcchh}. 

\subsection{Beyond the Standard Model Higgs boson searches}

The clean environment at the FCC-ee enables searches for additional Higgs bosons and non-Standard-Model decays of the Standard Model Higgs boson. Some examples are discussed in Section~\ref{sec:bsm}.

\section{Precision electroweak physics}
\label{sec:ewk}
\textcolor{red}{Editors: J. Alcaraz, A. Blondel, J. de Blas, A. Freitas, J. Gluza, M. Hildreth,  and J.~Zhu}

Precision electroweak measurements at the FCC-ee will constitute an important part of the physics program, with a sensitivity to new physics that is very broad and largely complementary to that offered by measurements of the Higgs boson properties 
and flavor observables. 
Electroweak precision observables (EWPOs)~\cite{ALEPH:2005ab, ALEPH:2013dgf} are sensitive to the existence of new weakly-coupled particles via radiative corrections or mixing,
even if these particles cannot be directly produced or observed in current experiments. Historically, the first positive evidence for the existence of a heavy top quark was obtained from $\rm B-\bar{B} $ mixing, while the global fit of the SM predictions to electroweak precision data 
led to the correct predictions of the top quark~\cite{Clarke:1994xta,Pietrzyk:1994rx} and Higgs boson~\cite{ALEPH:2005ab} masses before their discoveries in 1995 and  2012 respectively. 
These predictions were made assuming the minimal SM (three families and a unique Higgs scalar), which, following these discoveries, no longer contains any unmeasured parameters. Any significant deviation would decisively point to the existence of new physics. 

The combination of large data samples at different center-of-mass energies 
from the Z to above the top quark pair threshold 
and continuous parts-per-million control of the beam energy at the Z and WW threshold~\cite{Blondel:2019jmp} will allow the 
experimental precision of many EWPOs to be improved by 1--3 orders of magnitude. 
A summary of the main EWPOs with their expected statistical uncertainties 
and a provisional set of  
systematic uncertainties is shown in Table~\ref{tab:fccee_EWPOs} and compared to  present uncertainties. 
The improvement in statistical uncertainties at the Z is typically a factor 500. Performing the measurements with systematic accuracy matching the available statistics requires proactive design of the detectors, of the analysis techniques and tools, and considerable development of theoretical calculation techniques.  
The resulting improved precision,
as well as the increased number of observables reaching interesting precision, 
equates to increased sensitivity to 
new physics, and hence to enhanced 
discovery potential.

\begin{table}
    \centering
        \caption{Expected statistical and systematic uncertainties for selected electroweak precision measurements at FCC-ee, compared with present precision and accuracy~\cite{Blondel:2021ema, AlcarazMaestre}. These precisions and accuracies can be obtained using the run plan shown in Table~\ref{tab:fcceesamplesize}. The systematic uncertainties are initial estimates; the aim is to improve them so that they reach the same level as the statistical errors.\vspace{0.25cm}}
    \label{tab:fccee_EWPOs}
    \resizebox{\textwidth}{!}{%
    \begin{tabular}{lllll}
    \toprule
    Observable & Present &  FCC-ee & FCC-ee & Comment and dominant exp. error\\ 
               & value $\pm$ error &  Stat. & Syst. & \\ \hline
    $m_\mathrm{Z}$ (keV) & $91,186,700 \pm 2200$  & 4 & 100 & From Z lineshape scan; beam energy calibration \\
    $\Gamma_\mathrm{Z}$ (keV) & $2,495,200 \pm 2300$ & 4 & 25 & From Z lineshape scan; beam energy calibration\\
    $R_\ell^\mathrm{Z}$ ($\times 10^3$) & $20,767 \pm 25$ & 0.06 & $0.2-1.0$ & Ratio of hadrons to leptons; acceptance for leptons \\
    $\alphas(m^2_\mathrm{Z})$ ($\times 10^4$)  & $1,196 \pm 30$ & 0.1 & $0.4-1.6$ & From $R_\ell^\mathrm{Z}$ above\\
    $R_b$ ($\times 10^6$) & $216,290 \pm 660$ & 0.3 & $<60$ & Ratio of $b\bar{b}$ to hadrons; stat. extrapol. from SLD \\
    $\sigma_{\text{had}}^0$ ($\times 10^3$) (nb) & $41,541 \pm 37$ & 0.1 & 4  & Peak hadronic cross section; luminosity measurement\\
    $N_\nu$ ($\times 10^3$) & $2,996 \pm 7$ & 0.005 & 1 & Z peak cross sections; luminosity measurement\\
    $\sin^2 \theta_{\text{W}}^{\text{eff}}$ ($\times 10^6$) & $231,480 \pm 160$ & 1.4 & 1.4 & From $A_{\text{FB}}^{\mu\mu}$ at Z peak; beam energy calibration \\
    $1/\alpha_{\text{QED}}(m^2_\mathrm{Z})$ ($\times 10^3$) & $128,952 \pm 14$ & 3.8 & 1.2 & From $A_{\text{FB}}^{\mu\mu}$ off peak\\
    $A_{\text{FB}}^{b,0}$ ($\times 10^4$) & $992 \pm 16$ & 0.02 & $1.3$ & $b$-quark asymmetry at Z pole; from jet charge \\
    $A_{e}$ ($\times 10^4$) & $1,498 \pm 49$ & 0.07 & $0.2$ & from $A_{\text{FB}}^{\text{pol},\tau}$; systematics from non-$\tau$ backgrounds \\
    $m_\mathrm{W}$ (MeV) & $80,350 \pm 15$ & 0.25 & 0.3 & From WW threshold scan; beam energy calibration \\
    $\Gamma_\mathrm{W}$ (MeV) & $2,085 \pm 42$ & 1.2 & 0.3 & From WW threshold scan; beam energy calibration \\    
    $N_\nu$ ($\times 10^3$) & $2,920 \pm 50$ & 0.8 & Small & Ratio of invis. to leptonic in radiative Z returns\\
    $\alphas(m^2_\mathrm{W})$ ($\times 10^4$)  & $1,170 \pm 420$ & 3 & Small & From $R_\ell^W$\\
   \bottomrule
    \end{tabular}
    }
\end{table}

\subsection{Electroweak program around the Z pole}
The typical proposed Z-pole operation plan contains runs taken at 
three energy points: 87.7~GeV, 91.2~GeV, and 93.9~GeV. 
These points correspond to half integer spin tunes to ensure good energy calibration  by resonant depolarization.  
About half the data will be taken at 91.2~GeV. The integrated luminosity will be $2 \times 10^5$ times the integrated luminosity produced by  LEP at the Z pole, which means that $5 \times 10^{12}$ Z bosons will be available for study. Exquisite center-of-mass energy calibration and excellent detector performances will further improve the experimental accuracy. 

Z lineshape scans are expected to provide measurements of the Z boson mass $m_\mathrm{Z}$ with an accuracy of 100 keV and the Z boson width $\Gamma_\mathrm{Z}$ with an accuracy of 25 keV~\cite{Abada:2019lih, AlcarazMaestre:2021ssl}. The uncertainty on $m_\mathrm{Z}$ is dominated by the uncertainty on the collision energy, which can be determined with $\sim 100$ keV accuracy using resonant depolarization of the transversely polarized beams~\cite{Blondel:2019jmp}; 
the FCC-ee is designed explicitly with only a single RF accelerating section at lower energies to ensure that this precision is achievable. 
The smaller uncertainty on $\Gamma_\mathrm{Z}$ results from  possible systematic deviations from a linear calibration of the center-of-mass energies (the so-called ``point-to-point" uncertainties~\cite{Blondel:2019jmp}).

Cross-section dependent measurements like $\sigma_{\text{had}}^0$ or $N_\nu$ are limited by luminosity uncertainties. The absolute luminosity can be measured at FCC-ee with a relative precision of better than $10^{-4}$ using low-$Q^2$ Bhabha scattering, with point-to-point relative uncertainties an order of magnitude smaller. This is more than a factor of three better than the best LEP luminosity measurement~\cite{OPAL_Lumi:2000}. A phenomenological study of using large-angle two photon production ($\epem \to \gamma \gamma$) as an alternative can be found in Ref.~\cite{Carloni:2019ejp}. 

The ratio between hadronic and leptonic cross sections $R_\ell^\mathrm{Z}$ is an essential observable for the extraction of the invisible Z width and the individual leptonic and hadronic partial widths. It also provides one of the most precise ways to measure $\alphas(m^2_\mathrm{Z})$ (assuming negligible new physics contributions at these energies)~\cite{dEnterria:2020cpv}. This measurement is independent of luminosity uncertainties, and therefore a relative precision below $10^{-4}$ (one order of magnitude better than today) or better can be achieved~\cite{FCC-ee-accelerator}; systematic errors at LEP were dominated by the lepton angular acceptance; reducing them to the statistical error level will require specific detector design in the low angle region. More details on the $\alphas(m^2_\mathrm{Z})$ measurement can be found in Section~\ref{sec:qcd}. The individual leptonic partial widths will also be a powerful test of lepton coupling universality. A measurement of the relative invisible width $\Gamma_\text{inv}/\Gamma_\ell$ together with the measurement of the hadronic Z peak cross section~\cite{ALEPH:2005ab, Voutsinas:2019hwu, Janot:2019oyi} $\sigma^0_{\text{had}}$ can be used to determine the number of light neutrino species $N_\nu$ with a relative precision of $3 \times 10^{-4}$. 

A measurement of the forward-backward asymmetry using $\epem \to \mu^+\mu^-$ events around the Z pole $A^{\mu\mu}_{\text{FB}}$ allows the determination of the leptonic effective weak mixing angle $\sin^2 \theta_\mathrm{W}^{\text{eff}}$ to $\pm 2 \times 10^{-6}$.
In the $\epem \to \tau^+\tau^-$ channel, the measurement of the polarization asymmetry as a function of the scattering angle provides uncorrelated determinations of the left-right asymmetry parameters $A_e$ and $A_\tau$. The determination of $A_e$ is not affected by systematic uncertainties on the knowledge of $\tau$ polarization distributions, so  $\sin^2\theta_\mathrm{W}^{\text{eff}}$ can be extracted with a precision similar to that of $A^{\mu\mu}_{\text{FB}}$. The vector and axial couplings of each lepton species can be determined using the $\tau$ asymmetry measurements together with $A^{\mu\mu}_{\text{FB}}$ and the three leptonic partial widths.

Similarly, individual measurements of heavy-quark vector and axial-vector couplings (for $b$ quarks, $c$ quarks and, possibly $s$ quarks) can be determined from heavy-quark forward-backward asymmetries together with the corresponding Z decay partial widths.
These observables are of particular interest given the special relationship of the $b$ quark with the top quark, and, in case of supersymmetry models, to the sbottom and stop particles. The precision increase with respect to LEP is further enhanced by the reduction in beam pipe radius and the consequent improvement in $b$-tagging efficiency up to 80--90\%. 

The direct measurement of the electromagnetic coupling constant at the Z energy scale $\alpha_{\text{QED}} (m_\mathrm{Z})$ is an example of 
a fundamental parameters
whose precision can be improved by the projected large data samples.
It can be measured with a relative statistical uncertainty of the order of $3 \times 10^{-5}$ via the muon forward-backward asymmetry measurement at two off-peak center-of-mass energies~\cite{Janot:2015gjr}. Most systematic uncertainties are common to both energies and almost cancel in the slope determination. This direct determination of ${\rm \alpha_{{QED}}}$ is insensitive to sources of systematic uncertainties that affect the classical method from the dispersion integral over low energy ${\rm \epem \to hadrons} $ data.
Further improvements might be possible with four experiments and using the other lepton asymmetries. 

The unprecedented numbers of Z bosons and $\tau$ leptons produced allow us to perform many other searches: examples include Z boson charged-lepton-number violating decays and $\tau$ decays to three muons or to a muon and a photon.  Details can be found in Section~\ref{sec:flavor}. 

\subsection{The W pair production threshold}

\begin{figure}[hbtp]
\centering
\resizebox{0.42\textwidth}{!}{\includegraphics{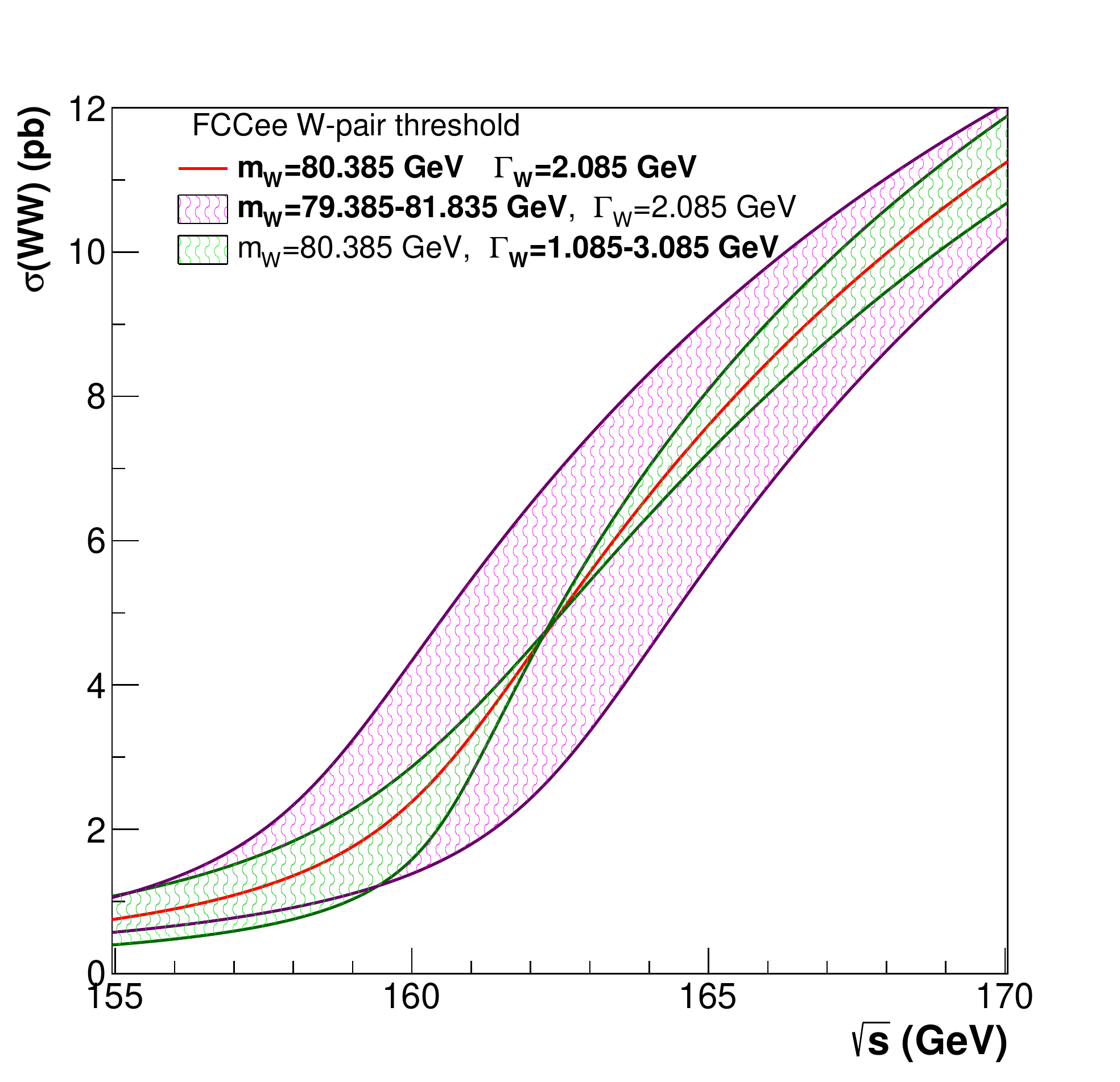}}
\resizebox{0.55\textwidth}{!}{\includegraphics{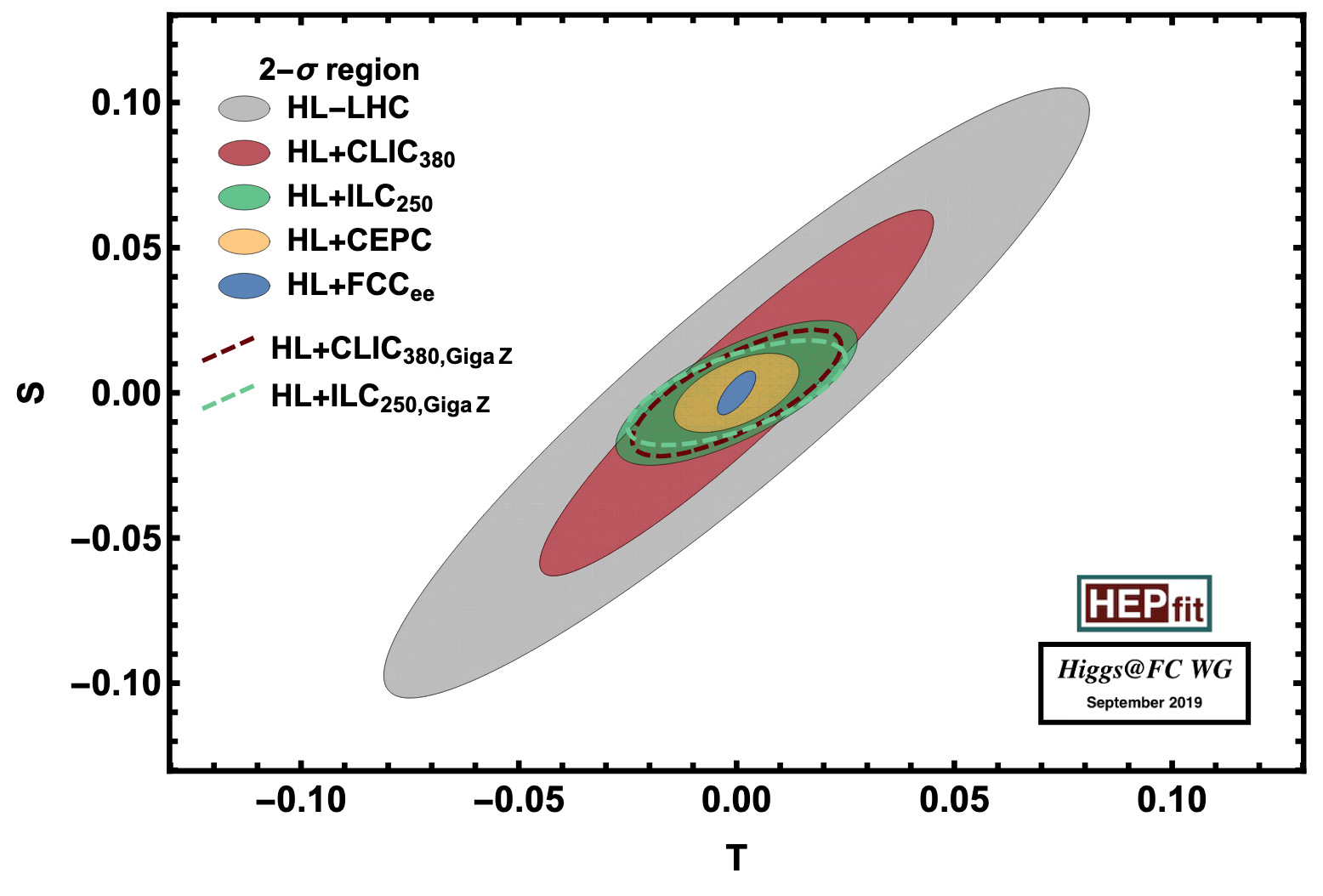}}
\caption{Left: W$^+$W$^-$ production cross section as a function of the $\epem$ collision energy~\cite{Abada:2019lih}. The central curve corresponds to the predictions obtained with $m_\mathrm{W}=80.385$~GeV and $\Gamma_\mathrm{W}=2.085$~GeV. Purple and green bands show the cross section curves obtained varying the W mass and width by $\pm 1$~GeV. Right: Expected uncertainty contour for the $S$ and $T$ parameters for various colliders in their first energy stage~\cite{eps_strategy,deBlas:2019rxi}.}
\label{fig:EW-precision}
\end{figure}

The W boson mass ($m_\mathrm{W}$) and width ($\Gamma_\mathrm{W}$) can be determined from the W-pair threshold cross section lineshape~\cite{ALEPH:2013dgf}. The left plot of Fig.~\ref{fig:EW-precision} shows the W-pair cross section as a function of the $\epem$ collision energy with $m_\mathrm{W}$ ($\Gamma_\mathrm{W}$) set at the central value of 80.385~GeV (2.085~GeV), and with $\pm1$~GeV variation bands around the mass and width central values. Sensitivities to $m_\mathrm{W}$ and $\Gamma_\mathrm{W}$ are different at different energies. The width dependence shows a crossing point at $\sim 162.3$~GeV, where the cross section is insensitive to $\Gamma_\mathrm{W}$. The values of the two energy points, $E_1$ and $E_2$, as well as the data luminosity fraction $f$ delivered at the higher energy, can be optimized to obtain the best
results.
Studies indicate that with $E_1=157.5$~GeV, $E_2=162.5$~GeV and $f=0.5$ for an integrated luminosity of 12 ab$^{-1}$, a simultaneous fit of $m_\mathrm{W}$ and $\Gamma_\mathrm{W}$ to the $\epem \to \mathrm{W^+W^-}$ cross section measurements yields an accuracy of 0.5~MeV on $m_\mathrm{W}$ and 1.2~MeV on $\Gamma_\mathrm{W}$~\cite{Azzurri:2021yvl}. To reach this sensitivity, the center-of-mass energies must be measured with an accuracy of 
0.3~MeV, 
which is considered doable given that (unlike at LEP) a good enough level of transverse polarization should be achievable up to above the W pair threshold~\cite{Blondel:2019jmp}. 
The point-to-point variation of the detector acceptance (including that of the luminometer) and the W$^+$W$^-$ cross section prediction must be controlled within ${\cal{O}}(10^{-4})$. In addition, the background must be known at the few per-mil level. These conditions are less stringent than the requirements at the Z pole.  

It can also be useful to take data at additional energy points, especially the points where $(\mathrm{d}\sigma/\mathrm{d}m_\mathrm{W})^{-1}$, $(\mathrm{d}\sigma/\mathrm{d}\Gamma_\mathrm{W})^{-1}$, $\sigma(\mathrm{d}\sigma/\mathrm{d}m_\mathrm{W})^{-1}$, and $\sigma(\mathrm{d}\sigma/\mathrm{d}\Gamma_\mathrm{W})^{-1}$ are equal, to reduce background and acceptance uncertainties~\cite{Azzurri:2021yvl}. These additional measurements can also serve to disentangle possible new physics effects.  For example, the dependence of the W$^+$W$^-$ cross section on the center-of-mass energy is different for the $t$-channel neutrino exchange process that dominates in the threshold region and TGC processes. Anomalous TGC contributions would therefore lead to distinctive differences in the W$^+$W$^-$ cross section lineshape. 

The W boson mass and width can also be determined from the kinematic reconstruction of the W$^+$W$^-$ decay products using fully-hadronic ($qqqq$) and semileptonic ($qq\ell\nu$) decay events~\cite{Azzurri:2021yvl}. In both cases the reconstructed W mass values are obtained by imposing the constraint that the total four-momentum in the event should be equal to the known initial center-of-mass energy and zero momentum. The four-momentum constraints are implemented by means of a kinematic fit where the measured energies of jets and leptons are adjusted within their measurement uncertainties. 
The reconstructed W mass resolution can be further improved with the additional constraint of equal mass for both W bosons in each event. With this methodology, it is  estimated that the combined statistics of all FCC-ee data would deliver a precision of around 1~MeV for $\Gamma_\mathrm{W}$ and below 0.5~MeV for $m_\mathrm{W}$, matching the precision delivered by the threshold scan. 

A simultaneous kinematic fit of WW, ZZ and Z$\gamma$ events, can lead to a determination of $m_\mathrm{W}/m_\mathrm{Z}$ ratio where many systematic uncertainties common to the three channels can cancel, and the W mass can be derived given the independent precision on the Z mass from the Z peak data~\cite{Azzurri:2021yvl}.

\subsection{Global electroweak fits and theoretical challenges}
With the discovery of the Higgs boson, all the parameters defining the SM have been measured, and global electroweak fits can be used to test the internal consistency of the theory, 
while any deviation will constitute a definite BSM signal. 
In addition, high-precision measurements 
will enlarge the phase space over which significant  deviations from the SM predicted values can be  observed in one or several observables.
This 
search method can be remarkably robust, sensitive to a wide variety of new physics models 
and potentially able to distinguish between them. 
The right plot of Fig.~\ref{fig:EW-precision} shows the projected uncertainty on the $S$ and $T$ parameters from a global fit of EWPOs for different $\epem$ colliders~\cite{ESPP}. The FCC-ee measurements at the Z pole, at the W$^+$W$^-$ threshold, and at the $t\bar{t}$ threshold, give the best sensitivity, 
even considering that the systematic errors are still very conservative. 

The large number of precision observables and the unknown nature and scale of the new physics that would be responsible e.g., for dark matter and the baryon asymmetry of the Universe, leads to several different ways to fit the data. In this chapter the Standard Model Effective Field Theory (SMEFT) fit is described, which is tailored to be sensitive to decoupling new physics at high energy scale. A specific fit (which could be included in the SMEFT framework, as in~\cite{deBlas:2021jlt}) can be made to enhance sensitivity to e.g., neutrino mixing with heavy (above the Z mass and up to 1000~TeV) Neutral Leptons, for which lepton family universality violation and energy-independent mixing with high mass states would be introduced~\cite{Antusch:2017pkq}. Such a fit is described in the BSM section, leading to limits on the mixing in the range of 10$^{-4}$ to 10$^{-5}$, independent of the heavy neutrino mass.
Similarly, other fits can be tailored for a variety of specific models. A systematic investigation in this direction remains to be explored. 

\begin{figure}[hbtp]
\centering
\resizebox{0.7\textwidth}{!}{\includegraphics{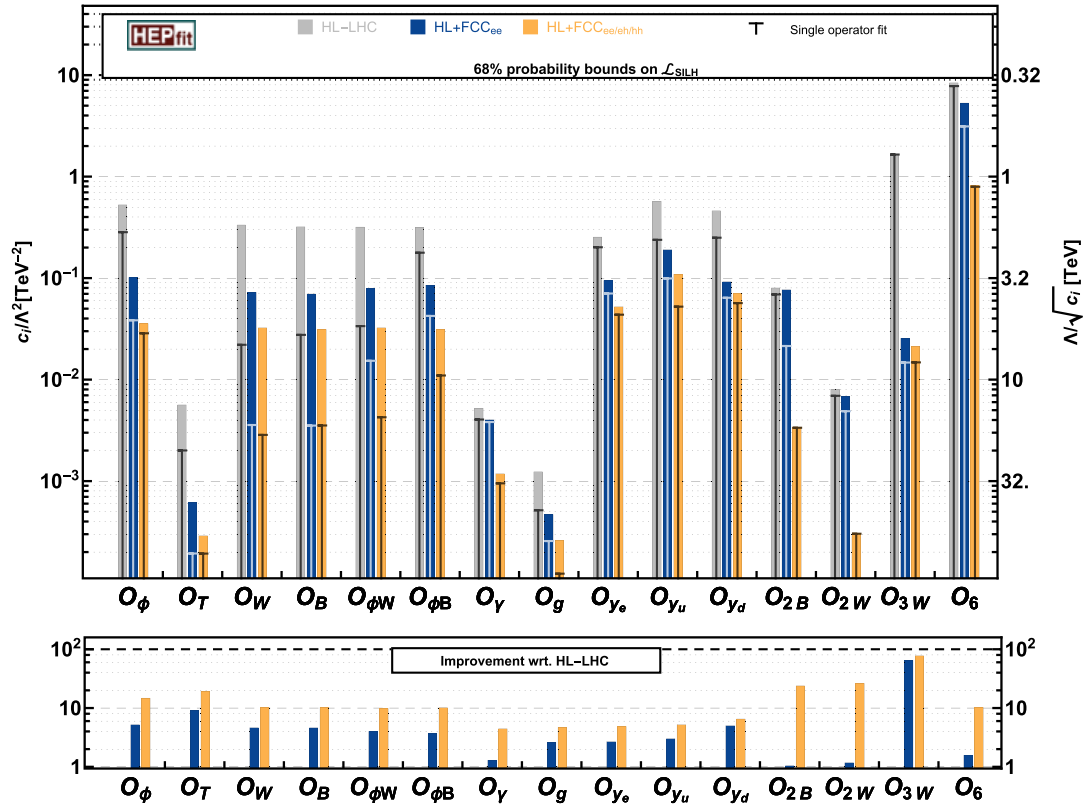}}
\caption{The 68\% probability reach for $c_i/\Lambda^2$ from a fit to the EFT Lagrangian in Eq.~(3.19) of ~\cite{deBlas:2019rxi}.
The right axis shows the corresponding bound on the new physics interaction scale.
}
\label{fig:EW-EFT}
\end{figure}

In the SMEFT, the renormalizable interactions
in the SM are supplemented by higher-dimensional operators. These are composed of
all possible combinations of SM fields that respect the $SU (3)_C \times SU (2)_L \times U (1)_Y$ gauge
symmetries and Lorentz invariance. The leading lepton-number-conserving effects are 
parametrized by dimension $d = 6$ operators with unknown Wilson coefficients that could
be generated by new particles with masses far beyond the EW scale, assuming that these also respect
the SM gauge symmetries. We then consider just the dimension-6 SMEFT Lagrangian:
$${\cal L}_{\rm SMEFT} = {\cal L}_{\rm SM} + \sum_i \frac{c_i}{\Lambda^2}{\cal O}_i$$
where the ${\cal O}_i$ are the dimension-6 operators,
$\Lambda$ represents the scale of new physics, and the coefficients $c_i$ depend on the details of
its structure. The projected constraints on $c_i/\Lambda^2$ for several dimensional-6 operators in the SMEFT fit are shown in Fig.~\ref{fig:EW-EFT}. 
The complementarity between the Higgs and EWPOs are clearly shown: constraints on some Wilson coefficients are greatly improved using the combined Higgs and electroweak measurements. 
Furthermore, information on the nature of the new physics might be revealed from the pattern of deviations between the different coefficients. 

The anticipated experimental accuracy on EWPOs has to be matched with a theory prediction of at least the same level of accuracy to make maximum use of the experimental data. 
A highly dedicated and focused investment over the next one or two decades is needed by the theoretical community to bring the state-of-the-art theory to the necessary level~\cite{Blondel:2018mad,Freitas:2019bre,Heinemeyer:2021rgq}. On one hand, theory inputs are required to extract interesting quantities, such as $\sin^2 \theta_\mathrm{W}^{\text{eff}}$ or $m_\mathrm{W}$, from the data. In particular, next-to-next-to-leading (NNLO) electroweak corrections and partial higher-order corrections will be needed for various pair production processes, such as $\epem \to f\bar{f}$ or $\epem \to \mathrm{W^+W^-}$~\cite{Freitas:2019bre}, and these must be matched with new and improved Monte Carlo programs that simulate multiple photon emission~\cite{Jadach:2019bye}.

\begin{table}
\centering
\caption{Estimated precision for the direct determination of several EWPOs at FCC-ee (column 2), current intrinsic theory uncertainties for the prediction of these quantities within the SM as well as the main sources of theory uncertainties (column 3), and estimated projected intrinsic theory uncertainties when leading 3-loop corrections become available (column 4)~\cite{Freitas:2019bre}.\vspace{0.25cm} \label{tab:EW_theory}}
\resizebox{\textwidth}{!}{%
 \begin{tabular}{l|lll}
    \toprule
    Quantity & FCC-ee & Current intrinsic theory uncertainty & Projected intrinsic theory uncertainty \\  \hline 
    $m_\mathrm{W}$ (MeV) & $0.5-1$ & 4 ($\alpha^3$, $\alpha^2\alphas$) & 1  \\
    $\sin^2 \theta^\ell_{\text{eff}}$ ($10^{-5}$) & 0.6 & 4.5 ($\alpha^3$, $\alpha^2\alphas$) & 1.5  \\
    $\Gamma_\mathrm{Z}$ (MeV) & 0.1 & 0.4 ($\alpha^3$, $\alpha^2\alphas$, $\alpha \alpha^2_s$) & 0.15  \\
    $R_b$ ($10^{-5}$) & 6 & 11 ($\alpha^3$, $\alpha^2\alphas$) & 5  \\
    $R_\ell$ ($10^{-3}$) & 1 & 6 ($\alpha^3$, $\alpha^2\alphas$) & 1.5  \\
   \bottomrule
    \end{tabular}
}
\end{table}

On the other hand, the ultimate goal of electroweak precision measurement is the search for possible new physics effects. To do this, the measurements need to be compared to at least equally accurate theory predictions for the relevant quantities within the SM. For Z-pole physics, it is estimated~\cite{Freitas:2019bre} that one will need N$^3$LO and leading N$^4$LO QED corrections for this purpose, which is at least one order beyond the current state of the art. As an example, the estimated precision for several EWPOs at FCC-ee, current and projected intrinsic theory errors are shown in Table~\ref{tab:EW_theory}.

To achieve these goals, new and improved calculational techniques for multiloop computations and new developments for Monte Carlo event generators are called for, but recent developments on this front are encouraging~\cite{Mandal:2018cdj,Liu:2021wks,Dubovyk:2022frj}.

\section{Top quark physics}
\label{sec:top}
\textcolor{red}{Editors: P. Azzi,  R. Demina,  F. Simon, L. Skinnari, and M.~Vos}

It is expected that about a million $t\bar t $ events will be produced at the FCC-ee near and above the production threshold. It does not sound like much compared to the order of magnitude larger top sample already recorded by the LHC but it is the clean environment, the precise knowledge of the collision energy and total momentum, and the ability to scan the center-of-mass energy that make all the difference. Hence, lepton colliders provide the ultimate precision in the determination of many particles' properties. Precise measurements of the properties of the top quark, the heaviest known elementary particle, are of great interest. The value of the top quark mass, $m_t$, was constrained first by the observation of $B\bar{B}$ mixing by ARGUS and CLEO, later by the EW measurements at LEP/LEPII, and was finally confirmed with the top quark discovery at the Tevatron~\cite{CDF:1995wbb,D0:1995jca}.
Precision measurements of $m_t$ and $m_\mathrm{W}$ in turn helped to constrain the value of the Higgs boson mass. With both particles discovered, the SM becomes overconstrained and further improvements in the precision of its parameters allow us to probe for new physics. Yet, being discovered in $p\bar{p}$ collisions, and later studied in even greater detail at the LHC, the top quark has never been produced using an $\epem$ machine. The FCC-ee will close this crucial gap. Precision in the top quark mass achievable at the FCC combined with  the precise knowledge of the top quark Yukawa coupling   will provide an important test of the Higgs mechanism.

\subsection{Precise determination of the top quark mass, width, and Yukawa coupling}  

The top quark mass is one of the  most important parameters of the SM. At hadronic machines there are two primary approaches to determining $m_t$: using the sum of the Lorentz vectors of the top quark decay products and using the total or differential production cross section of $t\bar{t}$ pairs or single-top quarks (see Ref.~\cite{Nason:2017cxd} for a review). Both approaches suffer from significant systematic uncertainties. The uncertainty in the renormalization scale is important for both approaches. In the former case, jet energy determination and modeling of the hadronization  and color connection of the decay products dominate the list of systematics, while in the latter case, since the initial state particles are proton constituents, i.e.\ quarks or gluons, the knowledge of the parton density functions is one of the leading limiting source of the systematic uncertainties. Moreover, the theoretical interpretation of the result has some ambiguities~\cite{Hoang:2020iah}. The method that relies on the production cross section and its behaviour near the threshold arguably is evaluating the pole mass. The kinematic reconstruction on the other hand relies on the comparison of the observed distributions with the ones predicted from Monte Carlo simulation, and thus even though the measured  parameter retains a  relation with the pole mass,  an additional uncertainty must be assigned as suggested in Ref.~\cite{Hoang:2020iah}.
For these reasons it is unlikely that the precision in $m_t$ can be pushed much below 0.5~GeV using data from a hadron machine.  
 
Already before the discovery of the top quark, it was recognized that an $\epem$ machine has the potential for a precise measurement of $m_t$ \cite{Gusken:1985nf, Fadin:1987wz,Fadin:1988fn,Strassler:1990nw} by comparing the measured pair-production cross section to theory predictions. Today, these predictions have reached N$^3$LO QCD precision~\cite{Beneke:2015kwa} and an NNLL resummation~\cite{Hoang:2001mm}.  Here, the interpretation of the $m_t$ measurement is straightforward since it is determined from the production cross section behaviour at the threshold and thus is performed in a well-defined mass scheme, with results that can be transformed into other definitions, such as $\bar{\rm MS}$ or the pole mass with small theoretical and parametric uncertainties \cite{Marquard:2015qpa}. Phenomenological studies of such measurements have been performed in the context of different $\epem$ colliders \cite{CLICdp:2018esa,Seidel:2013sqa,Horiguchi:2013wra,Martinez:2002st}, extending to parameters beyond the mass alone, and recently also considering theoretical uncertainties \cite{Simon:2016htt}. The potential in the context of FCC-ee is discussed in the following.  
  
\begin{figure}[hbtp]
\centering
\resizebox{0.5\textwidth}{!}{\includegraphics{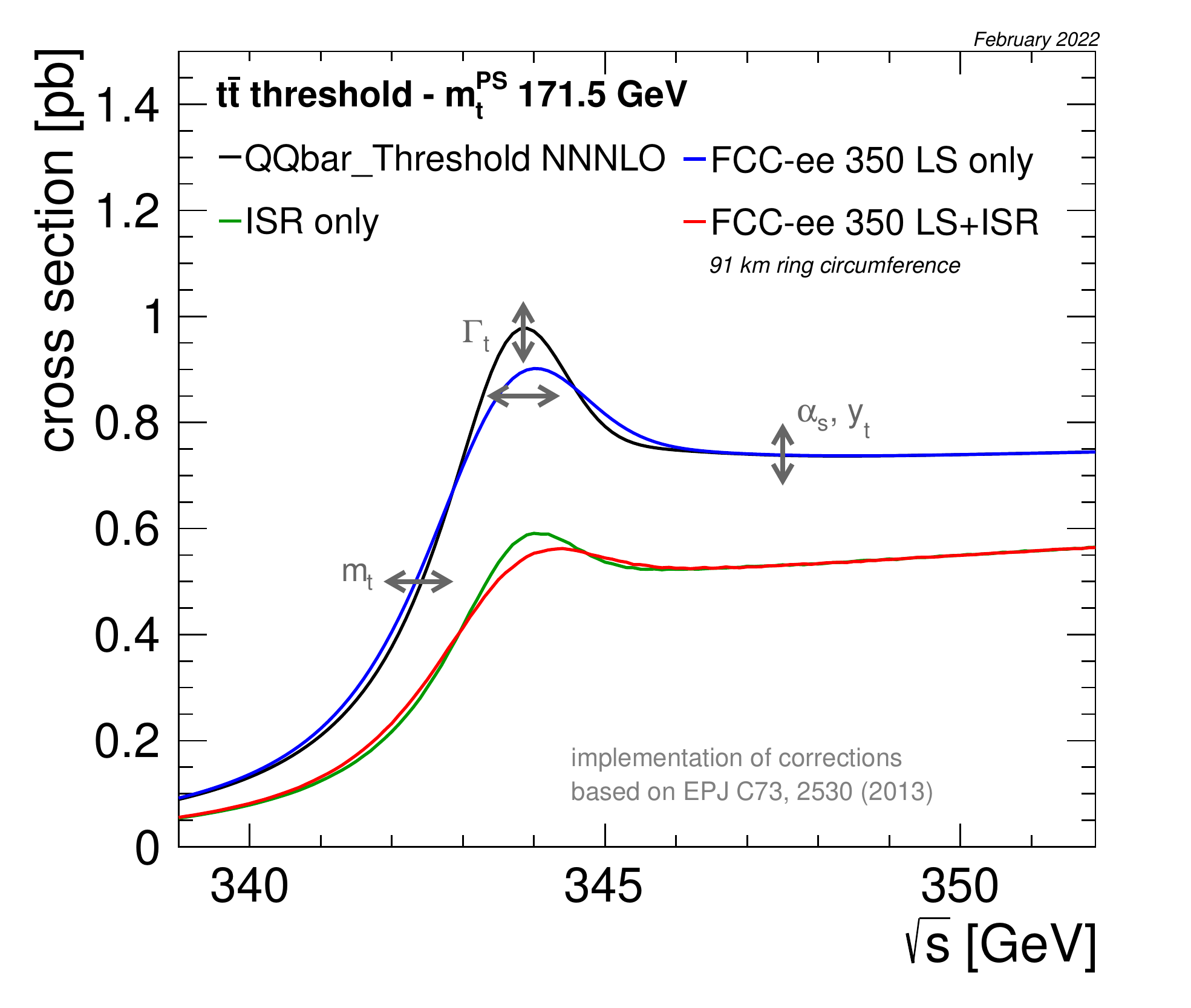}
}
\caption{$t\bar{t}$ production cross section vs the center-of-mass energy near the threshold. The effects of the top quark mass, width, and the top quark Yukawa coupling and the strong coupling constant on the theory cross section are indicated by the arrows. The effects of ISR (green) and the collider luminosity spectrum (LS) (blue) are also shown. The observable cross section is given by the combination of both effects (red). Figure taken from Ref.~\cite{Simon:2022FCCTop}.
}\label{fig:top_threshold}
\end{figure}

 In addition to the top quark mass, the behaviour of the production cross section at the $t\bar{t}$ threshold is dependent on the top quark width $\Gamma_t$, top quark Yukawa coupling $y_t$, and the strong coupling constant $\alphas$. Figure~\ref{fig:top_threshold} illustrates the dependence of the $t\bar{t}$ production cross section on these important SM parameters, calculated at N$^3$LO in QCD. Also shown are the corrections to the theoretical cross section from initial state radiation (ISR) and due to the collider luminosity spectrum (LS). The latter is where differences between different $\epem$ colliders appear, with FCC-ee providing an advantage over linear colliders due to the absence of a pronounced beamstrahlung tail in the luminosity spectrum. The final observable cross section is obtained by the combination of both ISR and LS. 
The effects of $y_t$ and $\alphas$ on the cross section are highly correlated. The sensitivity to these parameters arises due to the virtual exchange of a Higgs boson or a gluon between top quarks and anti-quarks, which have similar effects on the dependence of the cross section on the invariant mass of the system. Luckily, $\alphas$ can be constrained to its value precisely evaluated at the Z peak and its running established at the WW production threshold (\autoref{sec:qcd}).

A study based on the techniques developed in \cite{Seidel:2013sqa,Simon:2016htt,Simon:2019axh} shows that with a total integrated luminosity of 200 fb$^{-1}$ evenly split across eight different center-of-mass energies around the $t\bar{t}$ production threshold (340, 341, 341.5, 342, 343, 343.5, 344, and 345~GeV), $m_t$ can be determined with a statistical precision of better than 9\,MeV when assuming SM values for $\Gamma_t$ and $y_t$ \cite{Simon:2022FCCTop}. Uncertainties in the strong coupling constant lead to a parametric uncertainty of around 3.2~MeV for an $\alphas$ uncertainty of 0.00012, as expected from the Z-pole program described in \autoref{sec:qcd}. Similarly, the 3.1\% uncertainty in $y_t$ for the combination of HL-LHC and FCC-ee shown in \autoref{tab:h-couplings} translates into a parametric uncertainty in the mass of approximately 5\,MeV. Systematic uncertainties are expected to be on the order of a few tens of MeV, with theory uncertainties being a dominant source. Currently, scale uncertainties of N$^3$LO QCD calculations translate to a systematic uncertainty of 45~MeV in $m_t$. 

Additional parameters can be determined from multidimensional fits of the top threshold. For example, $m_t$ and $\Gamma_t$ can be determined simultaneously with a statistical precision of $\pm$17\,MeV and $\pm$45\,MeV, respectively.




\subsection{Measurement of top quark electroweak couplings}
The electroweak couplings of the top quark remain relatively unconstrained, as the previous generation of electron-positron colliders did not reach sufficient energy to produce top quarks. The Tevatron and LHC did probe the charged-current interaction vertex in top quark decays and single-top-quark production~\cite{Birman:2016jhg}. The rare associated production processes of top quarks with a photon, Z boson, or a Higgs boson observed at the LHC directly probe the neutral current interactions of the top quark~\cite{Miralles:2021dyw}. At the FCC-ee, top quark pair production $\epem\to \gamma^*/Z \to t \bar t$ is mediated by a photon or a Z boson. Thus, measurements of the  $t \bar t$  cross section can probe the electroweak couplings $t\bar{t}\gamma$ and $t \bar{t}$Z at the production vertex.

The sensitivity of $\epem$ colliders operated above the $t\bar{t}$ production threshold to anomalous electroweak couplings of the top quark is well-established~\cite{Amjad:2015mma,Bernreuther:2017cyi,Durieux:2018tev, CLICdp:2018esa, Durieux:2019rbz}. Ref.~\cite{janot2015precision} has demonstrated that the couplings to the photon and the Z boson can be effectively disentangled at or slightly above the $t\bar{t}$ production threshold by measuring the top quark polarization, using the charged leptons from the top quark decay as polarimeters.

Ref.~\cite{janot2015precision} projects a precision of $1(3)\times 10^{-3}$ for the anomalous vector coupling of $\gamma$(Z), and of $1(2)\times 10^{-2}$ for the anomalous axial coupling. Any deviation of these couplings from the SM values would signal the presence of new physics. 
An analysis of a circular-collider-like scenario in Ref.~\cite{Durieux:2018tev} in the SMEFT confirms that the sensitivity to top quark electroweak couplings exceeds that of the HL-LHC by an order of magnitude and demonstrates the added value of $\epem$ collision data at a center-of-mass energy well above the $t\bar{t}$ production threshold to disentangle four-fermion and two-fermion operators.

The precise measurement of top quark couplings to a photon or the Z boson are essential to precisely determine the top quark Yukawa coupling at the FCC-hh~\cite{janot2015precision}. 
While the top quark Yukawa coupling can be determined with high statistical accuracy at hadron colliders, it must be normalized to the $t\bar{t}$Z process to reduce theoretical systematic uncertainties that would otherwise dominate~\cite{Blondel:2021ema}. 

\subsection{Searches for FCNC interactions}

Flavor-changing neutral current (FCNC) interactions constitute an excellent tool for constraining the SM and probing new physics in the top quark sector. FCNC interactions in the top quark sector involve the exchange of a neutral boson ($\mathrm{H, \gamma, g, Z}$) rather than a W~boson. FCNCs are forbidden at tree level and highly suppressed in the SM, with branching ratios of the order of $10^{-12}-10^{-17}$. However, numerous extensions to the SM are expected to significantly enhance these rare FCNC interactions by contributing additional loop diagrams mediated by new beyond-SM particles. Depending on the particular model, enhancements to the order of $10^{4}-10^{7}$ are possible~\cite{Abada:2019lih}. Therefore, signs of FCNC interactions involving top quark production or decay would be a clear indication of physics beyond the SM. 

At the FCC-ee, searches for top quark FCNCs can proceed using data collected above the $t\bar{t}$ threshold of 365~GeV (1.5 ab$^{-1}$). Such searches would probe FCNCs in the decay of a top quark from $t\bar{t}$ production. In addition, searches for anomalous single-top quark production processes can profit from the 5~ab$^{-1}$ run at $\sqrt{s}=240$~GeV, nominally dedicated to HZ production. Preliminary studies have been performed in the $\epem \to Z/\gamma \to t\bar{q} (\bar{t}q)$ channel, where the W boson from the top quark decays leptonically~\cite{Abada:2019lih,SingleTop-FCNC}. These studies showed an expected sensitivity to FCNC branching ratios ($t \to q\gamma$, $t \to q$Z) of about $10^{-5}$, which might further improve by considering the combination of the analysis for single-top quark anomalous production in the leptonic and hadronic final state, plus the results obtained from the FCNC in the decay. Comparison of expected limits at 95\% confidence level on FCNC interactions in top quark production or decay are shown in Table~8.2 from Ref.~\cite{ESPP}. 

 
Improved sensitivity beyond the preliminary studies mentioned above are expected to be possible. These searches rely on, e.g., precise jet reconstruction and bottom/charm quark tagging algorithms. As pointed out in Ref.~\cite{azzi:higgs-top-reco}, using kinematic constraints in the event reconstruction can improve reconstructed invariant masses and utilizing new techniques, including machine-learning with particle-flow reconstruction, can provide enhanced jet and event reconstruction. These will contribute to increase the sensitivity in FCNC searches compared to the existing limit projections that utilize more traditional analysis approaches~\cite{azzi:higgs-top-reco}.

\subsection{Top program summary}
To summarize, the US physicists played a crucial role in the top quark discovery at the Tevatron  and measurement of its properties at the same collider followed with even more detailed studies performed at the LHC. There is no doubt that the opportunity to perform the exquisite measurements of the top quark properties enabled by the FCC will excite the US HEP physics community.
\clearpage

\section{Beyond the Standard Model}
\label{sec:bsm}
\textcolor{red}{Editors:  S. Antusch, A. Blondel,  S. Heinemeyer, R. Gonzalez Suarez,  E. Thomson, S.~Willocq}



While the FCC-ee will provide new insights into electroweak and Higgs physics through the precise measurement of Z- and Higgs-boson properties ---thereby providing indirect sensitivity to physics beyond the Standard Model (BSM)--- it will also open up significant new phase space in the direct search for BSM physics. In particular, the FCC-ee program will give access to feebly-interacting light particles in the mass range between $\sim 1$ and 100~GeV. Such particles and interactions are predicted in a broad range of BSM theories, addressing some of the key unanswered questions in particle physics~\cite{Agrawal:2021dbo}. These include models addressing the electroweak hierarchy problem (e.g., neutral naturalness or relaxion models), the origin of dark matter, the genesis of the baryon asymmetry of the Universe (BAU), or the origin of neutrino masses. A dedicated Snowmass white paper on long lived particle (LLP) searches at the FCC-ee~\cite{Verhaaren:2022ify} contains many more details. 

The clean environment at the FCC-ee will allow for a considerably larger acceptance for these searches than can be obtained at a hadron collider. For example, in channels with associated Z-boson production as in Higgs-strahlung production, nearly all final states can be utilized, instead of just $\mathrm{Z} \to \epem$ or $\mu\mu$ decays (with a corresponding branching fraction of only about 7\%). The searches benefit from the inclusion of hadronic decay channels and channels with missing transverse momentum, which severely limit the acceptance of BSM searches for new light particles at hadron colliders due to trigger requirements and huge QCD multijet backgrounds. Moreover, due to the small couplings being probed, many of the searches involve LLPs and thus require experimental techniques to identify displaced vertices and other unconventional signatures that further benefit from the clean environment.

Given the lack of BSM discoveries so far, it is critical that the model-focused search program of future colliders be complemented by a model-agnostic effort.  Machine learning tools are well-suited for this task, as they can efficiently explore high-dimensional feature spaces.  A variety of new methods have been proposed to do this \textit{anomaly detection} beyond the standard classification approach (see e.g., Ref.~\cite{Karagiorgi:2021ngt}).  In the context of $e^+e^-$ collisions~\cite{Gonski:2021jek}, there are multiple ways to implement such searches, and they may motivate aspects of the detector design (e.g. forward energy resolution) and computing environment (e.g. a large number of graphic processing units or GPUs) at FCC-ee. Such tools will be especially useful in an electron-positron collider environment, where trigger criteria are not necessary.

\subsection{Dark Sector / Hidden Sector}

\begin{figure}[hbtp]
  \centering
\resizebox{0.85\textwidth}{!}{\includegraphics{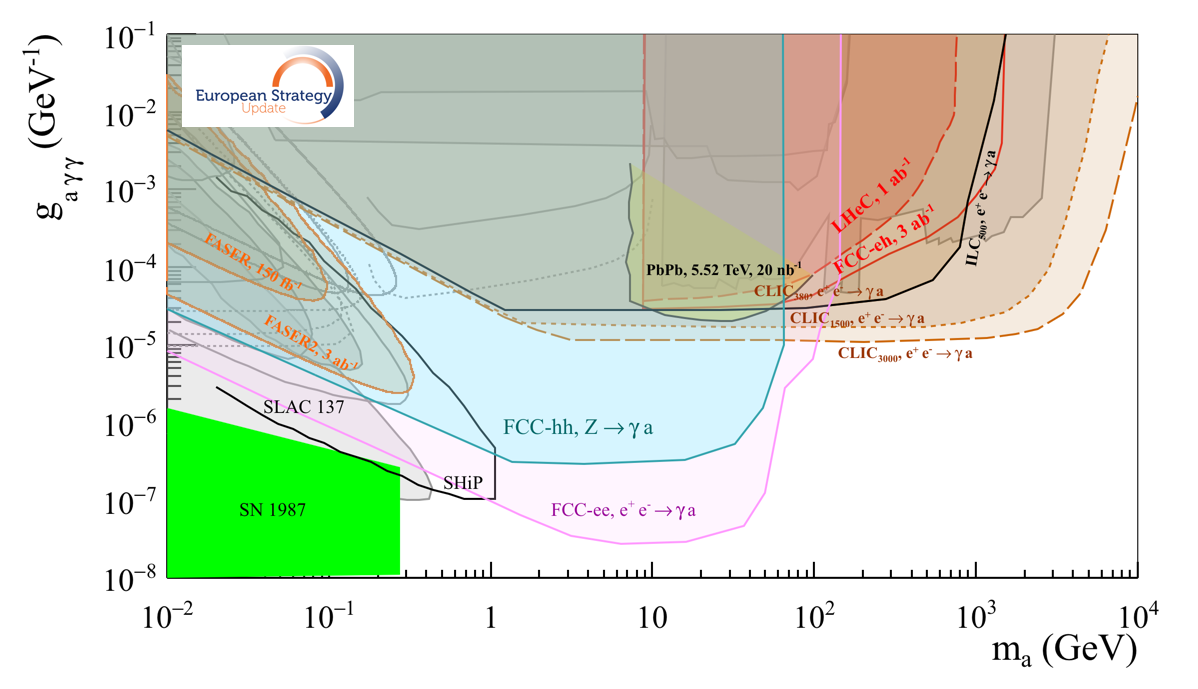}}
  \vspace{-5mm}
\caption{Figure 8.18 from Ref.~\cite{eps_strategy}, showing 
exclusion limits on the coupling between ALPs and photons as a function of the ALP mass. All curves correspond to 90\% CL exclusion limits, except for LHeC/FCC-eh (95\% CL exclusion limits), FCC-ee (observation of four signal events) and FCC-hh (observation of 100 signal events). The FCC-ee estimate corresponds to the combined results for integrated luminosities of 145, 20, and 5 ab$^{-1}$ at $\sqrt{s}= 91$, 161, and 250~GeV, respectively~\cite{Bauer:2018uxu}. 
}\label{fig:BSM_ALPS}
\end{figure}

Models featuring axion-like particles (ALPs) are extensions of QCD axion models initially proposed to solve the strong CP problem. In these models, ALPs can be considerably heavier than the QCD axion and provide a mediator between particles of the dark matter sector and SM particles. Whereas the ALP-photon coupling is rather strongly constrained by astrophysics and beam-dump experiments at lower mass, the constraints in the ALP mass range $\sim 1-100$~GeV are much weaker~\cite{dEnterria:2021ljz} and potentially accessible at the FCC-ee via the process $\epem \to a \gamma$, with the ALP decaying according to $a \to \gamma\gamma$. The sensitivity reaches couplings below $10^{-7}$ in much of this range, as shown in Fig.~\ref{fig:BSM_ALPS}, extending current limits by up to four orders of magnitude~\cite{Bauer:2018uxu}. Much of the sensitivity comes from Z-boson decays collected during the Tera-Z data taking phase. At small values of the coupling, the ALP lifetime is large enough to give rise to macroscopic decay lengths. The study in Ref.~\cite{Bauer:2018uxu} requires the decay to occur before the ALP reaches the electromagnetic calorimeter. The ALP coupling to Z and Higgs bosons can also be probed via the process $\epem \to \mathrm{H}a$ with $\mathrm{H} \to b\bar{b}$ and either $a \to \gamma\gamma$ or $a \to \ell^+\ell^-$. Decays to SM particles other than photons are less constrained and provide an additional opportunity for ALP discovery at the FCC-ee.
  
Many BSM scenarios involve a dark (or hidden) sector that features its own matter and field content. In these scenarios, interactions between dark sector and SM particles may occur via a neutral portal. Of particular interest are Higgs and vector portals. In the latter case, a dark photon may mix kinetically with the SM hypercharge gauge boson with strength~$\epsilon$. This possibility can be explored with the radiative-return process $\epem \to A' \gamma$ where the dark photon decays according to $A' \to \mu^+\mu^-$. The sensitivity to small $\epsilon$ values extends down to $\sim 2 \times 10^{-4}$ in the $m_{A'}$ range between 10 and 80~GeV~\cite{Karliner:2015tga}. Other possible avenues to search for a dark photon or dark Z' boson have been investigated using $\epem \to$~A'H or Z'H at $\sqrt{s} = 160$ and 240~GeV~\cite{Giffin:2020jtl}.


In addition to the processes discussed above, the FCC-ee provides opportunities to search for a wide range of other exotic Z-boson decays, in particular those motivated by dark sector models. This takes advantage of the Tera-Z run and the ability to fully reconstruct final states with soft decay products or including particles escaping detection. Analysis of such final states requires the presence of high-momentum initial state radiation to be able to trigger on these events at a hadron collider, thereby significantly reducing the acceptance. Decays into hadronic final states also suffer from high backgrounds at a hadron collider. Depending on the final state, the FCC-ee reaches 95\% CL. exclusions on the branching fraction that are 2-9 orders of magnitude lower than those that can be achieved at the HL-LHC~\cite{Liu:2017zdh}. For example, the branching fraction exclusion for Z\,$ \to \phi_d A' \to \chi\bar{\chi}jj$ in a vector-portal model improves from about $10^{-2}$ to as low as $10^{-11}$, depending on model parameters. For the fully-hadronic Z\,$ \to \phi_d A' \to jj jj$ decay, the improvement is from $10^{-2}$ to $10^{-7}$.

Searches for exotic Z-boson decays into LLPs can also be carried out at the FCC-ee. In particular, the sensitivity to long-lived light neutralinos predicted in $R$-parity-violating Supersymmetry exceeds what can be achieved at hadron colliders or even future experiments dedicated to the search for LLPs. A coupling strength $\lambda'_{112}/m^2_{\tilde{f}}$ as low as $1.5 \times 10^{-13}$~GeV$^{-2}$ for $m(\tilde{\chi}_1^0) = 40$~GeV can be reached for ${\cal B}(\mathrm{Z} \to \tilde{\chi}_1^0\tilde{\chi}_1^0) = 10^{-5}$, a limit at least an order of magnitude stronger than can be achieved in any other future experiment~\cite{Wang:2019orr}. Other studies,   as that of Z-boson decays into two-body hidden meson final states~\cite{Cheng:2019yai}, exhibit similar gains in sensitivity.

\begin{figure}[hbtp]
\centering
\resizebox{0.95\textwidth}{!}{\includegraphics{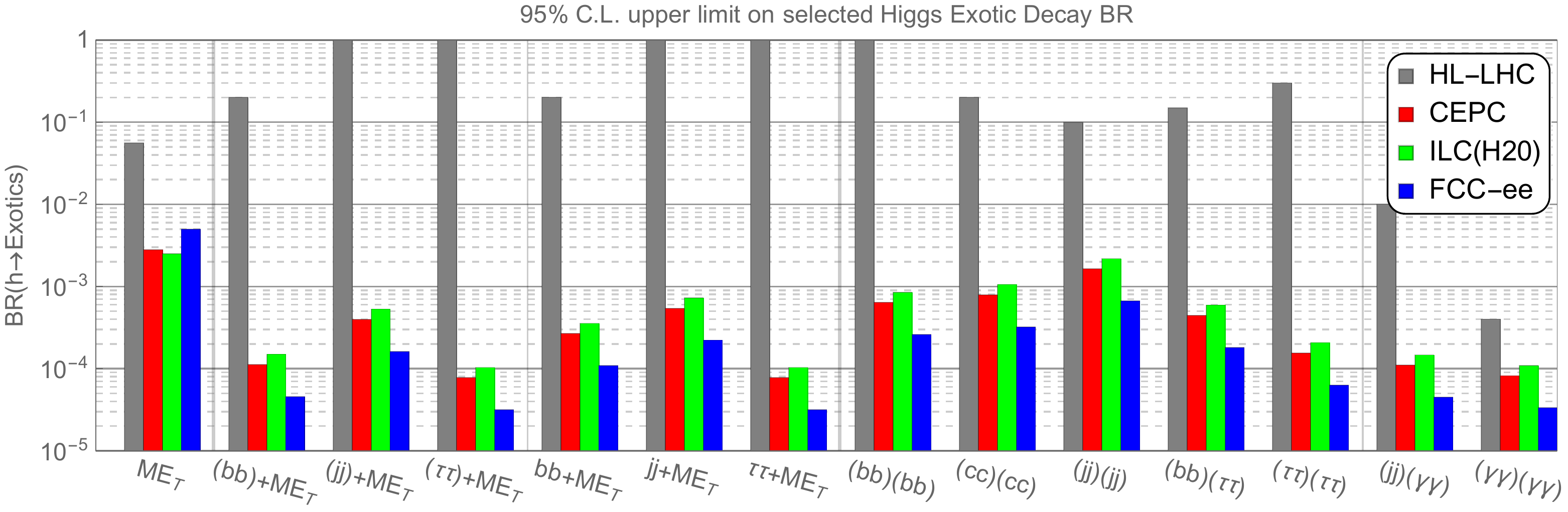}}
\caption{Figure 12 from Ref.~\cite{Liu:2016zki}, showing 
exclusion upper limits on the branching fraction of Higgs boson decays to different final states. Limits for different future accelerator projects are compared. The integrated luminosity at the FCC-ee is assumed to be 30~ab$^{-1}$ at $\sqrt{s} = 240$~GeV (5~ab$^{-1}$ is assumed in the CEPC case).
}\label{fig:BSM_exotic_Higgs}
\end{figure}

One of the main goals of the FCC-ee program is to measure precisely Higgs boson couplings to fermions and gauge bosons in order to find deviations from SM predictions and thus hint at new degrees of freedom. Another approach is to look directly for exotic decays of the Higgs boson, taking advantage of the large sample of Higgs bosons produced. A systematic search is therefore warranted. Exotic decays are predicted by many BSM theories~\cite{Curtin:2013fra,Cepeda:2021rql}, including extended scalar sectors, SUSY, Higgs or vector portal to dark sector models. The search takes advantage of the full 4-momentum conservation available at FCC-ee to reduce backgrounds in $\epem \to$\,ZH events using the recoil mass. Compared with the reach at the HL-LHC, the FCC-ee will probe exotic Higgs branching fractions that are one to four orders of magnitude lower, down to several times $10^{-5}$, as shown in Fig.~\ref{fig:BSM_exotic_Higgs} from Ref.~\cite{Liu:2016zki}. The gain in sensitivity is most pronounced in hadronic final states (e.g., $b\bar{b}+\ensuremath{E_{\text{T}}^{\text{miss}}}$). In several of these there is no sensitivity at the HL-LHC. It should be noted that in this study, only Z\,$ \to \ell^+\ell^-$ decays are used and further gains in sensitivity are possible by exploiting other Z decay modes.
  
Traditional solutions (e.g., SUSY and Little Higgs) to the electroweak hierarchy problem feature the presence of TeV-scale top-quark partners to control quadratic
contributions to the Higgs mass. However, no evidence for supersymmetric or vector-like top quarks has been found at the LHC so far, up to masses of $\sim 1.2-1.6$~TeV. This has motivated neutral naturalness models in which the top partners are neutral under the SM gauge group (or at least QCD) so that their cross section is small enough to evade searches at the LHC. In this scenario, partner particles may form scalar bound states that mix with the SM Higgs boson and thus decay into SM particles. Evidence for such states can be searched for in $\epem \to \mathrm{Z H}$ events with Z\,$ \to \ell^+\ell^-$ and the Higgs boson decaying into the exotic state, yielding a displaced hadronic vertex signature in the detector. In the particular case of the Folded SUSY model, stop masses up to about 800~GeV can be tested~\cite{Alipour-Fard:2018lsf}, a result that well exceeds indirect constraints from precision Higgs measurements (${\cal B}(\mathrm{H} \to \gamma\gamma)$ specifically). Other models addressing the EW hierarchy problem like relaxion ($\phi$) models can also be probed at the FCC-ee via the process $\epem \to \phi f\bar{f}$ with $\phi \to b\bar{b}$. While precision Higgs measurements at the FCC-ee will exclude heavy relaxions down to $m_\phi\sim 8$~GeV, the Tera-Z run will extend the sensitivity down to about 3~GeV~\cite{Frugiuele:2018coc,Fuchs:2020cmm}.

\subsection{Heavy Neutral Leptons}

Three mysteries stand after the discovery of the Higgs boson: (i) the origin of the masses of the neutrinos; (ii) the origin of the baryon asymmetry in the universe; and (iii) the nature of dark matter.  The FCC-ee provides an exciting opportunity to resolve these mysteries with the discovery of heavy neutral leptons (HNLs), in particular using the large sample of Z bosons produced in early running at the Z resonance.  
 
Experimental observations of neutrino flavor oscillation show that neutrinos have mass.  If the masses of the neutrinos are at least partly generated by the Higgs mechanism, then right-handed neutrinos must exist.  The right-handed neutrino may also have a Majorana mass term $M$ since it has no weak isospin or charge in the Standard Model.  The presence of both mass terms leads to “seesaw” models that have eigenstates with low masses (the observed neutrinos) and with heavier masses (heavy neutral leptons). The HNLs allow generation of the baryon asymmetry, and the lightest HNL is even a candidate for dark matter.  If kinematically allowed, any process that involves “ordinary” neutrinos $\nu$ can also occur with heavy neutral leptons $N$, but suppressed by the mixing $U_{ai}$ between the right-handed neutrino with index $a$ and the left-handed neutrino with index $i$.
 
\begin{figure}[ht!]
\centering
\includegraphics[width=0.7\textwidth]{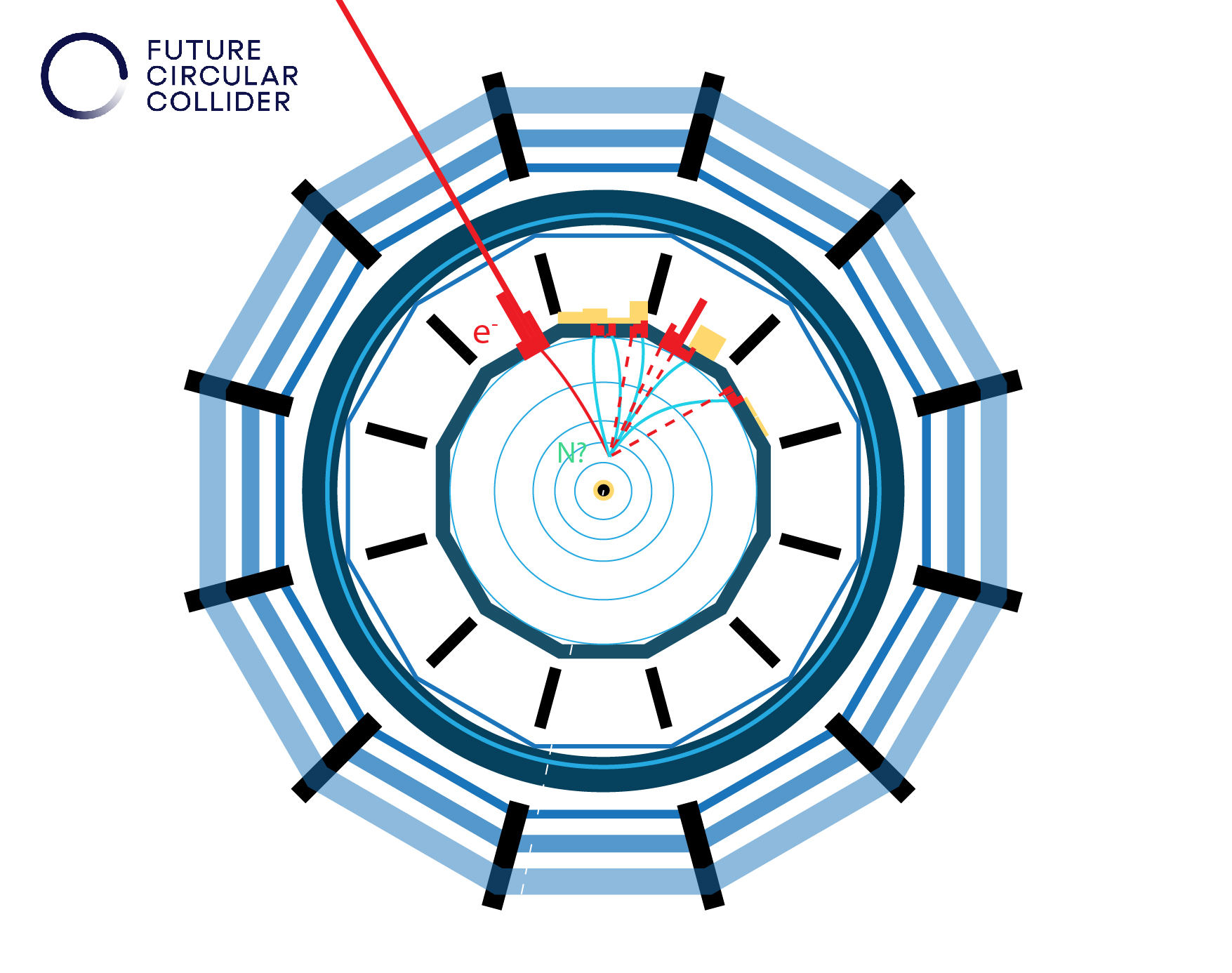}
\caption{Representation of an event display at an FCC-ee detector of a HNL decay into an electron and a virtual W decaying hadronically. Courtesy of the FCC collaboration.}
\label{fig:display}
\end{figure}

A lepton collider can search for HNLs using a large sample of Z bosons produced resonantly. The $s$-channel production of Z\,$ \to \nu N$ followed by the decay of the long-lived heavy neutral lepton $N \to \ell\,\mathrm{W}^*$ leads to a spectacular signature of a displaced vertex with significant mass, which a first analysis~\cite{Blondel:2014bra} indicated should be essentially background free.  Figure~\ref{fig:display} shows how such a possible decay of $N$ at a future FCC-ee experiment would look like. 

\begin{figure}[hbtp]
\centering
\resizebox{0.45\textwidth}{!}{\includegraphics{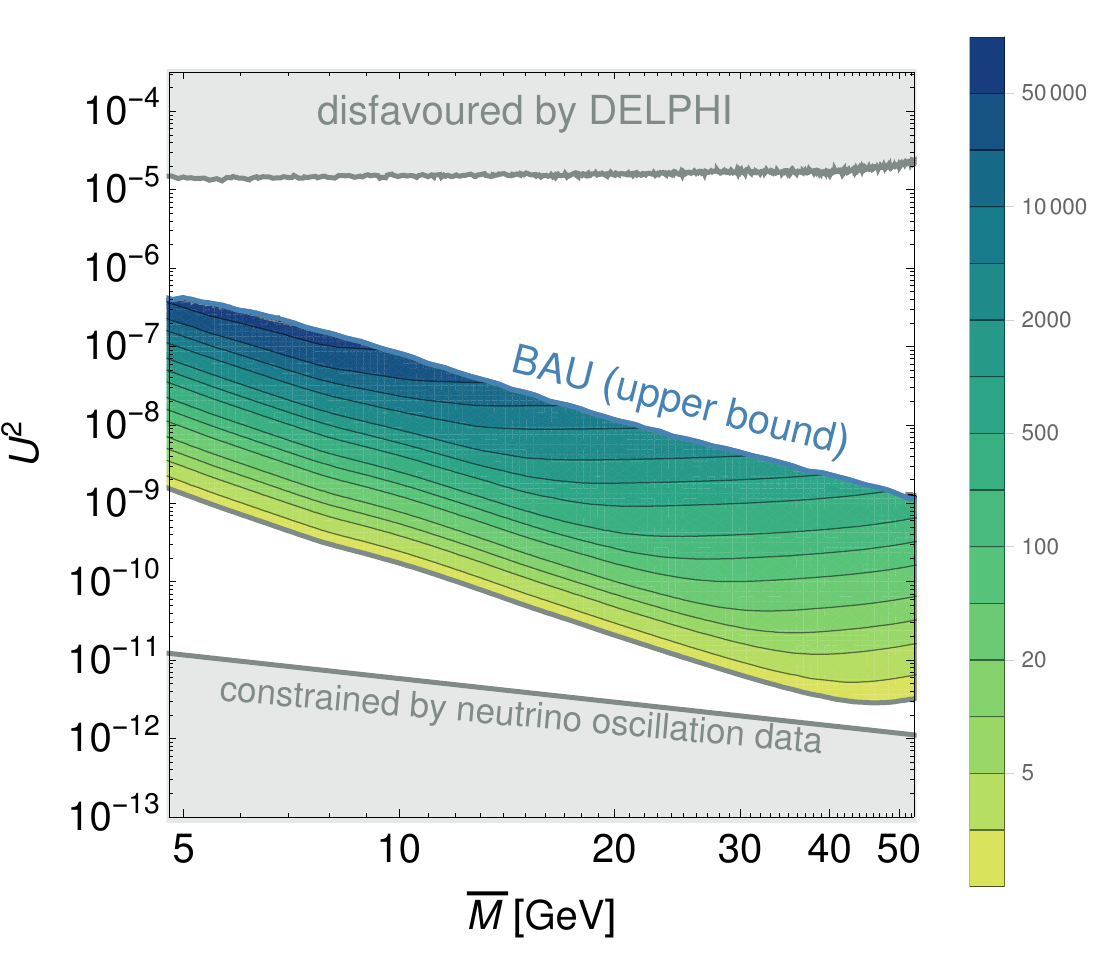}}
\resizebox{0.45\textwidth}{!}{\includegraphics{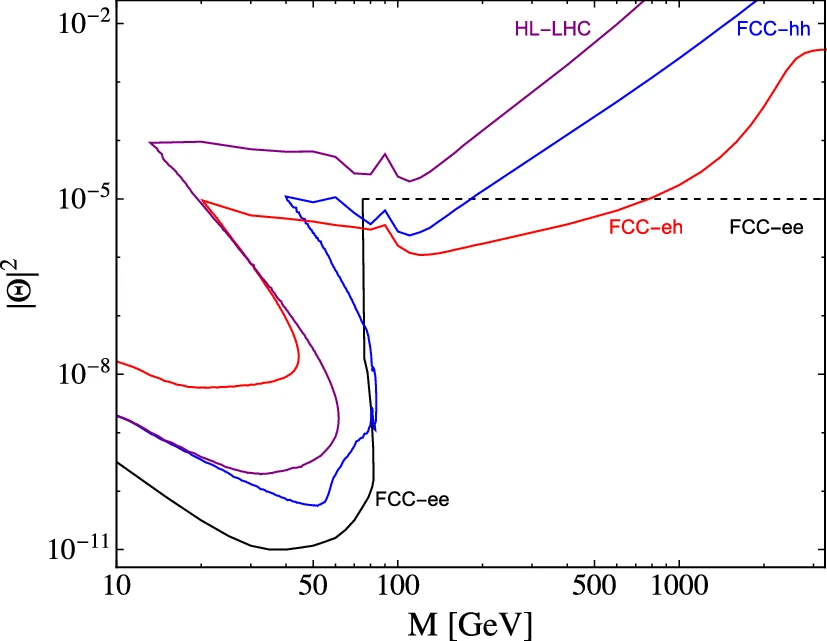}}
\resizebox{0.85\textwidth}{!}{\includegraphics{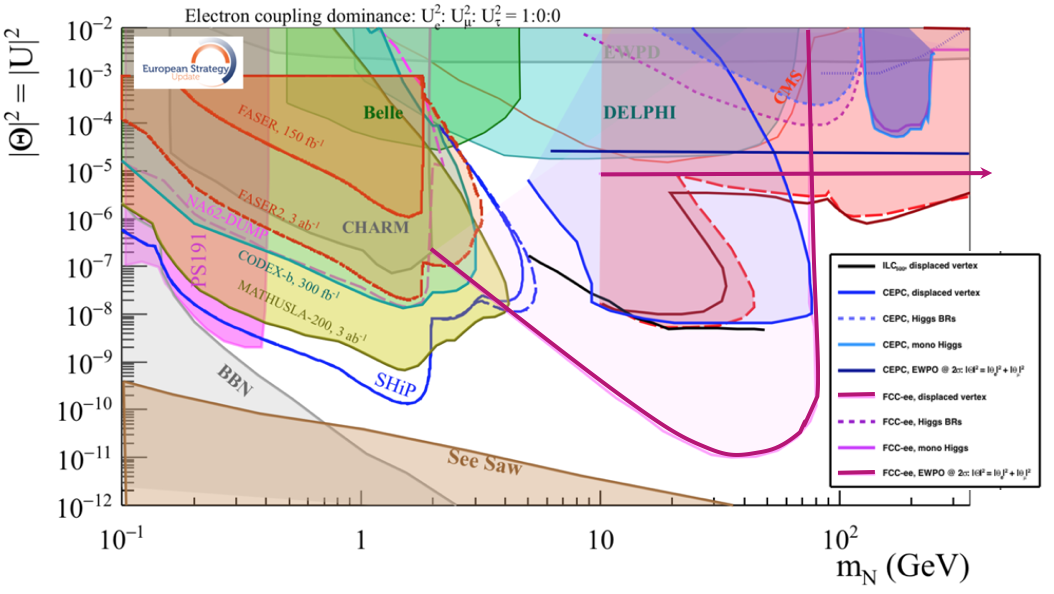}}
\caption{Top left: Figure 11 from Ref.~\cite{Antusch:2017pkq}, showing the number of expected HNLs in the Tera-Z run at the FCC-ee with 20~ab$^{-1}$ for parameters consistent with leptogenesis in models with two HNLs. Results are shown for the normal neutrino mass ordering. 
Top right: Figure 13.2 from Ref.~\cite{Abada:2019lih} showing a summary of HNL search prospects at all FCC facilities. Solid lines are shown for direct searches and the dashed line denotes the constraint from EW precision measurements.  These limits are taken from Ref.~\cite{Antusch:2016ejd} where further details of the signatures and searches can be found. Bottom: Fig~8.19  from the ESPP 2020 briefing book~\cite{ESPP}, showing the FCC-ee sensitivity and complementarity with a compilation of experiments on neutrino beams, beam dumps, at the LHC, and B factories. The direct sensitivity curve at FCC-ee is the  95\% exclusion limits in case of no observation in a sample of $10^{12}$ Z decays. The limit from precision measurements extends to high masses, up to 1000~TeV.  
}\label{fig:BSM_HNL}
\end{figure}

Discovery is possible over a large range of the parameter space of interest, as shown by Fig.~\ref{fig:BSM_HNL} from Ref.~\cite{Antusch:2017pkq} where the color scale shows the potential for a large number of events when the HNL decays inside the detector between 10~$\mu$m and 1.22~m. The blue line is the upper bound on the parameters that would explain the BAU from leptogenesis with two Heavy Neutrinos.  The grey bands show the current bounds from previous searches at LEP and from neutrino oscillations data.  Further, the mass of the HNL could be precisely reconstructed from the mass of the displaced vertex, or, for sufficient decay length, from the time-of-flight, and the flavor of the lepton would determine the flavor-dependent mixing $U^{2}_{a}/U^2$.  The limits from the FCC-ee are complementary to fixed target experiments like SHiP that have sensitivity at lower masses and larger values of $U^2$.  For higher masses, a direct search is also possible at higher collider energies using the $t$ channel exchange of a W boson, though this depends on the size of the coupling to electrons $U_{ae}^2$. 

The measurement of the invisible width of the Z boson is exquisitely sensitive to HNLs that have higher masses or do not decay in the detector. Surprisingly, the run at the WW pair production threshold can exploit the “radiative return” to the Z to obtain an extremely precise measurement of the number of neutrino species, by taking the ratio of the number of events with only the ISR photon to the number of events with an ISR photon and a leptonic Z decay (see Sec.~3.2.3 of Ref.~\cite{Abada:2019lih}).   The excellent overall sensitivity of FCC-ee compared to other FCC colliders is shown in Figure~\ref{fig:BSM_HNL}.

An essential prediction from the seesaw models is that the heavy neutrinos are Majorana particles. The Majorana mass term is in effect a particle to antiparticle transition, which is one of the Sakharov conditions for the BAU generation. The process Z\,$\to \nu N$ or $\bar{\nu} N$  is at first sight insensitive to such an effect since both leptonic charges are produced. Closer investigation, however, shows that, on a statistical basis, differences exist that allow differentiation between the two hypotheses; this possibility is discussed in detail in the FCC-LLP Snowmass white paper~\cite{Verhaaren:2022ify}.

\begin{figure*} 
\centering
\includegraphics[width=0.99\textwidth]{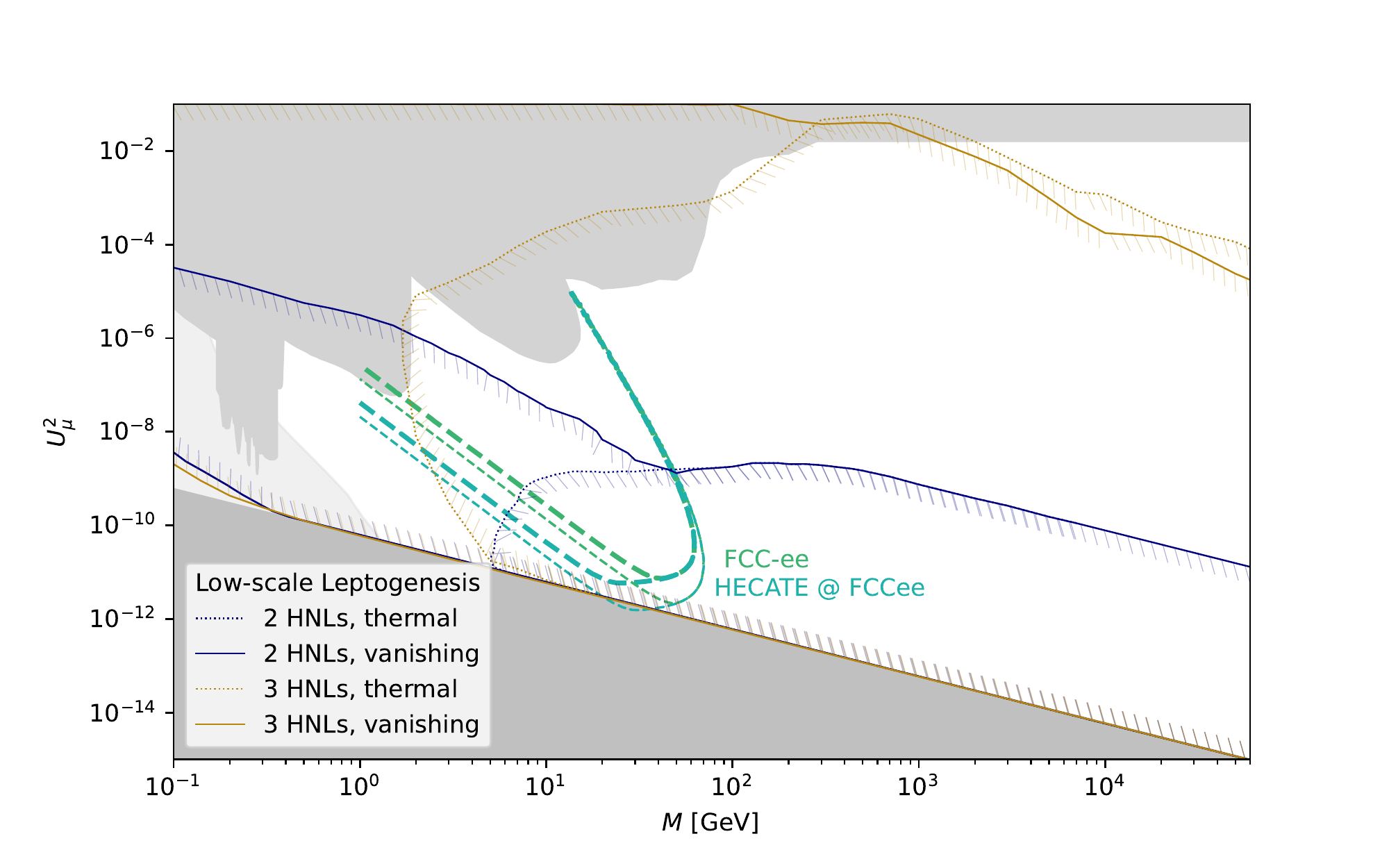}
\caption{Regions in the $(U_{ai},M)$ plane where BAU generation can occur from decays into two and three HNL particles. The contours for 4 events (bold lines) and 1 event (thinner lines) are shown for the LLP displaced vertex analysis only, and an FCC-ee exposure of $5 \times 10^{12}$ Z  decays. The 4-event lines corresponds to the 95\% CL limit in case of no signal observed. The 1-event line defines the contour within which there is a better than 63\% chance to see one or more event -- which for a background-free signal might constitute a discovery.  Curves are shown for observation at a 1.2~m radius volume. In addition, curves are also shown for a putative 5~m radius volume as in the HECATE~\cite{Chrzaszcz:2020emg} setup, increasing the sensitivity for the low mass and couplings regions of the parameter space (Fig. from ~\cite{Verhaaren:2022ify}, where more details are given on the BAU limits). 
}
\label{fig:HNLsummary_1event}
\end{figure*}

The most sensitive access to this question is obtained when the HNL is long lived, so that on an event-by-event basis the decay length and the time-of-flight can be determined. This, thanks to the two-body kinematics of the process, allows determination of both the mass and the proper decay time. For HNL mass which is well above the tau mass, the inclusive $ {\rm Z \to N\nu}$ branching ratio and the average HNL decay rate  are both proportional to the same combination of light to heavy neutrino couplings, $|U_N|^2 \equiv \sum_{\ell = e, \mu, \tau} |U_{\ell N}|^2 $. For a given HNL mass and given the observed production rate, the average HNL lifetime can be thus be predicted, but is twice shorter for a Majorana HNL than for a Dirac particle, because the Majorana particle can decay both in the lepton-number conserving and violating channels, thus in twice as many final states. 

The Dirac vs.\ Majorana information can also be retrieved from the spin conservation combined with the intrinsic parity violation in Z production and decay. Both the charge forward-backward asymmetry and the final state lepton spectrum (from the HNL spin-polarization) provide analysis power even in the case of a prompt HNL decay~\cite{Blondel:2021mss}. Also, in the case of the   HNL decay $N \to \nu \ell^{\pm} \ell^{\mp}  $, the interference between the Z- and W-induced decay leads to observable effects sensitive to the Dirac or Majorana nature of the HNL~\cite{Petcov:1984nf}.

A detailed simulation and reconstruction of the process will be necessary and  of great interest to understand how much statistics will be required, first to establish the existence of the new particle, and then to establish the possible existence of a lepton number violating process (Majorana or Dirac nature). This will also lead to the identification of specific detector requirements to optimize the discovery potential. Compared to the detectors that have been so far proposed for the Higgs studies, the most straightforward improvement appears to be an increase in the detector size with an HNL catcher as in the HECATE proposal~\cite{Chrzaszcz:2020emg} taking advantage of the large size of the caverns in provision for the FCC-hh detectors. The gain in sensitivity is shown in Fig.~\ref {fig:HNLsummary_1event}. Another evident requirement will be to ensure adequate time-of-flight coverage for HNL decays occurring anywhere in the detector, from the vicinity of the luminosity diamond to deep into the calorimeters or magnetic flux return volumes. Finally, it will be essential that the data acquisition be trigger-free.

\section{QCD physics}
\label{sec:qcd}

\textcolor{red}{Editors: D. d'Enterria, and S. Eno}

A precise knowledge of the strong force is a prerequisite to properly interpret all collider measurements in precision SM studies and BSM searches. From color-connection between quark-jets in the decays of weak bosons, to form-factors that influence tests of flavor universality in heavy-quark hadron decays, to virtual loops of quarks (and at higher order, gluons) affecting even measurements with only leptons in the initial and final states, understanding QCD is key to reducing the systematic experimental and theoretical uncertainties on an enormous variety of observables. And of course, studies of the strong force are interesting in their own right, QCD being arguably the most dynamically rich and computationally challenging of the SM forces. As new techniques are developed to further improve the precision of the theoretical calculations including higher fixed-order, log-resummed, and mixed QCD-EW corrections, increasingly clean and high statistics data sets are required to validate the predictions. 
Important questions include: 
what are configurations where perturbative calculations appear to be unexpectedly inadequate; can we improve the understanding of soft and collinear parton evolution in jet showers, as well as any other multiscale observable affected by logarithmic resummation corrections, to a level of precision and accuracy matching those of the fixed-order calculations; is there important physics missing from current phenomenological hadronization models; can we reduce the uncertainty of the QCD coupling constant $\alphasmZ$ well below the $\mathcal{O}(1\%)$ current precision; can we improve mixed perturbative expansions in both $\alphas$ and $\alpha$; are there currently undiscovered stable configurations of quarks and gluons;
will perturbative QCD factorization continue to be an adequate assumption at FCC-eh and FCC-hh energies;
and are there other charged heavy stable particles that affect the running of $\alphas$ at the very high energy scales probed by FCC-hh (the way the top quark influence
on $\alphas$ sets the scale of the proton mass~\cite{quigg2013dis})?

The clean environment and huge data samples available at FCC-ee will deeply probe all these these questions. The FCC-ee offers unique possibilities for high-precision studies of the strong interaction in the clean environment provided by $\epem$ collisions, thanks to its broad span of center-of-mass energies ranging from the Z pole to the top-pair threshold, and its huge integrated luminosities yielding $10^{12}$ and $10^{8}$ jets from Z and W bosons decays, respectively, as well as $10^5$ pure gluon jets from Higgs boson decays. In this chapter, we summarize studies detailed in e.g.~\cite{FCC-ee-accelerator,Skands2016bxb,Proceedings:2017ocd,Proceedings:2019pra,QCDatFCC-ee}
on the impact the FCC-ee will have on our knowledge of the strong force.  The main QCD physics possibilities include: (i) $\alphasmZ$ strong coupling extractions with permil uncertainties, (ii) parton radiation and parton-to-hadron fragmentation functions (splitting functions at NNLO, small-$z$ NNLL resummations, global FF fits,  parton shower MC generators, \dots), (iii) jet properties (light-quark--gluon discrimination, $\epem$ event shapes and multijet rates at NNLO+N$^{\rm n}$LL, jet broadening and angularities, jet substructure, jet charge determination, $\epem$ jet reconstruction algorithms, \etc), (iii) heavy-quark jets (dead cone effect, charm-bottom separation, gluon-to-$\ccbar,\bbar$ splitting, \etc); and (iv) nonperturbative QCD phenomena (color reconnection, baryon and strangeness production, Bose-Einstein and Fermi-Dirac final-state correlations, color string dynamics: spin effects, helix hadronization, \dots). A few physics cases are just briefly discussed below.

\subsection{Strong coupling constant}

The strong coupling constant is one of the fundamental constants of nature. The precision of all calculations of observables (cross sections, decays,...) that are sensitive, either directly or virtually, to quarks and gluons depend on its accurate determination.
However, the current uncertainty on $\alphasmZ$ is $\pm$0.85\%~\cite{pdg}, making it the least well-known of the four SM interaction couplings (g, g', ${\rm G_F}$, and $\alphas$). The most precise extraction is that from lattice QCD~\cite{flag} (0.1182$\pm$0.0008, with a 0.6\% uncertainty) where experimental input is used to set the ``physical scale''. Scale sources depend on the calculational method used, but various lattice collaborations have used $f_{\pi}$, $f_{K}$, the proton mass, the $\Omega$ mass, quarkonium spectra or measurements of $\alphas$ from colliders, among others~\cite{flag}). The value of $\alphasmZ$ dominates the parametric uncertainties in the theoretical predictions of electroweak observables, limiting our ability to probe high energy scales via precision electroweak fits.  Table~2 of Ref.~\cite{freitas2019theoretical} indicates these uncertainties dominate theory uncertainties on the Z width and the ratio of the \bbar to total and lepton to total Z hadronic cross sections. Table~5 of this reference also indicates that uncertainties in $\alphas$ are an important uncertainty in the branching fraction of Higgs to gluons, and to $\ccbar$ pairs. 
An precise knowledge of $\alphas$ is also mandatory for an accurate extraction of the top-quark mass by comparing the experimental threshold cross sections, $\ee \to \ttbar$, to the theoretical predictions~\cite{Hoang:2020iah}.

\begin{figure}[hbtp!]
\centering
\resizebox{0.48\textwidth}{!}{\includegraphics{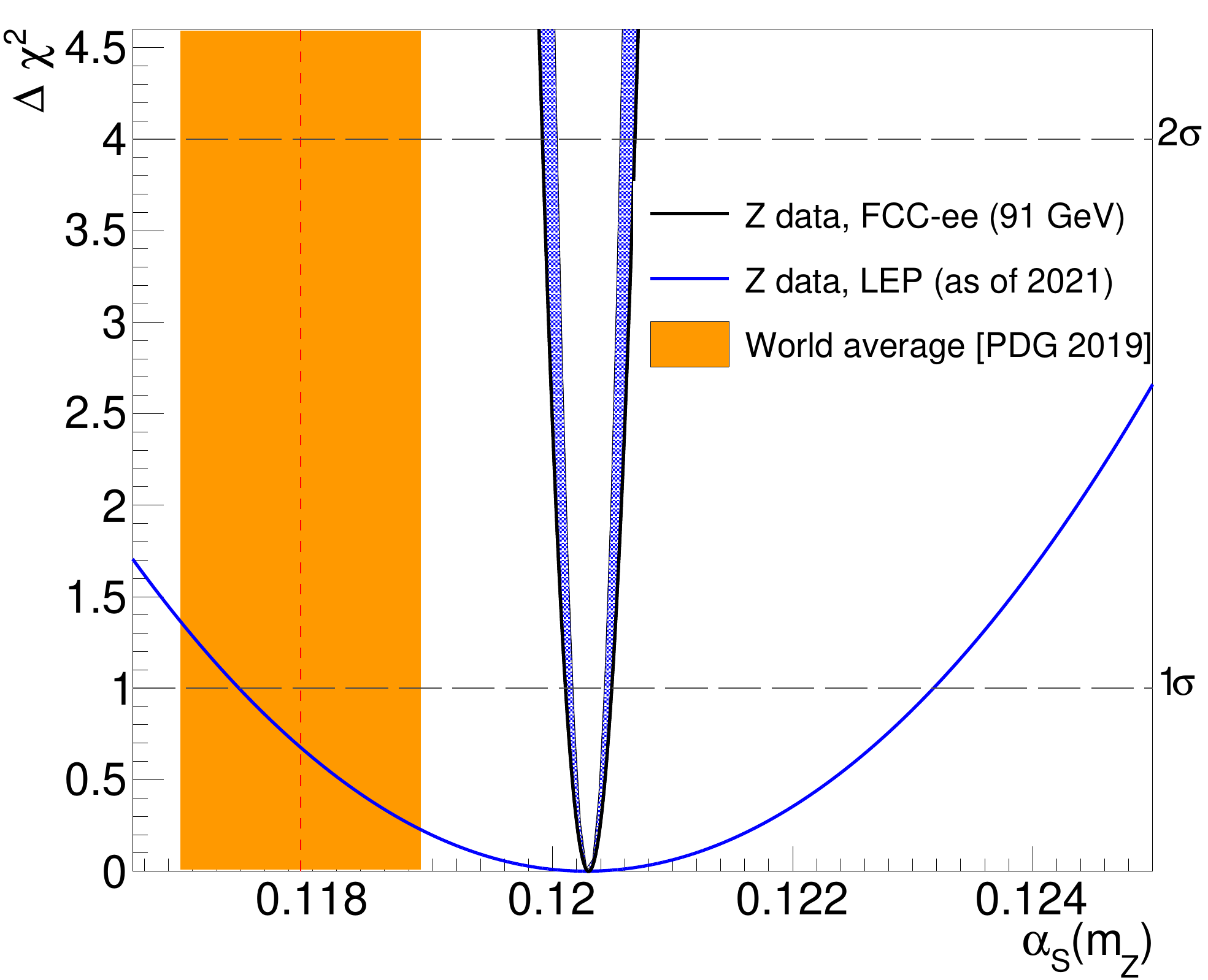}}
\resizebox{0.49\textwidth}{!}{\includegraphics{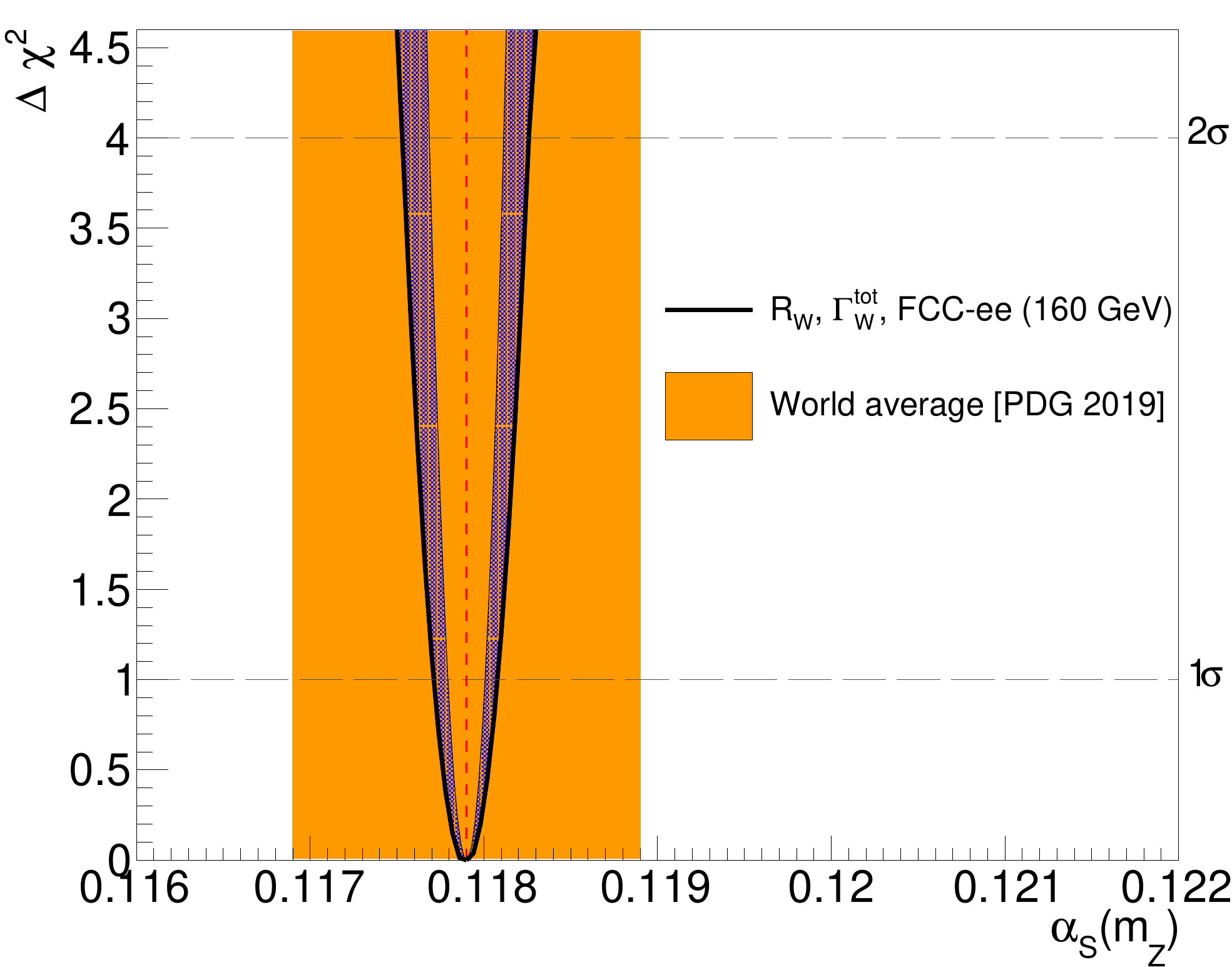}}
\caption{$\chi^2$-profiles of $\alphas(m_\mathrm{Z})$ extracted from the expected Z-boson (left) and W-boson (right) hadronic observables at FCC-ee (blue parabolas) compared to the PDG world average (orange band). The Z-boson result is centered at its current LEP-based value of $\alphasmZ = 0.1203$, whereas the W-boson result is arbitrarily set to the PDG world $\alphasmZ$ average. Figs.~from Ref.~\cite{dEnterria:2020cpv}.
\label{fig:alphas_FCCee}}
\end{figure}

A circular electron-positron collider running at the Z pole provides the ideal environment to reduce the $\alphasmZ$ uncertainty. The strong coupling can be measured to better than 0.1\%, with an integrated luminosity of $4.6 \times 10^{36}$\cms exploiting three Z-boson hadronic pseudo-observables~\cite{dEnterria:2020cpv} (Fig.~\ref{fig:alphas_FCCee}, left). The theoretical uncertainties can be arbitrarily reduced by systematically incorporating missing higher-order QCD and mixed QCD+EW corrections~\cite{astheory,Baikov:2016tgj,freitas2019theoretical}, leading to an almost guarantee of unprecedented precision.
The uncertainty can be further reduced by using measurements from the large samples of hadronic W boson~\cite{dEnterria:2016rbf,dEnterria:2020cpv} (Fig.~\ref{fig:alphas_FCCee}, right) and $\tau$-lepton~\cite{Pich:2020qna} decays, as well as studying jet production rates and event shapes~\cite{Verbytskyi:2019zhh}, especially as experimental systematics are understood and reduced to the level of the FCC-ee statistical uncertainties. Currently the systematic uncertainty on $\alphas$~\cite{pdg} extracted from $\tau$ decays is 1.6\%, and from hadronic final states at electron-positron colliders is 2.6\%~\cite{pdg}. The authors of Ref.~\cite{QCDatFCC-ee} note that jet grooming techniques may reduce hadronization and other nonperturbative uncertainties, although a systematic study of these techniques at FCC-ee energies is needed.
 
\subsection{Parton showering and nonperturbative QCD}

The clean environment of FCC-ee will also provide enormous (multi)jet data samples to improve our understanding of parton showers, higher-order logarithmic resummations, as well as hadronization and nonperturbative phenomena.
QCD effects, such as b-quark showering and hadronization, are for example the leading sources of systematic uncertainty in the measurement of the forward-backward asymmetry of b quarks in $\epem$ collisions at LEP, $\AFBb = 0.0992\pm0.0015 \pm 0.0007$, which remains today the electroweak precision observable with the largest disagreement (2.4$\sigma$) with respect to the SM prediction, $(\AFBb)_{_{\rm th}} = 0.1030 \pm 0.0002$~\cite{ALEPH:2005ab}. Confirmation or resolution of this long-term discrepancy requires a Tera-Z machine collecting orders-of-magnitude more data at the Z pole to significantly reduce the current statistical and systematic uncertainties~\cite{dEnterria:2020cgt}. Other examples discussed in Refs.~\cite{Skands2016bxb,dEnterria:2016rbf} include measurements of event shape variable, jet multiplicities, particle spectra by particle type, \etc~that typically had uncertainties of order 10\% at LEP.

Non-perturbative uncertainties from final-state hadronic effects linked to power-suppressed infrared phenomena such as color reconnection (CR), hadronization, and multiparticle correlations (in spin, color, space, momenta) --- which cannot be currently computed from first-principles QCD and often rely on phenomenological Monte Carlo modeling-- will be also precisely studied at the FCC-ee~\cite{Proceedings:2017ocd}. In $\epem \to \ttbar$, as the top quarks decay and hadronize close to each other, interaction and interference between them, their decay bottom quarks, and/or any radiated gluons affect the rearrangement of the color flow and thereby the kinematic distributions of the final hadronic state. Whereas the perturbative radiation in the process can be in principle theoretically controlled, there is a CR ``cross talk'' among the produced hadronic strings that can only be modelled phenomenologically~\cite{Khoze:1994fu}.
In the pp case, such CR effects can decrease the precision that can be achieved in the extraction of the top mass, and constitute 20--40\% of its uncertainty~\cite{Argyropoulos:2014zoa}. Color reconnection can also impact limits for CP-violation searches in $\rm H \to \mathrm{W^+ W^-}$ hadronic decays~\cite{Christiansen:2015yca}. Searches for such effects can be optimally studied in the process $\epem \to \rm \mathrm{W^+W^-} \to q_1\bar{q}_2 q_3\bar{q}_4$~\cite{Christiansen:2015yca}, where CR can lead to the formation of alternative ``flipped'' singlets $\rm q_1\bar{q}_4$ and $\rm q_3\bar{q}_2$, and correspondingly more complicated string topologies~\cite{Sjostrand:1993hi}. The combination of results from all four LEP collaborations excluded the no-CR null hypothesis at 99.5\% CL~\cite{ALEPH:2013dgf}, but the size of the WW data sample was too small for any quantitative studies. At the FCC-ee, with the W mass determined to better than 1~MeV by a threshold scan, the semileptonic WW measurements (unaffected by CR) can be used to probe the impact of CR in the hadronic WW events~\cite{Proceedings:2017ocd,Abada:2019lih}. 

The enormous data sets available at the Tera-Z will certainly lead to negligible statistical (and improved systematic) uncertainties of such observables, sensitive to the geometric pattern of soft QCD interference that are instrumental to our understanding of the strong interaction at all orders. As an example, in Ref.~\cite{ALEPH:2003obs} (from 2004), the systematic uncertainty due to modeling of charged particles was obtained by varying the track quality selection requirements in the simulated samples until the observed and predicted number of selected events agreed.  Especially, as noted in the section on determination of charged multiplicity, uncertainties in the tracking efficiency at low momentum were important.
However, even at the LHC, with larger uncertainties on particle production due to underlying event, multiple parton scattering, \etc~and denser and more-forward (and therefore more challenging to model) track production, the data-simulation agreement for a basic quantity such as number of selected events is usually better than this.  Further improvement in generators, trackers with less material, higher reconstruction
efficiency and wider angular and momentum acceptance, and  improvements in the detector simulation GEANT4 code~\cite{Agostinelli:2002hh}  will only reduce the systematic uncertainties.
State-of-the-art detectors ---with advanced hadron identification capabilities (e.g., via timing, \dEdx, and/or \dNdx, see Section~\ref{sec:instrumentation})--- and modern data analysis methods, e.g., by exploiting the jet substructure and the Lund plane approaches~\cite{Marzani:2019hun,Dreyer:2021hhr}, should further reduce these uncertainties, 
leading to much improved tests and understanding of strong force calculations.

In summary, the FCC-ee provides a unique opportunity for reducing the uncertainty of the strong coupling constant to the permil level, and for improving our theoretical understanding of all QCD dynamical regimes (fixed-order, resummation, nonperturbative, \etc) to a similar level of precision. In particular, {\bf a ten-times improved determination of the strong coupling requires a Tera-Z facility.} FCC-ee, FCC-hh, and FCC-eh provide a complementary frontier to studies that probe the high density QCD regime at the future electron-ion collider (EIC) at Brookhaven National Laboratory.  It is hard to imagine a better set of laboratories for studying the strong force.
\section{Flavor physics}
\label{sec:flavor}
\textcolor{red}{Editors: W. Altmannshofer, M. Dam, D. Hill, G. Isidori, G. Landsberg, R. Novotny,  A.~Pich}

The Z pole run of the FCC-ee will provide unprecedented data samples of $\mathcal{O}(5\times10^{12})$ Z bosons decaying into $\mathrm{Z}\to b\bar{b}$, $\mathrm{Z}\to c\bar{c}$, and $\mathrm{Z} \to \tau^+ 
\tau^-$ events that will be recorded without any triggers or prescales.  
This provides the opportunity to enrich the knowledge of the flavor physics of quarks and leptons both quantitatively and qualitatively.
The flavor program of FCC-ee will be a natural continuation of the upgraded LHCb experiment run at the LHC \cite{LHCb:2018roe} and the Belle II experiment \cite{Belle-II:2018jsg}.
The anticipated production yields of heavy-flavored particles at FCC-ee are compared to the Belle II experiment in Table \ref{Tab:Flavor:ProductioYields}.

\begin{table}[!htp]
\centering
\caption{Expected production yields of heavy-flavored particles at Belle II ($50~\mathrm{ab^{-1}}$) and FCC-ee (Z pole). The $X/\bar{X}$ represents the production of a $B$-hadron or its charge conjugated state. The Z branching fractions and hadronization rates are taken from \cite{pdg}.\vspace{0.25cm}}
\begin{tabular}{c c c c c c c c}
\hline
Particle production ($10^9$)& $B^0/\bar{B}^0$&$B^+/B^-$& $B_s^0/\bar{B}_s^0$&$B_c^+/\bar{B}_c^-$& $\Lambda_b/\bar{\Lambda}_b$ &$c\bar{c}$& $\tau^+\tau^-$\\\hline
Belle II& 27.5 &27.5 &n/a &n/a &n/a &65 &45\\
FCC-ee &620 &620  &150 &4 &130 &600 &170\\\hline
\end{tabular}
\label{Tab:Flavor:ProductioYields}
\end{table}

\subsection{Decays of b-flavored hadrons} \label{sec:b_decays}

As shown in Table~\ref{Tab:Flavor:ProductioYields}, the Z pole run of the FCC-ee will produce $\mathcal O(10^{12})$ neutral and charged $B$ mesons, $\sim 10^{11}$ $B_s$ mesons and $\Lambda_b$ baryons, as well as few$\times 10^9$ $B_c$ mesons. The huge sample of $b$-flavored hadrons in an experimentally clean $\epem$ environment combined with the large boost from the decay of the Z boson will enable a rich $b$ decay program at FCC-ee with unique sensitivities.

\paragraph{Leptonic and semileptonic flavor changing neutral current decays:} In recent years, the semileptonic decays $B \to K^{(*)} \epem$ and $B\to K^{(*)}\mu^+\mu^-$ have attracted considerable attention due to a number of persistent $2\sigma - 3\sigma$ tensions between data and SM expectations, in particular in the lepton flavor universality ratios $R_{K^{(*)}}$~\cite{LHCb:2017avl,LHCb:2019hip,LHCb:2021trn}, and the angular observable $P_5^\prime$~\cite{LHCb:2020lmf}. 
With an expected $B$ meson data set that exceeds that of Belle II by an order of magnitude, FCC-ee is well-positioned to provide an independent confirmation of these so-called ``$B$ anomalies'', should they indeed be signs of new physics. If confirmed, the anomalies in the rare $B$ decays establish a generic new physics scale of $\sim 35$~TeV~\cite{Altmannshofer:2017yso,DiLuzio:2017chi} and thus provide clear targets for direct searches at FCC-hh.

\begin{figure}[tbh]
\centering
\includegraphics[width=0.495\linewidth]{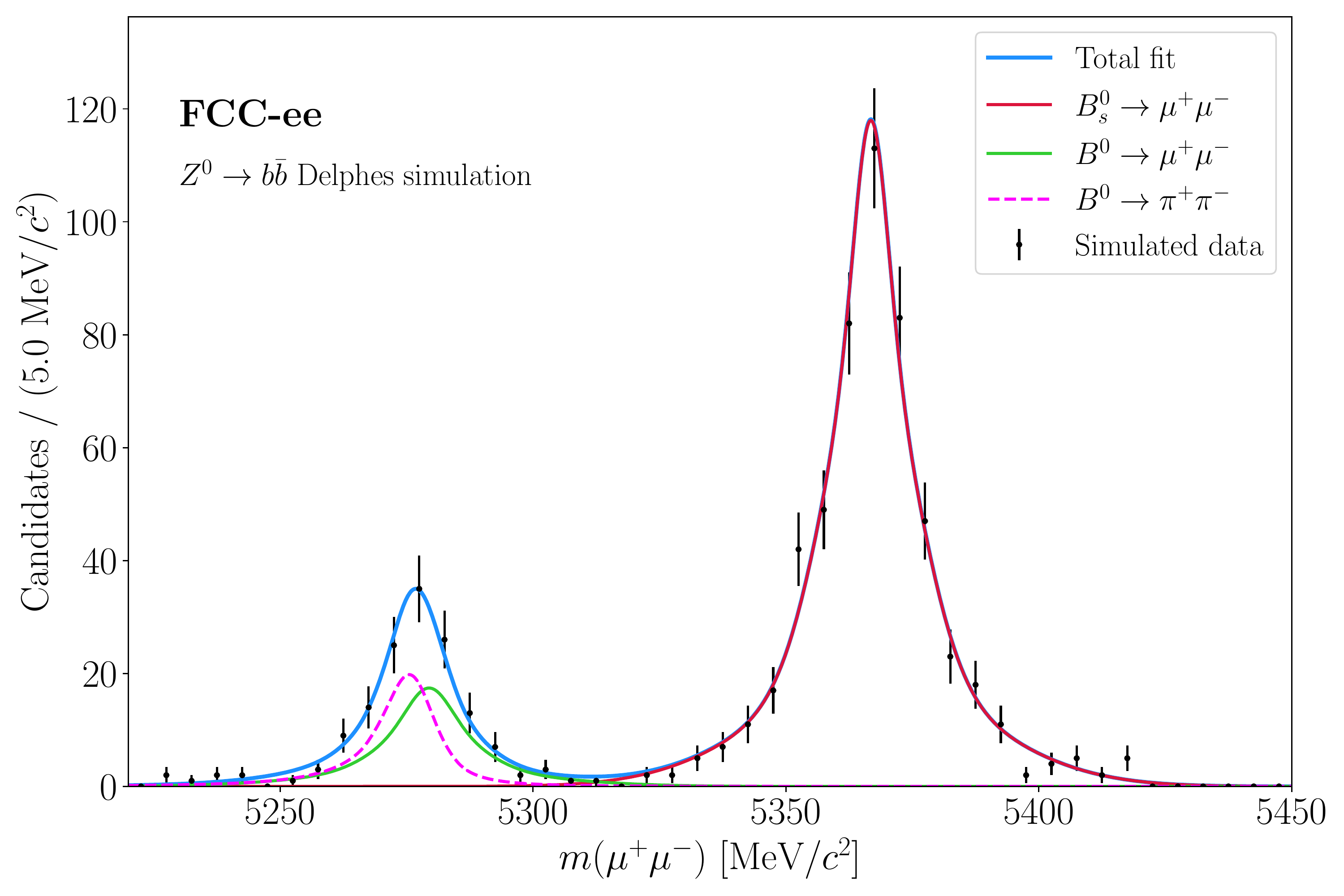} \quad 
\includegraphics[width=0.465\linewidth]{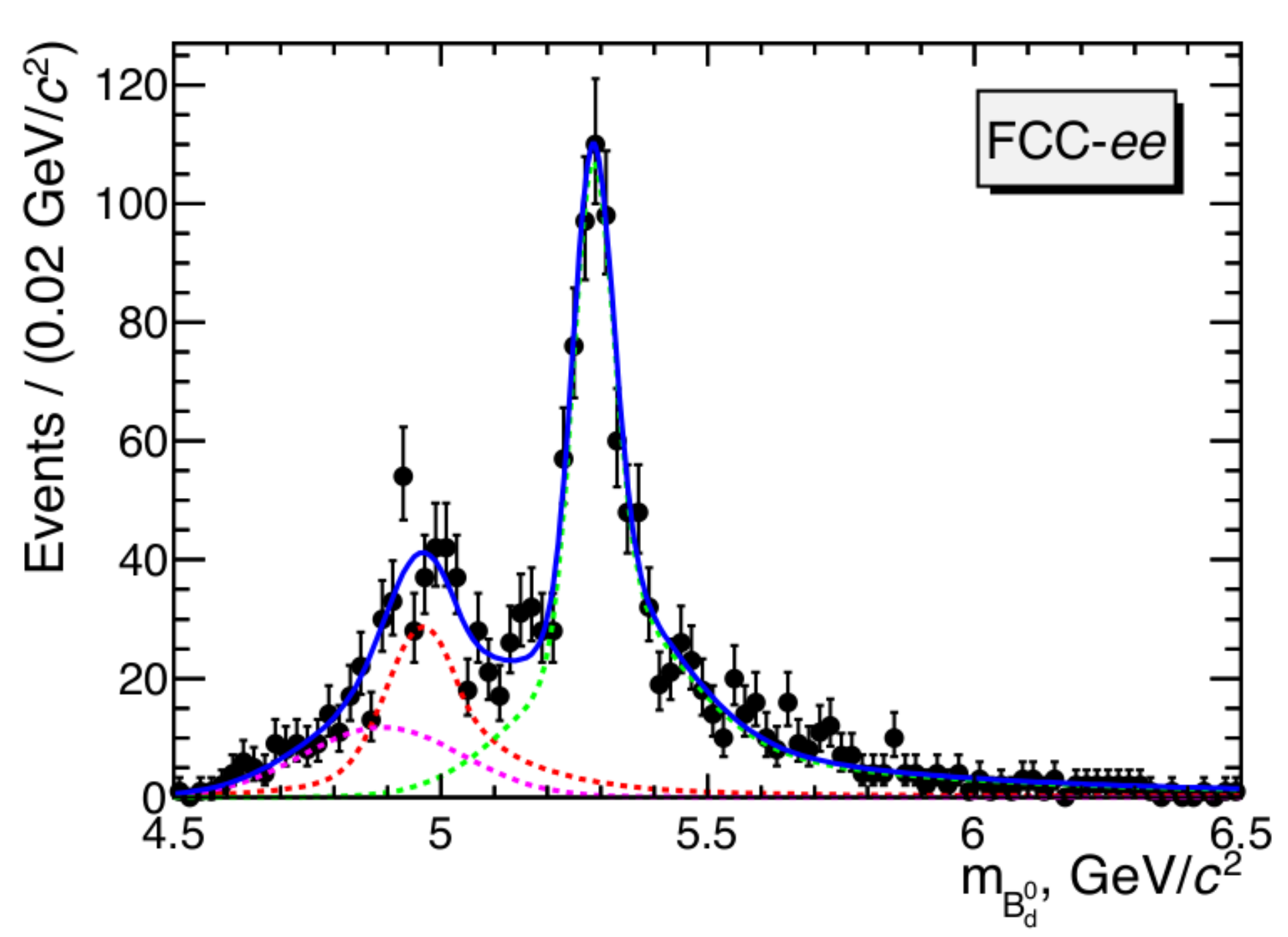}
\caption{Left: Reconstructed dimuon invariant mass of $B^0 \to \mu^+ \mu^-$ and $B_s \to \mu^+ \mu^-$ candidates for $5 \times 10^{12}$ Z decays (from~\cite{Monteil:2021ith}). Right: Reconstructed $K^* \tau^+ \tau^-$ invariant mass of $\bar B^0 \to K^{*0} \tau^+ \tau^-$ candidates based on $10^{13}$ Z decays. The signal component is shown in green. Backgrounds from $\bar B_s \to D_s^+ D_s^- K^*$ and $B^0 \to D_s^+ \bar K^{*0} \tau^- \nu$ are in red and pink, respectively (from~\cite{Kamenik:2017ghi}).}
\label{fig:rareB}
\end{figure}

Of particular interest are tauonic and semitauonic decays, for which FCC-ee has unique sensitivities. Current bounds on the branching ratios of decays like $B_s \to \tau^+ \tau^-$, $B^0 \to \tau^+ \tau^-$, and $B \to K^{(*)} \tau^+ \tau^-$, are still several orders of magnitude above the SM predictions~\cite{BaBar:2005mbx,BaBar:2016wgb,LHCb:2017myy,Belle:2021ndr}. 
Sensitivities will improve at the HL-LHC and at Belle II~\cite{LHCb:2018roe,Belle-II:2018jsg} but cannot reach the SM.
Precision measurements of these decays are highly motivated to complete the studies of lepton flavor universality in $b \to s \ell\ell$ and $b \to d \ell \ell$ decays. Many BSM scenarios predict characteristic effects in the decays with taus in the final state~\cite{Altmannshofer:2014cfa,deMedeirosVarzielas:2015yxm,Capdevila:2017iqn,Bordone:2018nbg}. The $B \to K^* \tau^+ \tau^-$ decay is a particularly rich laboratory to probe new physics. Using three-prong tau decays and assuming high precision vertex reconstruction, it has been shown that this decay can be fully reconstructed at Z-pole machines~\cite{Kamenik:2017ghi,Li:2020bvr}. At the FCC-ee, $\mathcal O(10^3)$ cleanly reconstructed SM events can be expected (Fig.~\ref{fig:rareB}, right). Such an event sample will not only allow a precision measurement of the $B \to K^* \tau^+ \tau^-$ branching ratio, but also opens up the possibility of measuring the angular distribution of the decay~\cite{Kamenik:2017ghi}.

The leptonic decays $B_s \to  \mu^+ \mu^-$ and $B^0 \to \mu^+ \mu^-$ have extremely small branching ratios in the SM of $(3.66 \pm 0.14)\times 10^{-9}$ and $(1.03 \pm 0.05)\times 10^{-10}$, respectively~\cite{Beneke:2019slt}.
Their well known tiny branching ratios make them highly sensitive probes of new physics~\cite{Li:2014fea, Arnan:2017lxi, Altmannshofer:2017wqy}. The $B_s \to  \mu^+ \mu^-$ decay has been observed at the LHC, but so far only upper bounds exist for $B^0 \to \mu^+ \mu^-$~\cite{CMS:2014xfa,ATLAS:2018cur,CMS:2019bbr,LHCb:2021vsc}. Observation of $B^0 \to \mu^+ \mu^-$ is in reach of the HL-LHC~\cite{Cerri:2018ypt}.
One advantage of FCC-ee over the LHC is the excellent mass resolution that allows a clear separation of the $B_s$ and $B^0$ signals in the dimuon invariant mass spectrum (Fig.~\ref{fig:rareB}, left). Starting from $5 \times 10^{12}$ Z decays, one expects $\sim 540$ reconstructed $B_s \to \mu^+\mu^-$ events and $\sim 70$ reconstructed $B^0 \to \mu^+\mu^-$ events in the SM~\cite{Monteil:2021ith}. Important for the the observation of $B^0 \to \mu^+\mu^-$ at FCC-ee will be the control of the $B^0 \to \pi^+ \pi^-$ background, using particle identification techniques such as $\mathrm{d}E/\mathrm{d}x$, time of flight, and Cherenkov detectors. With high performance flavor tagging, also measurements of CP asymmetries in $B_s \to \mu^+\mu^-$ might become possible at FCC-ee.

\paragraph{Decays with missing energy:} The FCNC decays $B \to K \nu\bar\nu$ and $B\to K^* \nu \bar\nu$ are well established probes of new physics~\cite{Altmannshofer:2009ma,Buras:2014fpa}. Belle II is expected to make first observation of these decays and measure their branching ratios with an uncertainty of $\sim 10\%$~\cite{Belle-II:2018jsg}. Given the expected number of $B$ mesons produced at a Z-pole run, FCC-ee should be able to further improve these measurements, which is highly motivated given that these decays are theoretically well understood. Moreover, FCC-ee has the unique opportunity to measure the related decays $B_s \to \phi \nu\bar\nu$, $\Lambda_b \to \Lambda \nu\bar\nu$, and even $B_c \to D_s \nu\bar\nu$ with high precision. A precision of $\mathcal O(\%)$ might be possible for BR$(B_s \to \phi \nu\bar\nu)$~\cite{Li:2022tov}. Combining the results from the whole family of $b \to s \nu \nu$ decays, i.e. the pseudoscalar to pseudoscalar transitions $B \to K \nu\bar\nu$ and $B_c \to D_s \nu\bar\nu$, the pseudoscalar to vector transitions $B\to K^* \nu \bar\nu$ and $B_s \to \phi \nu\bar\nu$, and the fermion to fermion transition $\Lambda_b \to \Lambda \nu\bar\nu$, will be a powerful way to probe BSM physics.

\paragraph{$B_c$ physics:} The $B_c$ meson is still largely uncharted territory. While several of its decay modes have been observed by LHCb, little quantitative information exists about its decay branching ratios due to the lack of established normalization modes. Very interesting are the theoretically clean leptonic decays $B_c \to \tau \nu$ and $B_c \to \mu \nu$ that have new physics sensitivity that complements the well studied decay modes $B \to \tau \nu$ and $B \to D^{(*)} \tau \nu$. The ratio $\mathcal{B}(B_c \to \mu \nu)/\mathcal{B}(B_c \to \tau \nu)$ is of particular interest in view of the existing anomalies in the lepton flavor universality ratios $R_{D^{(*)}}$. $B_c$ mesons are not produced at Belle II, and the limited final state information that is available renders a measurement of $B_c \to \tau \nu$ infeasible at hadron colliders. FCC-ee is in a unique position to make precision measurements of $B_c$ decays. While the $B_c$ was not observed at LEP, one expects as many as few$\times 10^9$ $B_c$ mesons from the Z-pole run of FCC-ee. Starting from $5 \times 10^{12}$ Z decays and assuming the SM branching ratio of $\mathcal{B}(B_c \to \tau \nu) = (1.95 \pm 0.09)\%$~\cite{Amhis:2021cfy} one can expect a huge SM signal yield. Using three-prong tau decays, the study~\cite{Amhis:2021cfy} finds that a high purity signal sample containing $\sim 4000$ $B_c \to \tau \nu$ decays is achievable. Such a sample would allow the $B_c \to \tau \nu$ branching ratio to be determined with a few percent precision. Similar conclusions have been reached in~\cite{Zheng:2020ult}.
Compared to $B_c \to \tau \nu$, the branching ratio of $B_c \to \mu \nu$ is suppressed by a factor $\sim m_\mu^2/m_\tau^2 \sim 3 \times 10^{-3}$. Still, one can expect $\mathcal O(10^5)$ $B_c \to \mu \nu$ events at FCC-ee. A precision measurement of $\mathcal{B}(B_c \to \mu \nu)$ will require good control over the background from $B_c \to \tau \nu$ with $\tau \to \mu \nu \nu$.

\subsection{Precise CKM and CP-violation parameters studies}
The measurement of CKM angles and the CP-violation parameters are of high interest and since many of the observables in CP-violation studies are very precisely predicted they warrant continued experimental attention.
All current experimental results exhibit remarkable agreement with the SM predictions, however, they still leave room for BSM contributions to CP-violating transitions. 
The measurements, that become limited by systematic biases at the LHC, will benefit from the cleaner and very different environment of the FCC-ee, especially decay modes involving $B_s$, $B_c$ or b-baryons with neutral final state particles.
Due to better particle identification, it is expected that flavor-tagging efficiency will be significantly higher than in the LHC era.
This will be a large advantage for any time-dependent measurement.

The measurement of CP violation parameters in the $B_\mathrm{d,s}$ meson mixing have been well established in many b hadron decays.
These systems are very sensitive to any new BSM contribution because the box diagrams that drive the oscillations and carry CP-violating phases are the neutral entry point for any new BSM particle. 
To probe deviations from the SM predictions and test the BSM models, increased precision
on $|V_{ub}|$ and $\gamma$ is required \cite{Charles:2015gya}.
Considering the sample size collected at FCC-ee, it will be possible to measure the relevant observables with similar or better precision to previous experiments \cite{Abada:2019lih,Barbieri:2021wrc}.

A particular strength of the FCC-ee flavor program will be the ability to make very sensitive studies of decays containing neutral particles. 
This possibility will enable to measure various CP-violating asymmetries such as time-dependent CP asymmetry in $B^0 \to \pi^0 \pi^0$ decay or measurement of CP asymmetry in $B^-\to DK^-$ (where D indicates a superposition of $D^0$ and $\bar{D}^0$) and $B_s \to D_s^{(*)\pm} K^\mp$. 
Another benefit of the FCC-ee environment will be the possibility to measure semileptonic CP-violating asymmetries and determinations of the $|V_{ub}/V_{cb}|$ performed with $B_s^0$ mesons and $\Lambda_b$ baryons that are not accessible at current experiments with enough precision.

All these measurements are sensitive to various parameters of the CKM triangle.
It is expected that FCC-ee will provide precise determination of $|V_{cb} |$, and the semileptonic asymmetries $a^d_{sl}$ and $a^s_{sl}$ that are key parameters for the CKM.
It is expected that first observation of CP-violation in B mixing will be within reach and the analysis of the BSM contributions in box mixing processes will provide valuable test of BSM physics. 
Assuming the minimum flavor violation scenario, where the new flavor structures are aligned with the SM Yukawa couplings, the energy scale up to 20~TeV can be probed.

\subsection{Charged-lepton flavor violating decays}

The unprecedented number of Z bosons ($5 \times 10^{12}$) and a large sample of Higgs bosons ($\sim 10^6$) expected to be produced at FCC-ee open up a very rich program of searches for charged-lepton flavor violation (CLFV). In the following, we review the expected sensitivity coming from a number of dedicated studies, as well as a few ideas that have not been explored in detail yet.

\subsubsection{Higgs boson flavor violating decays to leptons}

The present best direct limits on the branching fractions of the $\mathrm{H} \to e\mu$, $\mathrm{H} \to e\tau$, and $\mathrm{H} \to \mu\tau$ decays are $6 \times 10^{-5}$~\cite{ATLAS:2019old}, $2.2 \times 10^{-3}$~\cite{CMS:2021rsq}, and $1.5 \times 10^{-3}$~\cite{CMS:2021rsq} at 95\% CL, respectively. Indirect limits from searches for $\mu \to e\gamma$ conversion by the MEG experiment~\cite{MEG:2013oxv} can be interpreted as indirect limits on ${\cal B}(H \to e\mu) < 10^{-8}$. Present indirect limits from analogous $\tau \to e\gamma$ or $\tau \to \mu\gamma$ decays on the corresponding branching fractions of CLFV Higgs boson decays are $\sim 10\%$, i.e. weaker than the direct limits.
While the direct limits will continue improving as (HL-)LHC accumulates more data, the improvement will be less than an order of magnitude.

With about one million Higgs bosons produced in association with the Z boson at FCC-ee, about the same sensitivity in the $e\mu$ channel and about a factor of two better sensitivity in the other two channels as after full HL-LHC running can be obtained: $1.2 \times 10^{-5}$, $1.6 \times 10^{-4}$, and $1.4 \times 10^{-4}$, respectively~\cite{Qin:2017aju}. In these studies, only the dominant $\mathrm{Z}(jj)$ decay channel was considered, which is not background-free in the $\tau$ channels. In addition, the $\tau$ decay is reconstructed only in the muonic (electronic) channel for the $\mathrm{H} \to e\tau$ ($\mathrm{H} \to \mu\tau$) decay, which corresponds to only 17\% branching fraction. Considering other decay channels of the $\tau$ lepton, in particularly hadronic decay mode, as well as cleaner decay modes of the Z boson, may result in a further improvement of the FCC-ee sensitivity to these LFV processes.

\subsubsection{Z boson charged-lepton flavor violating decays}

The improvement compared to the HL-LHC is expected to be significantly better for the LFV Z boson decays. The most stringent limits on the branching fractions of the $\mathrm{Z} \to e\mu$, $\mathrm{Z} \to e\tau$, and $\mathrm{Z} \to \mu\tau$ to date come from the LHC experiments and amount to $7.5 \times 10^{-7}$~\cite{ATLAS:2014vur}, $5.0 \times 10^{-6}$~\cite{ATLAS:2021bdj}, and $6.5 \times 10^{-6}$~\cite{ATLAS:2021bdj} at 95\% CL, respectively. The sensitivity is expected to improve to $10^{-7}$ ($10^{-6}$) level in the $e\mu$ ($\ell\tau$) channels at the HL-LHC.

With $5 \times 10^{12}$ Z bosons produced at FCC-ee, the sensitivity to the branching fractions of the $\mathrm{Z} \to e\tau$ and $\mathrm{Z} \to \mu\tau$ decays is expected to reach $10^{-9}$ level~\cite{Dam:2018rfz}, while for the $\mathrm{Z} \to e\mu$ decay it is expected to be in the range $10^{-10}$ to $10^{-8}$, depending on to which degree the major background from $\mathrm{Z} \to \mu\mu$ decays with a muon being misreconstructed as an electron can be controlled (e.g., using $\mathrm{d}E/\mathrm{d}x$ information)~\cite{Dam:2018rfz}. This corresponds to up to three orders of magnitude improvement achievable at FCC-ee.

\subsubsection{Decays of the tau to three muons and to a muon and a photon}

The current best limit on ${\cal B}(\tau \to 3\mu)$ comes from Belle~\cite{Hayasaka:2010np} and is $2.1 \times 10^{-8}$ at 90\% CL. The LHCb experiment set a limit of $4.6 \times 10^{-8}$ using $\tau$ leptons copiously produced in $B$ and $D$ meson decays, based on an integrated luminosity of 3 fb$^{-1}$ at 7--8~TeV~\cite{LHCb:2014kws}. Given that the upgraded LHCb detector at HL-LHC is expected to record about 100 times that amount of data, and at a higher production cross section at 14~TeV, the sensitivity is expected to improve by only about an order of magnitude (as in the Belle case, there is sizable background). The CMS measurement~\cite{CMS:2020kwy} using both heavy-flavor and $\mathrm{W} \to \tau\nu$ channels based on 36 fb$^{-1}$ of data at 13~TeV set a limit of $8.0 \times 10^{-8}$  at 90\% CL on this branching fraction. With full HL-LHC data set, an improvement by about an order of magnitude is expected, similar to that in LHCb.

Projected sensitivity from Belle II on this branching fraction will reach $3.3 \times 10^{-10}$~\cite{Belle-II:2018jsg}. Given that the number of $\tau$ leptons produced at Belle II is about 4 times less than in the $\mathrm{Z} \to \tau^+\tau^-$ at FCC-ee (as shown in Table~\ref{Tab:Flavor:ProductioYields}), FCC-ee is expected to reach sensitivity $\sim 10^{-10}$ to this branching fraction~\cite{Dam:2018rfz}, which is a few times better than the expected sensitivity at Belle II.

The sensitivity to the $\tau \to \mu\gamma$ decay branching fraction at FCC-ee is expected to reach about $2 \times 10^{-9}$~\cite{Dam:2018rfz}. Preliminary studies done for Belle II expect about 5 times better sensitivity, thanks to the beam energy constraint that helps eliminating the dominant $\epem \to \tau^+\tau^-\gamma$ background~\cite{Moore:2016vhk,Belle-II:2018jsg}. Nevertheless, subsequent FCC-ee studies~\cite{Dam:2021ibi} suggested that the $2 \times 10^{-9}$ sensitivity may be a conservative estimate, thus bringing FCC-ee and Belle II sensitivities closer.

A number of other CLFV tau decays can be studied at FCC-ee, similar to what has been achieved by the B factories. Particle identification, which will be available in the FCC-ee detectors, should make these measurements highly competitive with the ultimate precision achievable in Belle II.

\subsubsection{Other measurements with tau leptons}

\begin{figure}[tbh]
\centering
\includegraphics[width=0.495\linewidth]{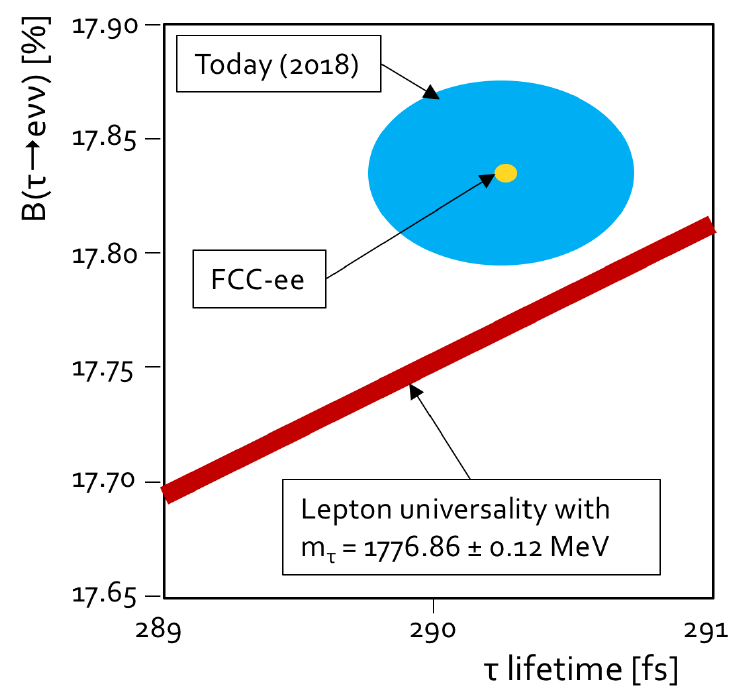} 
\caption{Branching fraction of $\tau\to e\bar\nu_e\nu_\tau$ vs.\ $\tau$ lepton lifetime. The current world averages of the direct measurements are indicated with the blue ellipse, while the projected FCC-ee accuracy is given by the yellow ellipse (from Ref.~\cite{Dam:2018rfz}). The red line corresponds to the prediction based on lepton universality given the present world-average value $\tau$ lepton mass, which may be further improved by the current and proposed charm factories in the future.}
\label{fig:LFU}
\end{figure}

Finally, the large $\tau$ samples expected at FCC-ee, should allow to measure the $\tau$ lepton lifetime to an absolute precision of 0.04~fs ($10^{-4}$ relative precision) and leptonic branching fractions to an absolute precision of $3 \times 10^{-5}$ ($2 \times 10^{-4}$ relative precision)~\cite{Dam:2018rfz}. This would allow to measure the Fermi constant in $\tau$ decays to a similar or even higher precision (potentially as good as $10^{-5}$ if the systematic uncertainties can be kept at the same level as the statistical ones). Comparing this number with the canonical $G_F$ measurement based on the muon lifetime~\cite{pdg}, offers another way of probing new physics possibly responsible for non-flavor-universal couplings, as shown in Fig.~\ref{fig:LFU}~\cite{Dam:2018rfz}.

Another direct test of lepton favor universality can be achieved at FCC-ee by the precision simultaneous measurement of the branching fractions of the $\tau \to \mu\bar\nu_\mu\nu_\tau$ and $\tau \to e\bar\nu_e\nu_\tau$ decays. The present $2 \times 10^{-3}$ precision~\cite{pdg} in these branching fractions is still unchallenged since the LEP times, where it was possible to control the systematic uncertainties very well. The FCC-ee will be able to achieve the statistical precision of $\sim 10^{-5}$ in these branching fraction measurements~\cite{Dam:2021ibi}. Detailed studies are needed to demonstrate that the systematic uncertainties in the ratio of the two branching fractions can be kept under control. This might require not just the experimental improvements, but also full understanding of radiative corrections to leptonic tau decays. These studies will be continued in the future, and have a potential to provide most stringent lepton universality tests achievable.

Several other important measurements in the tau sector, presently dominated by the LEP data, will be significantly improved at the FCC-ee, e.g., the measurement of the spectral functions needed to determine the strong coupling $\alpha_s$ and the Cabibbo mixing $V_{us}$  from hadronic tau decays. These would allow precision tests of QCD at a qualitatively new level of accuracy.

\section{FCC-hh and FCC-ep physics}
\label{sec:fcchh}
\textcolor{red}{Editors: E. Barberis,  R. Harris,  M. Mangano, M. Selvaggi, and L.-T. Wang }

The FCC project begins with an exploration of the precision frontier at FCC-ee and culminates with the deepest possible probe of the energy frontier at FCC-hh~\cite{Benedikt:2018csr}.  A 100~TeV proton-proton collider, within the tunnel built for FCC-ee, is a pragmatic and familiar plan to attain a machine with the highest discovery potential.  The crucial motivations of FCC-hh are outlined below, but first we point out that that these two stages of FCC are coherent.  FCC-hh builds on the precision Higgs measurements of FCC-ee, complementing and extending them. The FCC-hh would be needed to understand the nature of any new physics indirectly observed at FCC-ee, providing many additional channels for measurement and discovery and a broad program of physics, resulting from the many types of partons of varying momentum within colliding protons, interacting both strongly and weakly.


\subsection{FCC-hh}

\begin{figure}[ht]
\centering
\includegraphics[width=0.58\hsize]{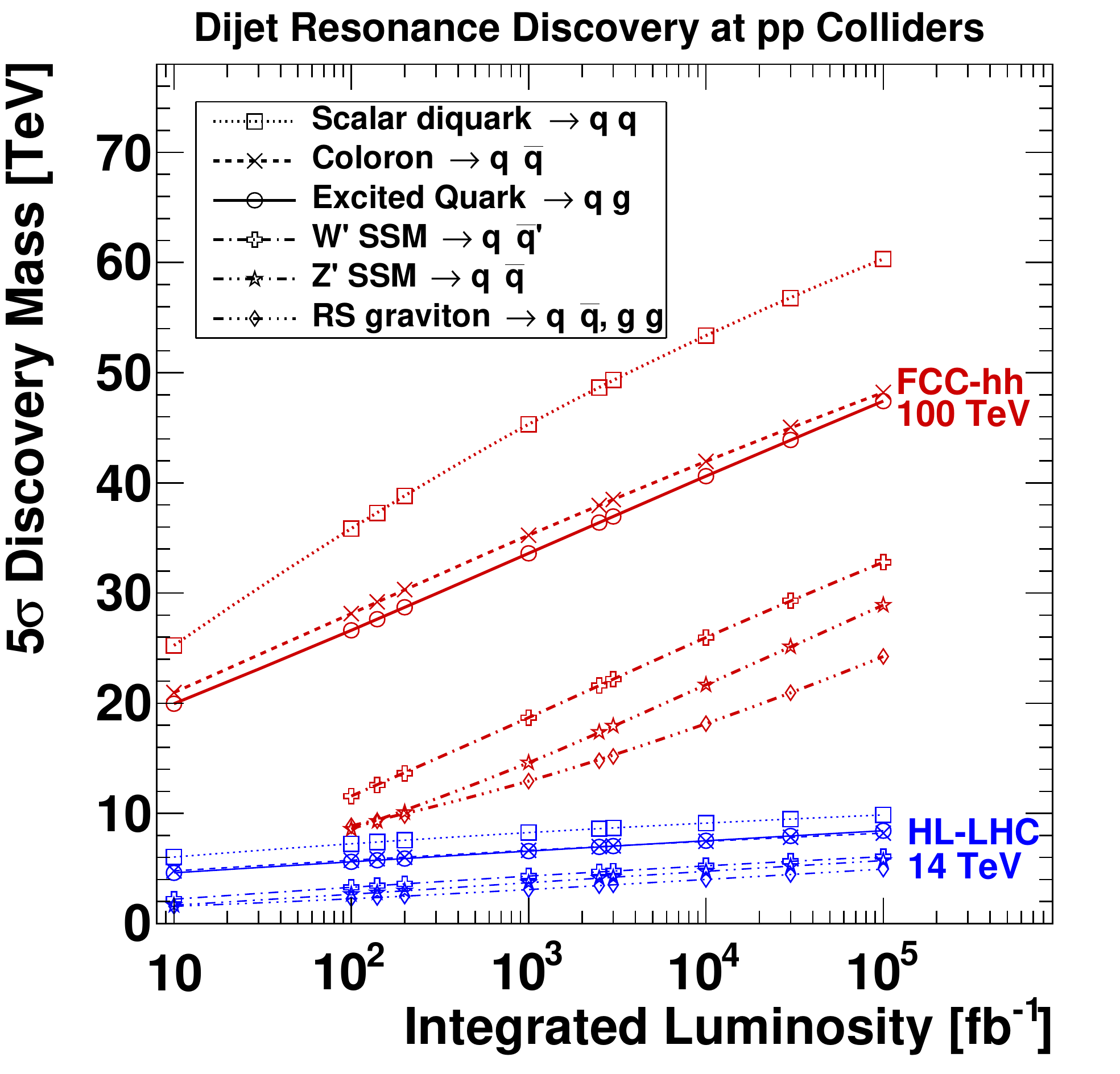}
\includegraphics[width=0.41\hsize]{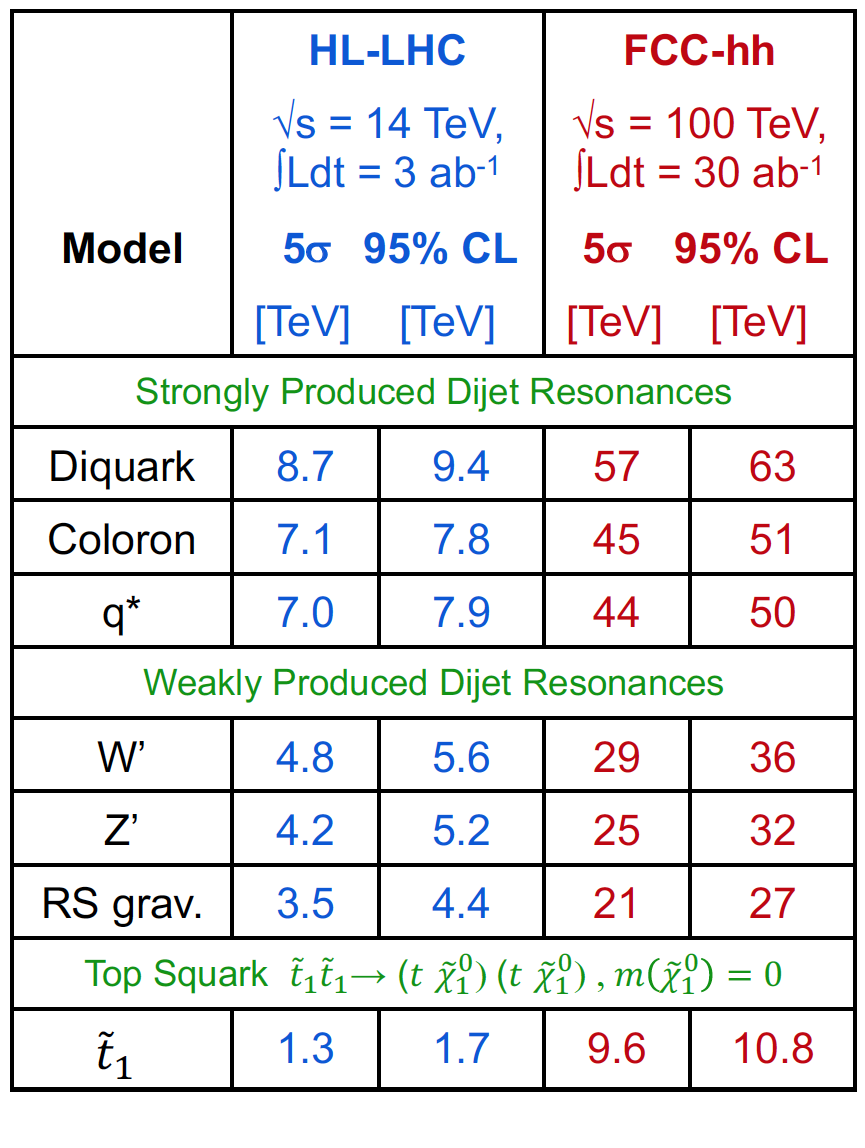}
\caption{Comparison of the sensitivity of HL-LHC and FCC-hh for models of new physics. Left: The $5\sigma$ discovery sensitivity for dijet resonances as a function of integrated luminosity from Ref.~\cite{harris2022sensitivity}. Right: The $5\sigma$ discovery and 95\% CL exclusion sensitivities at baseline integrated luminosity for dijet resonances and for a top squark from Ref.~\cite{harris2022sensitivity,Abada:2019lih,CidVidal:2018eel,eps_strategy}.}
\label{figFCC-LHC-Dijet}
\end{figure}

\begin{figure}[ht]
\centering
\includegraphics[width=0.8\hsize]{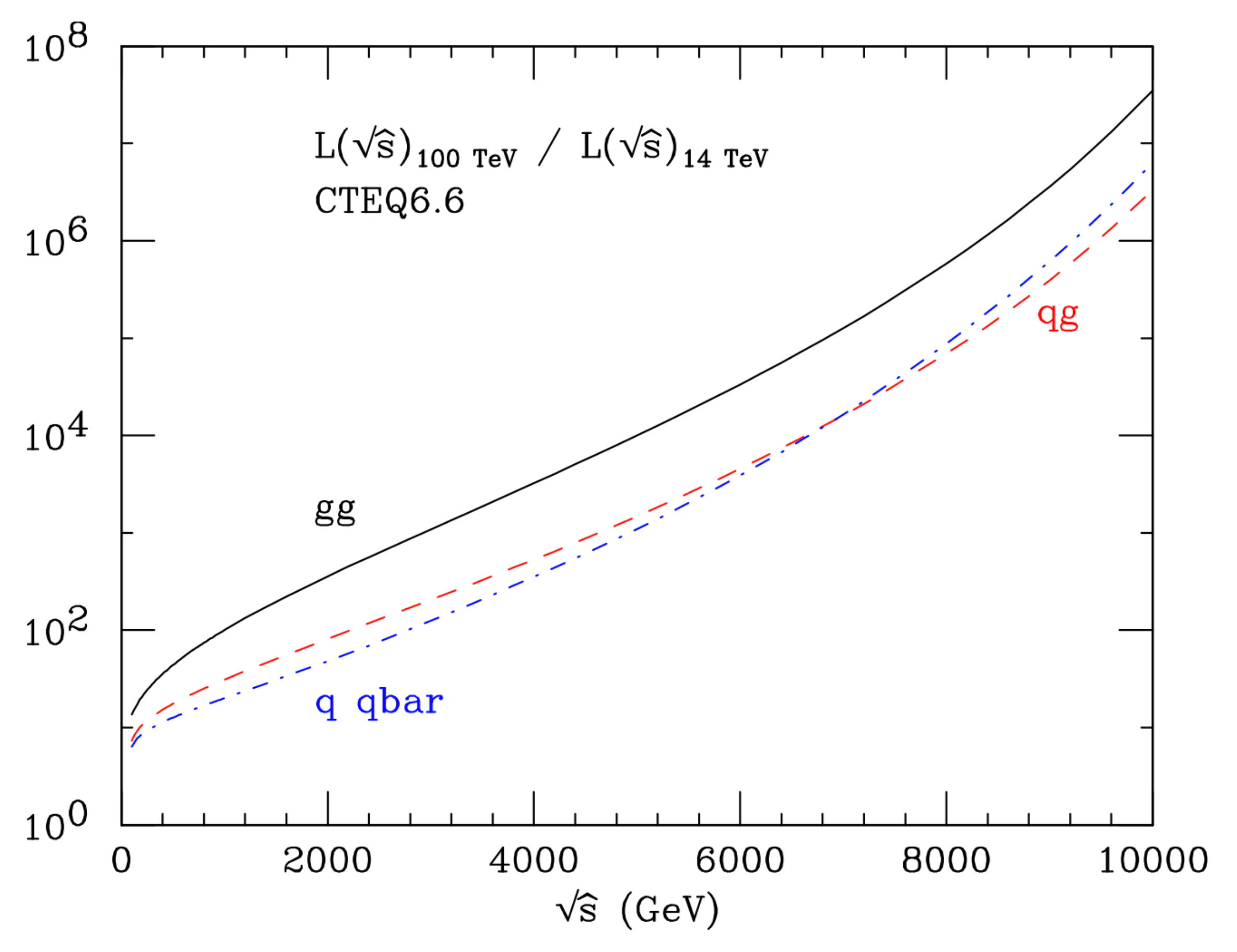}
\caption{Ratio of partonic luminosities at 100 and 14~TeV, as a function of partonic center-of-mass energy
$\hat{s}$, for different partonic initial states. From Ref.~\cite{Hinchliffe:2015qma}.}
\label{figFCC-hh-pdflumi}
\end{figure}

The experimental motivation for FCC-hh is the exploration of the energy frontier. By virtue of the increased collision energy compared to LHC, with sufficient integrated luminosity~\cite{Hinchliffe:2015qma} FCC-hh can proportionally extend the mass reach for discovery of many new phenomena. For example, Fig.~\ref{figFCC-LHC-Dijet} taken from Ref.~\cite{harris2022sensitivity} shows the sensitivity to a dijet resonance, typically the earliest search for new physics conducted at hadron colliders, and demonstrates that the mass reach of FCC-hh with 30~ab$^{-1}$ is roughly six times that of HL-LHC with 3~ab$^{-1}$.  The advantage of energy is realized soon after turning on FCC-hh, as shown in Fig.~\ref{figFCC-hh-pdflumi} by the huge ratio of parton-luminosities compared to HL-LHC, for final states with large masses for HL-LHC which are relatively small masses for FCC-hh.  With just 10 fb$^{-1}$ of integrated luminosity, accumulated during the first year of data taking, the mass reach for strongly produced objects like a coloron or an excited quark, is already three times that of HL-LHC with a full 3~ab$^{-1}$, and well beyond the HL-LHC total collision energy. A  modest additional mass reach beyond the baseline can be reached by running at higher luminosity. For this phenomena, with 100~ab$^{-1}$ at FCC-hh, a factor of seven increase in mass reach is achieved over HL-LHC, fully proportional to the increase in collision energy.

FCC-hh will help to address some of the most important outstanding questions in particle physics and cosmology, such as the mechanism of electroweak symmetry breaking and the nature of dark matter. 
Additional particles are needed to extend the standard model to answer these questions. 
The 100~TeV collision energy of FCC-hh not only increases the sensitive mass range for the discovery of such new particles, but it is necessary and sufficient to give clear answers in critical areas. For example, the nature of the Higgs potential, and whether thermal weakly interacting massive particles (WIMPs) are the dark matter.

FCC-hh will measure the properties of the Higgs boson with unprecedented precision. With 30~ab$^{-1}$ it will produce over 20 billion Higgs particles, and over 30 million Higgs pairs, more than a hundred times the HL-LHC. This will allow the measurement of Higgs couplings within $\sim$1\% as tabulated in Table~\ref{tab:Hhh}.  The large center-of-mass energy of the FCC-hh will also enable the precise
measurement of the differential cross section of Higgs bosons produced with large transverse momentum, $p_\mathrm{T}$. These measurements could reach a precision of up to 10\%, for $p_\mathrm{T}$ values in excess of 1~TeV, allowing FCC-hh to probe new physics via the presence of higher-dimension operators within an EFT framework.

FCC-hh will explore the electroweak phase transition (EWPT) that occurred in the early universe and created the mass of the Higgs boson (for a review see Ref.~\cite{Ramsey-Musolf:2019lsf}). At the FCC-hh the measurement of the Higgs self-coupling $\delta\lambda/\lambda$ will not be limited by statistics. Depending on the assumed detector performance and systematic uncertainties, the Higgs self-coupling will be measured with a precision in the range 3.4–7.8\% at 68\% confidence level~\cite{hhsensitivity}. For these estimates, the detector performance was defined by scenarios ranging from established LHC Run 2 (optimistic), to HL-LHC performance projected using today's algorithms (pessimistic). The 10$\%$ precision threshold could be achieved already with 3~ab$^{-1}$, assuming the expected systematics can be achieved during this early phase of data taking. The self-coupling is related to the nature of the EWPT, and the possibility in BSM scenarios that it is strongly of first order, and then may have played a role in the matter/antimatter asymmetry of the universe. The expected precision of the FCC-hh measurement of the self-coupling, together with its discovery reach for new particles in BSM scenarios, have the potential to conclusively test scenarios leading to a strong first-order EWPT. 

There is both complementarity and synergy between FCC-hh and FCC-ee in Higgs physics. First, FCC-ee provides the absolute and most precise measurement of the Higgs coupling to the Z boson and of the total Higgs width, as well as of the largest rate decays. It has the greatest sensitivity to elusive SM hadronic decays, such as $\mathrm{H}\to gg$ and $\mathrm{H}\to \ccbar$, and others within some BSM scenarios~\cite{Liu:2016zki,Curtin:2013fra}. Then, following the FCC-ee, the FCC-hh provides the most precision for the rare SM decays $\mathrm{H}\to\mu\mu$, $\gamma\gamma$, $\mathrm{Z}\,\gamma$, as well as the ttH coupling and the important Higgs self-coupling.  The critical synergy between the two colliders is that FCC-ee input is essential to reduce the parametric systematics of the aforementioned FCC-hh measurements. For example, Table~\ref{tab:Hhh} shows that for the rare SM decays $\mathrm{H}\to\mu\mu$, $\gamma\gamma$, $\mathrm{Z}\,\gamma$, FCC-hh measures precisely the ratio (R) of branching fractions ($\mathcal{B}$) in each channel ($xx$) relative to the branching fraction into four leptons, $\mathcal{B}(\mathrm{H}\to xx)/\mathcal{B}(\mathrm{H}\to 4l$). The absolute determination of the branching fractions for these rare SM decays then relies on the precise measurement of $\mathcal{B}(\mathrm{H}\to 4l)$ at FCC-ee.  Similarly, the extraction of ttH comes from the measurement of the ratio ttH/ttZ at FCC-hh and the precision measurement of ttZ at FCC-ee with $\sqrt{s}=365$~GeV. This joint measurement of ttH by FCC-hh facilitated by FCC-ee, in turns allows a precision extraction of the Higgs self-coupling, $\lambda$, from HH production at FCC-hh. This is because the box and triangle diagrams for HH production involve the top quark in the loop, which interfere and enter with different powers of the ttH coupling. Precise measurements of the ttZ coupling at FCC-ee allow us to avoid theoretical uncertainties in that coupling when extracting both ttH and the crucial Higgs self-coupling from measurements at FCC-hh.

One of the central open questions in the Standard Model is the origin of the electroweak scale. Any potential underlying mechanism would have to explain the vast difference between the EW scale and other fundamental scales, such as the Planck scale. Many solutions of this so-called hierarchy problem point to the presence of new physics not too far beyond the~TeV scale. Such new physics will continue to be a main physics target at the HL-LHC. FCC-hh will significantly enhance the reach, and provide a much more
stringent test of these ideas. In the unlikely case that a discovery is not made, the FCC-hh would force a dramatic rethinking, and significantly alter the direction of the quest for an explanation of the weak scale.  


\begin{table}[ht!]
\centering
\caption{\label{tab:Hhh}
Target precision for the parameters relative to the measurement of various Higgs decays and ratios thereof~\cite{Benedikt:2018csr}, and of the Higgs  self-coupling $\lambda$ with 30~ab$^{-1}$ at FCC-hh~\cite{hhsensitivity}. Notice that Lagrangian couplings have a precision that is typically half that of what is shown here, since all rates and branching ratios depend quadratically on the couplings. The range given for $\lambda$'s precision refers to the three scenarios mentioned in the text.\vspace{2.5mm}} 
\begin{tabular}{l|c|c|c} 
  \hline
   Observable & Parameter & Precision & Precision
   \\ &  & (stat) & (stat+syst+lumi) 
   \\ \hline 
  $\mu=\sigma$(H)$\times\mathcal{B}$(H$\to \gamma\gamma$)   & $\delta\mu/\mu$ & 0.1\% & 1.45\%
  \\
  $\mu=\sigma$(H)$\times\mathcal{B}$(H$\to\mu\mu$)   & $\delta\mu/\mu$ & 0.28\% & 1.22\%
  \\
  $\mu=\sigma$(H)$\times\mathcal{B}$(H$\to 4\mu$)   &  $\delta\mu/\mu$ & 0.18\% & 1.85\%
  \\
  $\mu=\sigma$(H)$\times\mathcal{B}$(H$\to \gamma\mu\mu$)   &  $\delta\mu/\mu$ & 0.55\% & 1.61\%
  \\
  $R=\mathcal{B}$(H$\to \mu\mu)/\mathcal{B}$(H$\to 4\mu$)   & $\delta R/R$ & 0.33\% & 1.3\%
  \\
  $R=\mathcal{B}$(H$\to \gamma\gamma)/\mathcal{B}$(H$\to 2e2\mu$)   & $\delta R/R$ & 0.17\% & 0.8\%
  \\
  $R=\mathcal{B}$(H$\to \gamma\gamma)/\mathcal{B}$(H$\to 2\mu$) & $\delta R/R$ & 0.29\% & 1.38\%
  \\
  $R=\mathcal{B}$(H$\to \mu\mu\gamma)/\mathcal{B}$(H$\to \mu\mu$)   & $\delta R/R$ & 0.58\% & 1.82\%
  \\
$R=\sigma$(t\={t}H)$\times\mathcal{B}$(H$\to$ b\={b})/$\sigma$(t\={t}Z)$\times\mathcal{B}$(Z$\to$ b\={b}) & $\delta R/R$ & 1.05\% & 1.9\%
   \\
$\mathcal{B}$(H$\to$ invisible) & $\mathcal{B}$@95\%CL & $1\times 10^{-4}$ & $2.5\times
  10^{-4}$
  \\
  HH production & $\delta\lambda/\lambda$ & 3.0--5.6\% & 3.4--7.8\%
\\
\hline
\end{tabular}
\end{table}

Thermal WIMPs are an important class of models of dark mater (DM) that could explain the large scale structure of the universe. Their mass is somewhere between a GeV and $10^2$~TeV. Any discovery or exclusion by the many planned direct or indirect detection experiments, in underground and/or space-based facilities, would require confirmation and detailed study by accelerator-based measurements to be conclusive, and these experiments do not cover all the possibilities. The LHC and prior accelerators covered a fraction of the mass range for thermal WIMPs. Also, dark matter searches for a broad range of models, in particular Higgs to invisible decays, can cover and in some cases go beyond the sensitivity of direct detection experiments past the neutrino floor. A 100~TeV FCC-hh facility is necessary to test several of the most compelling benchmark scenarios~\cite{Benedikt:2018csr}. In particular, in the simplest and most predictive scenario, DM is part of an electroweak multiplet and it interacts with the SM through the weak interaction. The
obvious examples of this class are an electroweak doublet or a Majorana electroweak triplet. These are the Higgsino and wino in the context of SUSY, although they can be perfect dark matter candidates without SUSY. FCC-hh energies are required to cover these cases. 
For example, wino DM has a thermal relic mass of 2.8~TeV, and only the FCC-hh is sensitive to masses this high. Further, many other weakly-interacting DM models would \emph{only} be discovered at FCC-hh, such as for example, ``Higgsinos'', ``bino-winos'', and ``co-annihilated singlets'', which contain two DM particles with different interaction strengths and masses. 

Within the discovery reach of the FCC-hh for BSM, there is also the possibility of detecting signs of lepton flavor violation, e.g., through the production of leptoquark (LQ) particles coupling to leptons and quarks of different generation. These LQ models can explain the anomalies currently observed by flavor experiments (e.g., LHCb and muon $g-2$). Both the FCC-hh and the FCC-ep will significantly extend the reach of the LHC and HL-LHC, and explore LQ masses in the multi-TeV range~\cite{Benedikt:2018csr}. 

\subsection{FCC-ep}

A FCC-ep~\cite{FCC-eh} machine will realize Deep Inelastic Scattering with unprecedented kinematic reach, allowing for the high precision study of the strong coupling constant $\alphas$, and the the parton distribution functions (PDFs) in the proton for all individual quarks and the gluon. High-precision determinations of the proton parton densities in the mid-$Q^2$ range are, first, critical to warrant the  calculation of the corresponding p-p cross sections for W, Z, and Higgs bosons and top quarks at FCC-hh, with reduced theoretical uncertainties. Second, measurements of PDFs at high $Q^2$ (up to $10^6$~GeV$^2$) will improve the precision of jet production cross sections at the FCC-hh, at the highest energies, improving the sensitivity to possible new contact interactions, or quark-substructure. Last but not least, measurements of the PDFs for $x$ down to $10^{-7}$, where gluon saturation and deviations from DGLAP QCD evolution equations can be tested. 
Beyond QCD studies, the FCC-ep machine will enable unique measurements of EW parameters, such as the effective weak mixing angle, and of the Wtb vertex in single top quark production, e.g., the measurement of the CKM matrix element $V_{tb}$ with a precision $<1\%$. 

In BSM physics, leptoquark searches are a particularly strong aspect of the FCC-ep program~\cite{Benedikt:2018csr}. Contact interactions and a lepton/quark substructure, up to scales in the 100~TeV range can also be tested with at a FCC-ep machine.

Last but not least, and in analogy to the LHC, the hadron collider complex opens the door to a varied and diverse physics program, ranging from the study of heavy ion collisions~\cite{Abada:2019lih,Dainese:2016gch}, to flavor physics, to  the use of the injector complex for fixed-target experiments~\cite{Goddard:2017nyi}.

\section{Instrumentation}
\label{sec:instrumentation}
\textcolor{red}{Editors: M. Aleksa, F. Bedeschi, P. Giacomelli, E. Perez,  J. Qian, and S. Seidel}


Two complementary detector design concepts have been proposed in the FCC-ee Conceptual Design Report (CDR)~\cite{FCC-ee-accelerator}, the ``CLIC-like Detector" (CLD)~\cite{Bacchetta:2019fmz} and the ``International Detector for Electron-positron Accelerator" (IDEA)~\cite{Bedeschi:2021nln}.  The concepts are evolution of the detectors for the past and current colliders incorporating the latest results from years of R\&D  as well as the newest technologies. They are the conceptual realizations of the detector requirements discussed in Ref.~\cite{azzi2021exploring} and serve as bases for further studies.

\begin{figure}[htb]
\centering
  \subfigure[]{\includegraphics[width=0.40\textwidth]{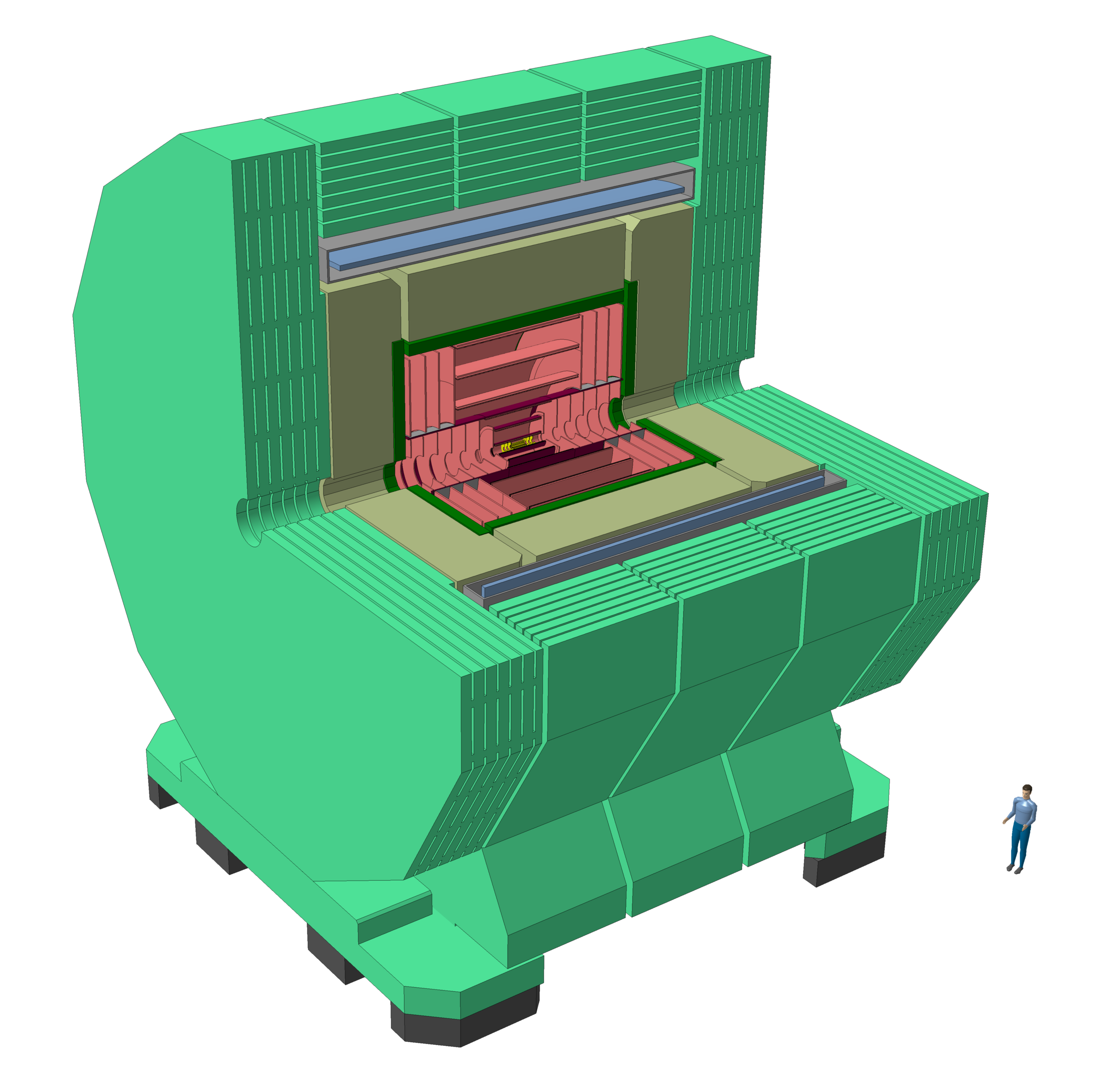}} \hspace*{1.0cm}
  \subfigure[]{\includegraphics[width=0.40\textwidth]{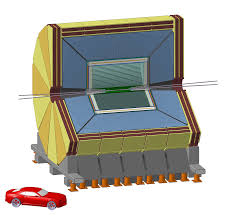}}
\caption{Schematics of the two detector concepts proposed in the FCC-ee Conceptual Design Report: (a) CLD and (b) IDEA.}
\label{fig:detector}
\end{figure}

As illustrated in Fig.~\ref{fig:detector}, the CLD detector features a silicon pixel vertex detector, a silicon tracker, followed by a highly granular calorimeters (a silicon-tungsten ECAL and a scintillator-steel HCAL) surrounded by a 2T superconducting solenoid and muon chambers interleaved with steel return yokes. The IDEA detector comprises a silicon vertex detector, a large-volume extremely-light drift chamber surrounded by a layer of silicon detector, a thin low-mass superconducting solenoid, a preshower detector, a dual-readout fiber calorimeter and muon chambers within the magnet return yoke. These detectors are designed to operate at bunch crossings from a minimum of 20~ns at the Z pole running to a maximum 7~$\mu$s at the highest energy, $\sqrt{s}=365\,\GeV$. Recently, work has started on a third detector concept~\cite{Francois}, comprising a silicon vertex detector, a light tracker (drift chamber or full Si are being studied), a thin low-mass superconducting solenoid, a noble-liquid based ECAL, a scintillator-iron HCAL and a muon tagger. 

All these detector concepts are still under evolution and still being optimized. There is ample room for further innovative concepts. Currently up to four interaction points are foreseen for the FCC-ee. 
This section summarizes the major subsystems and their expected performances and highlights areas for further R\&D and optimizations.

\subsection{Tracking and Vertexing}
The innermost subsystem~\cite{Bacchetta} of an FCC-ee experiment must provide excellent efficiency for tracking down to about 120\,mrad, for charged particles of transverse momentum as low as 100\,MeV.  A tracking volume of relatively large radius, with minimum mass and stable structure, will be needed to ensure that systematic uncertainties do not exceed the statistical uncertainties that will be achieved by the anticipated large data sets. The minimum tracking detector radius is limited by the detector solenoid strength, which must be restricted to 2T~\cite{Boscolo} for the Z pole running to avoid  blow-up of low-emittance beams crossing at 30\,mrad.  For a pixel pitch of $25\,\mu$m, an occupancy below $10^{-3}$ for cluster multiplicities of 5 (with safety factor 3) may be possible.  The attainable impact parameter resolution for individual tracks is $2 \oplus 10/(p_{\rm T} \sin^{1/2}\theta)~\mu$m~\cite{Bedeschi_ECFA}.
The intended momentum resolution is $\Delta(1/p_{\rm T})\sim 2\times10^{-5}\oplus1\times10^{-3}/(p_{\rm T}\sin^{1/2} \theta)$.  
The physics goals for the tracker imply a material budget for the innermost layer of $0.2\%\,X_0$ or better, with the budget for the entire vertex detector limited to less than $1\%\,X_0$ and individual hit resolution on the order of $3\,\mu$m.  This has direct implications for the power budget  and cooling mechanism of the tracker.  

Tracking in the CLD concept adapts the CLIC design~\cite{Linssen}, extending to radius 2.1\,m and introducing CO$_2$ evaporative cooling.  This all-silicon concept uses three double layers of silicon pixel sensors plus three double disks for the vertexing system, three double layers of short strip sensors plus seven double disks for the inner tracking volume, and three double strip layers plus four double disks for the outer tracking volume. The vertex detector implements $25\times 25\,\mu{\rm m}^2$ pixels of effective thickness $50\,\mu$m. The thickness per layer is being further optimized, current designs achieve $\sim 0.5\%\,X_0$ per layer~\cite{andreazza2022} but monolithic architectures are being studied to significantly reduce this value.

A variety of silicon technologies should be mature for consideration in the FCC-ee design.  Low-mass and low-power MAPS is being explored for large area coverage. Its commercial CMOS processes allow full circuitry inside the pixel cell with high resistivity wafers permitting operation in fully or partially depletion mode.  HVCMOS devices~\cite{Peric} embed NMOS and PMOS transistors in a single deep n-well that acts as a charge collection electrode; the substrate can be biased with a high negative voltage for depletion around the well. Looking ahead to the needs of FCC-hh, which require substantial R\&D,  driven by radiation hardness as well as fast readout and timing requirements, depleted MAPS (DMAPS) sensors use high resistivity substrate which, under reverse bias, allows signal charge to be collected through drift.  Some of its concepts have already been incorporated into ASICs such as the MALTA/Monopix chip~\cite{Snoeys}.

 Alternatives to DMAPS include those with split charge amplification and readout sectors: DEPFET and FPCCD.  Another potential direction uses heterogeneous vertical integration; a number of variations on the silicon-on-insulator (SoI) technology~\cite{Tsuboyama} are being pursued.  
 
 Associated mechanical technologies are also in development, including for example self-supported curved circuits that conform to small radii and minimize material overlaps.  Connections between this type of material and sensors have been demonstrated with aluminum wedge wirebonding.  Other interconnect technologies, such as anisotropic conductive film, are under study; these can also permit realization of serial powering with its accompanying reduction of services material. Microchannel cooling~\cite{Francisco} embedded in silicon, and microvascular networks~\cite{Pety} in carbon, hold promise as low-Z structurally stable options.  An ideal configuration might ultimately integrate support, electronics, and routing in a single ladder.

Hybrid technologies, for which the sensor and the readout are optimized separately and then brought together, provide a different approach. Research on low gain avalanche detectors (LGADs)~\cite{Sadrozinski} aims for a timing resolution of a couple of tens of picoseconds. Silicon sensors of the 3D geometry are achieving superior radiation hardness and, through their geometry, are intrinsically fast.  Parallel research on ASICs is proceeding.

The IDEA concept includes a silicon vertex detector~\cite{Pancheri} surrounded by a drift chamber, enclosed in a silicon wrapper.  The vertex detector implements monolithic active pixel sensor (MAPS) technology with fully depleted high-resistivity substrates, on-pixel sparsification, and data-driven time-stamped readout. The silicon wrapper encapsulates the drift chamber in both the barrel and disk regions.
The drift chamber~\cite{Tassielli:2021rjk} is designed to provide an efficient tracking, a high precision momentum measurement and an excellent particle identification.  
The distinctive element of the chamber is its high transparency, in terms of radiation lengths, obtained thanks to the novel approach for the wiring and assembly procedures adopted for the MEG2 drift chamber~\cite{Baldini}. Indeed, the total amount of material in radial direction, towards the barrel calorimeter is of the order of $0.016\, X_0$ and approximately $0.05\, X_0$ in the forward and backward directions, including the end-plates instrumented with front-end electronics.
The chamber is a unique volume, high granularity, all stereo, low mass cylindrical drift chamber, co-axial to the solenoid field filled with helium based gas mixture. 
The minimization of the amount of material in the end-plates is obtained thanks to an innovative system of tie-rods, which redirects the wire tension stress to the outer end-plate rim, where a cylindrical carbon fibre support structure bearing the total load is attached. The gas volume is enclosed by two thin carbon fibre domes, whose profile is suitably shaped in order to minimize the stress on the inner cylindrical wall, allowing for free deformations under gas pressure variations, without affecting the wire tension.
The choice of an extremely light gas mixture allows for the exploitation of the counting techniques for particle identification with a \dNdx resolution below 3\% (a factor 2 better than \dEdx)~\cite{Cascella:2014kca} and of the cluster timing to improve the spatial resolution under $100\, \mu$m.

\subsection{Calorimetry}
There have been considerable R\&D on calorimeter technologies over the past several decades, primarily driven by the physics needs of future $\epem$ colliders. Hadronic final states with two or more jets, either produced directly in the collisions or indirectly from the decays of W, Z, or H bosons, are expected to dominate the production. Therefore precise measurements of jet energies are critical for any physics program. One benchmark of particular interest is the performance in separating the hadronic $\mathrm{W}\to \qqbar'$ and $\mathrm{Z}\to \qqbar$ decays through the measurement of the dijet invariant mass. Given the mass difference of $\sim 12\%$ between the W and Z boson, a dijet mass resolution of 3--4\% or better is needed to adequately distinguish between the two decays. This sets a requirement of jet energy resolution of $\sim 30\%/\sqrt{E}$, approximately a factor of two better than those of the current and past calorimeters.  

Fluctuations in electromagnetic (EM) and non-EM energy deposits from hadronic showers and non-uniform detector responses to these deposits are the dominant sources affecting the jet energy resolution. Different technologies have been pursued to mitigate their impacts as discussed in Ref.~\cite{Aleksa:2021ztd}. Particle-flow (PF) and dual-readout (DR) sampling calorimeters, respectively incorporated in the CLD and IDEA detector concepts, are currently the two leading approaches. The PF method aims to identify and reconstruct all visible particles in an event while the DR method targets the separate measurements of the EM and non-EM energy deposits event by event. 

Prototypes of PF calorimeters with different absorbing materials and readout techniques have been tested with beams by the CALICE collaboration~\cite{Sefkow:2015hna}. The CLD calorimetry design calls for a silicon-tungsten ECAL with a transverse segmentation of $5\times 5\, {\rm mm}^2$ and a longitudinal segmentation of 40 identical layers with a total depth of $22 X_0$, and a scintillator-steel HCAL with a transverse segmentation of $30\times 30\,{\rm mm}^2$ and 44 longitudinal layers for a total depth of $5.5\lambda_I$.  This segmentation is deemed adequate to resolve energy deposits from nearby particles in jets. A PF-based calorimeter is being constructed for the Phase-2 upgrade of the CMS endcap calorimeter~\cite{CMS:2017jpq}. 

A fiber dual-readout calorimeter~\cite{Wigmans:2010zz} is proposed for IDEA. The calorimeter will read out both scintillation light (S) produced primarily by hadrons (non-EM component) and Cherenkov light (C) emitted mostly by relativistic electrons (EM component).
This technology has been developed by the RD52/DREAM collaboration~\cite{Akchurin:2014aoa, Wigmans:2016zfc,Antonello:2021tsz}. The proposed calorimeter for IDEA is 2\,m deep, corresponds to approximately $7\lambda_I$. It combines ECAL and HCAL functionalities in one uniform system and consists of a total of ${\cal O}(10^8)$ fibers.

Fine segmentation is a common feature for both PF and DR approaches, resulting in a far larger number of channels compared with traditional calorimeters. Silicon multiplier~(SiPM)~\cite{Renker:2006ay,Renker:2009zz} readout is a key component of both calorimeter designs. SiPMs with their high efficiencies, large gains, wide dynamic ranges along with their compact nature and cost effectiveness make them the ideal choice for calorimetry applications. However, the large number of readout channels will require robust monitoring and calibration system of detector responses. 

CALICE and RD52/DREAM collaborations have demonstrated that both designs can achieve a jet energy resolution of 3--4\% for jets expected from $\mathrm{W/Z}\to \qqbar$ decays~\cite{Sefkow:2015hna,Antonello:2021tsz}. However, the EM energy resolution is expected to be $\sim 15\%/\sqrt{E}$ for PF and $\sim 10\%/\sqrt{E}$ for DR, largely because of the small sampling fractions. These resolutions are significantly worse than those of crystal ECALs~\cite{L3BGO:1993tta,CMS:2013lxn}. It has been proposed that the combination of a DR crystal ECAL with a DR fiber HCAL can potentially maintain or even improve the jet energy resolution while attaining the EM resolution of $<3\%/\sqrt{E}$ expected from a crystal ECAL~\cite{Lucchini:2020bac}. A consortium of US teams is leading this R\&D. 

Highly granular noble-liquid sampling calorimetry was proposed for a possible FCC-hh experiment~\cite{Benedikt:2018csr,aleksa2019calorimeters, Aleksa:2020qdy}. It has been shown that, on top of its intrinsic excellent electromagnetic energy resolution, noble-liquid calorimetry can be optimized in terms of granularity to allow for 4D imaging, machine learning or, in combination with the tracker measurements, particle-flow reconstruction.  

Recently, a highly granular noble-liquid sampling calorimeter was also proposed as an electromagnetic calorimeter of an FCC-ee experiment~\cite{aleksa2021calorimetry}. It consists of a cylindrical stack of lead/steel absorbers, readout electrodes and active liquid Argon (LAr) gaps with an inner radius of 2.1\,m, compatible with a thin 2\,T solenoid integrated in the same cryostat.
Tungsten absorbers or liquid krypton (LKr) as active material are interesting options due to the resulting smaller radiation length and smaller Moli\`ere radius.
The granularity of each longitudinal compartment can be optimized according to the needs of particle-flow reconstruction and particle ID. Currently, a granularity of $\Delta\theta\times\Delta\varphi = 2.5\,\mathrm{mrad}\times 8.2\,\mathrm{mrad}$ ($5.4\,\mathrm{mm}\times 17.7\,\mathrm{mm}$) is foreseen in the first calorimeter compartment to optimize the $\pi^0$ rejection.
Single-particle simulations of electrons and photons have resulted in a stochastic term of 8.2\,\%~\cite{aleksa2019calorimeters,Aleksa:2020qdy} for the standalone electromagnetic energy resolution. Exploiting the full 4D imaging information, it was demonstrated that a deep-neural-network analysis can achieve a single-$\pi^-$ resolution stochastic term of 37\,\% ($B=4$\,T) for such a highly granular noble-liquid calorimeter complemented with a scintillator-iron HCAL.  It should be noted that this remarkable result relies on the calorimeter measurement alone, the planned PF reconstruction combined with the tracker will further substantially improve the hadronic resolution. R\&D on such a noble-liquid calorimeter is supported by the CERN EP R\&D program~\cite{CERN-OPEN-2018-006}.

The calorimeter performance could be further enhanced with the time-stamping of the energy deposits. This is particularly true for the fiber DR calorimeter that lacks adequate longitudinal segmentation. A resolution of 1\,ns or better could offer valuable information for the reconstruction of the longitudinal shower profiles and, for the FCC-hh, have the additional benefits of mitigating pileup impacts.

\subsection{Muon detection}
Compared with hadron colliders, muon detection at $\epem$ machines poses less of a challenge thanks to the relative cleanness of the $\epem$ collisions. Muon detectors at LEP and other past $\epem$ colliders were primarily aimed at tagging muons while relying on the inner tracking detectors for their momentum measurements.  The CLD detector concept envisions 6--7 layers of Resistive Plate Chambers (RPCs)~\cite{Santonico:1981sc,Santonico:1988qi} embedded in a steel return yoke to tag muons with a position precision of ${\cal O} (2-3\,{\rm mm})$, sufficient for matching with the tracks reconstructed in the inner tracking detector and therefore tagging muons. On the other hand, the IDEA concept proposes to use the new $\mu{\rm RWELL}$ technology~\cite{Bencivenni:2017wee,Bencivenni:2018syq} with a spatial resolution of ${\cal O}(100\,\mu{\rm m})$ to provide a standalone muon momentum measurement along the line of the L3~\cite{L3:1989aa}, ATLAS~\cite{ATLAS:2008xda}, and CMS~\cite{CMS:2008} detector designs. The more precise spatial resolution could be an interesting tool also for studying long-lived particles, LLPs. A detailed summary of the muon detector options can be found in Ref.~\cite{Braibant:2021wts}.

\subsection{Particle identification}
Particle identification (PID) capabilities are essential for the flavor physics program, can aid the searches for LLPs, and are also beneficial to other physics through for example the improvement of the jet flavor tagging. Candidate PID techniques~\cite{Wilkinson:2021ehf} include measurements of ionization energy (\dEdx), time-of-flight (TOF), and the Cherenkov radiation ring~(RICH). While a RICH detector is conceivable for a special flavor-oriented collider detector, \dEdx and TOF measurements are the most practical options for a general purpose FCC-ee detector.

The \dEdx measurement is a long-established technique for PID in a tracking detector for charged particles with momenta $\lesssim 10\,\GeV$. The performance of this traditional method, however, suffers from Landau fluctuations. An alternative method of counting the number of clusters, \dNdx, giving rise to the ionization can significantly improve the PID performance~\cite{Cascella:2014kca}. The drift chamber for the IDEA detector concept is well suited for PID using both the \dEdx and cluster counting methods. A three-sigma separation between $\pi$-$K$ is expected for charged particles with momenta up to $\sim$30~\GeV\, except for a narrow ``blind region" around 1~\GeV\, that can be covered with a minimal TOF detector~\cite{Bedeschi:2022rnj}. 

TOF is an effective PID method for charged particles with momenta up to a few GeV for a collider detector in which the flight distance is limited to $\sim 2\,{\rm m}$. Though a TOF detector is not incorporated explicitly in the current detector concepts, the advancement in SiPMs and semiconductor tracking detectors with good timing resolutions has made a TOF detector an attractive addition to extend physics reach, as exemplified by the MIP timing layer in CMS~\cite{Butler:2019rpu} and HGTD detector in ATLAS~\cite{Aboulhorma:2021xhs} for the HL-LHC upgrade. Both ATLAS and CMS timing detectors are expected to achieve a time resolution of ${\cal O}(30\,{\rm ps})$. With further R\&D, a timing resolution of ${\cal O}(10\,{\rm ps})$ could be within the reach at the time-scale of the FCC-ee, significantly improving the PID performance.  

Apart from the identifications of SM particles, both \dEdx and TOF techniques can improve the detectors' capabilities in tagging heavy long-lived particles predicted in many SM extensions. A compact TOF timing layer similar to those being constructed for the ATLAS and CMS upgrades but with improved time resolution will be particularly attractive for the FCC-ee. The physics potential of such a detector remains to be explored~\cite{Chekanov}.

\subsection{Luminosity monitoring}
The FCC-ee Z physics program calls for a precision of $10^{-4}$ on the absolute luminosity measurement and of $5\times 10^{-5}$ on the relative measurement between the energy points of the Z-lineshape scan~\cite{FCC-ee-accelerator}. These requirements are more stringent than those achieved previously, but more importantly for a much more challenging environment.
At $\epem$ colliders, the luminosity is traditionally determined from the rate of the small-angle Bhabha scattering, $\epem\to \epem$, due to its large cross section that can be calculated precisely. The Bhabha events are detected in a dedicated monitor, often called the luminosity calorimeter or LumiCal, covering small angles in the forward-backward regions.

Unlike LEP detectors, the forward-backward region, often referred to as the machine-detector interface (MDI) region~\cite{Boscolo:2021dxi}, of a FCC-ee detector is crowded with crossing beam pipes, the quadrupoles and compensating solenoids. It necessitates the integration of the LumiCal into the MDI design which positions the LumiCal at approximately 1\,m away from the interaction point. The design faces significant engineering and operational challenges as discussed in Ref.~\cite{Dam:2021sdj}. For example, the steep angular dependence of the Bhabha scattering cross section imposes a precision of ${\cal O}(1\mu {\rm rad})$ on the geometrical acceptance, correspond to a precision of ${\cal O}(1\,\mu {\rm m})$ on the radial dimension of the LumiCal.
Inspired by the second-generation of LEP LumiCals, a compact finely segmented sandwich calorimeter of tungsten-silicon (W-Si) layers is proposed for the FCC-ee. The OPAL collaboration showed that an experimental precision as low as $3.4\times 10^{-4}$ can be achieved with this design~\cite{OPAL:1999clt}. However, it will be considerably more difficult for the FCC-ee to achieve the required precision as the LumiCal will be deep inside the detector volume about 2.5 times closer to the IP than those at LEP.

The large angle $\epem\to\gamma\gamma$ production offers an alternative process for the luminosity determination. Despite of a cross section several orders of magnitude smaller than the Bhabha process, this process should be statistically sufficient to reach the required precision and enjoys entirely different sources of systematic uncertainties.

\subsection{Online, computing, and software}
The online, offline, and software requirements~\cite{Brenner:2021mxb,Helsens:2021diw,Ganis:2021vgv} for the FCC-ee are dominated by those of the Z pole running with a physics event rate of $\sim 200$~kHz. However, the data throughput rate, storage and CPU resource needs will be similar to those of the HL-LHC, in operation long before the FCC-ee. While there might be unique challenges specific to the FCC-ee and its detector designs, the requirements are not expected to be too different from those for the HL-LHC. Thus, the successful operation of the HL-LHC is the best R\&D for the FCC-ee. It is reasonable to expect that the models developed for and tested at the HL-LHC will serve as starting points for the FCC-ee, refined with experiences gained and improved with technological developments and targeted R\&D since.

A ``triggerless" DAQ system will likely be the preferred choice at the FCC-ee. A software-based online trigger provides a flexibility that cannot be matched by traditional multilevel hardware-based filtering systems. The triggerless system is already being implemented by the LHCb experiment for the LHC Run 3~\cite{LHCbCollaboration:2014vzo}. 

For what relates to software requirements, also in this case the developments and solutions that will be done and put in place for the HL-LHC will be very beneficial to the FCC-ee. FCC is currently fully committed in implementing its software around Key4hep~\cite{Ganis:2021vgv}, a software ecosystem which aims to become the common “Turnkey Software Stack” for future experiments. Key4hep is being actively developed and is expected to cover most, if not all, the needs of future colliders. It mostly builds on the experience and cumulative knowledge of the LHC, taking naturally advantage of the continuous ongoing developments required to meet the needs of the HL-LHC. Key4hep also aims to include software products emerging from R\&D programs, such as the AIDA series~\cite{AidaInnova}, addressing software aspects of interest for larger communities, such as the Pandora SDK for pattern recognition~\cite{Marshall:2015rfa}.

\subsection{FCC-hh}
The high energy and high luminosity of the FCC-hh impose stringent criteria on its detector design. The detector must be able to measure multi-TeV jets, leptons and photons from heavy resonances with masses up to ${\cal O}(50\,\TeV)$ in a harsh collision environment with the projected average value of 1000 $pp$ collisions per bunch-crossing. The significant higher level of radiation compared with the HL-LHC ---a factor of 20 to 30 higher radiation is expected--- will severely constrain technological choices for detector subsystems. A more detailed discussion of these challenges can be found in the FCC-hh conceptual design report~\cite{Benedikt:2018csr}. The report also presents one reference detector design without specific choices for subsystem technologies. These designs serve as concrete examples to facilitate identifications of areas where dedicated R\&D efforts are needed. 

\section{Executive Summary}
\label{sec:conclusion}

It is rare that a single investment can open up an entire wide-ranging physics program, allowing an unprecedented
investigation into the building blocks and forces of our Universe, and thousands of impactful measurements to be made.
The discovery of the Higgs boson marks the opening of a new era of exploration, for which the FCC program (the electron-positron collider FCC-ee, followed by the proton collider FCC-hh)  is the most effective answer, with its  complementarity and synergy with other domains of particle physics. It enables precision studies of the electroweak and strong forces, and extends our knowledge of potential new high energy physics coupling to the Higgs, the W/Z bosons,  the heavy 
quarks and leptons 
and even  neutrinos. 
The reach for the discovery of new particles is considerably extended both towards weaker couplings and higher masses.  

The FCC program starts with the construction of a 90\,km tunnel near CERN.  The first stage is a high luminosity $\epem$ collider  based on well established technologies with center-of-mass energies running from the Z boson mass, with the highest possible luminosity, allowing collection of $5 \times 10^{12}$ Z bosons in a few years, through the WW threshold, to  Higgs production in association with a Z, and up to top-quark pair production.

The FCC-ee run plan will give unprecedented measurements of Higgs couplings, the electroweak and flavor parameters, the top mass and the strong coupling constant, with ample discovery potential for light feebly interacting particles in BSM models addressing the origin of dark matter, neutrino masses, and of the EW scale.  
Record-setting control of the beam energy will allow measurements of boson masses to extreme precision. 
The development of new magnets will then allow a 100\,TeV proton-proton collider  complementing the FCC-ee for  precision studies of Higgs phenomena beyond the statistical reach of lepton colliders, including Higgs pair production and the associated study of the Higgs potential.  The contents of the proton can be further elucidated over many orders-of-magnitude in parton momentum fraction and energy scale with a subsequent electron-proton collider (FCC-ep).  A heavy ion run can yield truly hard probes to explore the 
only fundamental non-Abelian quantum field theory (QCD) whose collective properties can be studied in the lab.
It is a bold vision of a breathtaking scientific menu, bringing us ever closer to a true knowledge of the fundamental structure of the universe.

The program will allow measuring the Higgs boson mass to a few MeV, and its width to the percent level (40\,keV).  The Higgs branching fractions times production cross section will be measured with a precision at the percent level, reaching 0.3\% for the H $\to b\bar b$ decay in the ZH production mode. Most of the Higgs boson couplings determined at the HL-LHC with a number of assumptions will be measured in an absolute and model-independent manner at FCC-ee, and significantly improved to reach a precision better than 1\%, down to the per mil level for the HZZ coupling. All but the couplings to the first generation and the Higgs self-coupling (which will be eventually measured to better than 5\% precision) will be measured to that precision over the full program. With precision center-of-mass energy control, FCC-ee offers also the unique opportunity to probe the electron Yukawa coupling through resonant $\epem \to $\,H production. 

Precision measurements of electroweak observables such as forward-backward asymmetries at the Z pole, the W mass, and the top mass, will accurately test the closure of the Standard Model. Projected conservative reductions of uncertainties range from a factor of 90 ($\Gamma_Z$) or more, to a factor of 10 ($\sigma_\text{had}^0$). Studies above $\ttbar$ production threshold will reduce the uncertainty on the top mass by an order of magnitude compared to the HL-LHC reach.

All such ultra precise measurements will be able to indirectly discover new particles coupling to the Higgs and/or electroweak bosons up to scales of $\Lambda \approx 7$ and 50~TeV, respectively.

The program greatly extends our reach for observation of new weakly coupled or high mass particles.  For example, FCC-ee is unique in its sensitivity to axion-like particles with masses in the range $\sim$1--100\,GeV and SM couplings down to $10^{-7}$, and to heavy neutral leptons directly, all the way down to the seesaw limit for masses below the Z mass, or, indirectly from precision Z and $\tau$ measurements  up to 1000 TeV masses for mixing $U^2$ down to $10^{-5}$, giving a finite chance of a great step in our understanding of neutrino mass generation and of  the origin of the baryon asymmetry of the universe.
FCC-hh will extend the reach for heavy particles beyond that of LHC typically by a factor of six, up to masses of several tens of TeV, matching with direct measurements the indirect sensitivity to new phenomena obtained with precision FCC-ee measurements and, among other results, 
largely probing
weakly interacting massive particle scenarios of
thermal relic dark matter. 

Current results from flavor experiments  may indicate the first hints of new physics.  FCC-ee Z-pole running will increase samples well beyond those ultimately expected from BELLE-II by factors of 20 for B mesons containing light quarks, and allow precision studies of mesons containing charm and strange quarks.
Extremely large $\tau$ data samples will also be provided.
Many $\tau$ properties are still dominated by LEP measurements, and large improvements are guaranteed. Important examples are tests of lepton universality, where 1--3 orders of magnitude are possible via precise measurements of $\tau$ lifetime and leptonic branching fractions, and searches for lepton flavor violation, where enhancements as large as three orders of magnitude are expected.

 
The study of FCC physics is in its infancy, just 10 years old, and the above precision estimates are often very conservative. By design, in addition to high luminosity, FCC-ee provides superb experimental conditions.
Beam-induced backgrounds are small and easily filtered out, the beam spot is tiny, and a beam pipe radius of 10\,mm is being designed. Operation without trigger should be possible. The center-of-mass energy calibration will be at the ppm level, possibly better, based on the resonant depolarization of the beams. These experimental conditions promise huge leaps in sensitivity in the search for new phenomena, and in precision for measurements sensitive to new physics in loops or by mixing. This is especially true in the Z run, where the statistical uncertainty reduction by a factor 500 and sometimes greater, challenges equally: (i) the ingenuity and technological savvy of experimenters to imagine and design detectors, and (ii) the skill of theorists to organize the program of calculations; both with the aim of improving accuracy and sensitivity to match the available statistical uncertainty. Working groups are being formed to answer the challenge, several of which are led by US scientists.    

Finally, the FCC program builds on the large stable international community that has for more than 30 years been working at CERN on the LEP and LHC programs, the pioneering work on intense electron beams for VEPP-4M, DA$\Phi$NE, the SLAC and KEK B-factories, and the present SuperKEKB. CERN's stable funding structure, robust international participation and long-term commitment to the delivery of ambitious projects, together with CERN unparalleled infrastructure, give unique strength to this far-sighted project.
The U.S. has an opportunity to be a leader in this exciting, multifaceted physics program, continuing our contributions to humanity's 
exploration of fundamental physics.

\section{Acknowledgments}

We acknowledge support from the following funding agencies: DOE and NSF (USA), Istituto Nazionale di Fiscia Nucleare (Italy)

The work by UNM was made possible by support from the U.S. Department of Energy grant DE-SC0020255 and the National Science Foundation grant number 1906674.
U. Maryland is supported by DOE grant DESC0010072.
U. Mass Amherst is supported by DOE grant DE-SC0010004 

Research supported by the Fermi National Accelerator Laboratory, managed and operated by Fermi Research Alliance, LLC under Contract No. DE-AC02-07CH11359 with the U.S. Department of Energy

The work of Jorge de Blas has been supported by the FEDER/Junta de Andaluc\'ia project grant P18-FRJ-3735.

J. Gluza has been supported in part by the Polish National Science Center (NCN) under grant 2020/37/B/ST2/02371.

For Spain we must acknowledge: "MCIN/AEI/10.13039/501100011033 (Spain)" 

A. Pich is supported by MCIN/AEI/10.13039/501100011033, Grant No. PID2020-114473GB-I00,
Generalitat Valenciana, Grant No. Prometeo/2021/071,

The work of S.Heinemeyer is supported in part by
the grant PID2019-110058GB-C21 funded by
``ERDF A way of making Europe'' and by
MCIN/AEI/10.13039/501100011033, and in part
by the grant CEX2020-001007-S funded by
MCIN/AEI/10.13039/501100011033.

S.A. acknowledges support from the Swiss National Science Foundation (project grant number 200020/175502).

This work has been partially supported by the from the European Union's Horizon 2020 research and innovation programme under grant agreement No 951754.

\clearpage
\bibliographystyle{elsarticle-num}
\bibliography{theBIBGeneral,theBIBEWK, theBibQCD,theBibBSM,theBIBflavor,theBibDetector,theBibTop,theBibAccelerator,theBibFCC-hh}

\newpage
\appendix
\appendixpage  
\addappheadtotoc  


{\Large \bf Supporters of U.S. involvement in a future FCC program}

Brad Abbott	University of Oklahoma  \\
Kaustubh Agashe	University of Maryland \\
Nural Akchurin	Texas Tech University  \\
W. Altmannshofer University of California Santa Cruz \\
Giorgio Apollinari	Fermi National Accelerator Laboratory \\
Artur Apresyan	Fermi National Accelerator Lab.  \\
Howard Baer	University of Oklahoma \\
Ela Barberis Northeastern University \\
Lothar A. T. Bauerdick	Fermi National Accelerator Laboratory \\
Michael Begel	Brookhaven National Laboratory  \\
Alberto Belloni	University of Maryland  \\
Sergey Belomestnykh	Fermilab \\
Jeffrey Berryhill	Fermi National Accelerator Laboratory \\
Kenneth Bloom	University of Nebraska Lincoln  \\
Adi Bornheim	California Institute of Technology  \\
Dimitri Bourilkov	University of Florida  \\
Antonio Boveia	Ohio State University \\
Gustaaf Brooijmans	Columbia University  \\
Elizabeth Brost Brookhaven National Laboratory \\
Quentin Buat	University of Washington  \\
John Mark Butler	Boston University  \\
Yunhai Cai SLAC National Accelerator Laboratory \\
Anadi Canepa	Fermi National Accelerator Laboratory\\
Marcela Carena	Fermilab/UChicago \\
Cari Cesarotti	Harvard University \\
Zackaria Chacko	University of Maryland \\
Maria Chamizo Llatas Brookhaven National Laboratory	 \\
Sanha Cheong	SLAC National Accelerator  Laboratory  \\
Frank Chlebana	Fermi National Accelerator Laboratory \\
Dan Claes	University of Nebraska Lincoln  \\
Raymond Co University of Minnesota \\
Lucien Cremaldi	University of Mississippi \\
Tiesheng Dai	University of Michigan \\
Mariarosaria D'Alfonso	Massachusetts Inst. of Technology \\
Sridhara Dasu	University of Wisconsin Madison \\
Sally Dawson	Brookhaven National Laboratory \\
Andre Luiz De Gouvea Northwestern University \\
Regina Demina University of Rochester \\
Dmitri Denisov	Brookhaven National Laboratory \\
Bogdan Dobrescu	Fermi National Accelerator Laboratory \\
Javier Mauricio Duarte	Univ. of California San Diego \\
Kevin Frank Einsweiler	Lawrence Berkeley National Laboratory \\
Sarah Eno	University of Maryland  \\
Robin Erbacher	University of California Davis  \\
Jan Eysermans	Massachusetts Inst. of Technology  \\
JiJi Fan Brown University \\
Abhijith Gandrakota	Fermi National Accelerator Laboratory \\
Yongsheng Gao	California State University, Fresno \\
Yuri Gershtein	Rutgers State Univ. of New Jersey  \\
Tony Gherghetta University of Minnesota \\
Matthew Gignac	University of California,Santa Cruz  \\
Frank Golf	University of Nebraska Lincoln  \\
Dorival Gonçalves	Oklahoma State University \\
Julia Lynne Gonski	Columbia University  \\
Paul Grannis	Stony Brook University \\
Heather Gray, University of California Berkeley/LBNL \\
Lindsey Gray	Fermi National Accelerator Laboratory \\
Phillip Gutierrez	University of Oklahoma \\
Joe Haley Oklahoma State University \\
Tao Han	University of Pittsburgh \\
Mike Hance	UC Santa Cruz \\
Phillip Harris Massachusetts Inst. of Technology\\
Robert M. Harris	Fermi National Accelerator Laboratory\\
Nicole Michelle Hartman	SLAC National Accelerator  Laboratory  \\
Kenichi Hatakeyama	Baylor University  \\
Hannah Elizabeth Herde	SLAC National Accelerator  Laboratory \\
Christian Herwig	Fermi National Accelerator Laboratory \\
Mike Hildreth Notre Dame University \\
James Hirschauer	Fermi National Accelerator Laboratory \\
John Hobbs	Stony Brook University  \\
Julie Hogan	Bethel University  \\
Sungwoo Hong University of Chicago \\
Anson Hook University of Maryland \\
John Huth	Harvard University \\
Andreas Werner Jung	Purdue University  \\
Keti Kaadze	Kansas State University  \\
Michael Kagan	SLAC National Accelerator  Laboratory  \\
Boris Kayser Fermi National Accelerator Laboratory \\
Doojin Kim	Texas A\&M University \\
Boaz Klima	Fermi National Accelerator Laboratory \\
Markus Klute	MIT/KIT \\
Kyoungchul Kong	University of Kansas \\
Ashutosh Kotwal	Duke University  \\
Ilya Kravchenko	University of Nebraska-Lincoln \\
Eric Christian Lancon	Brookhaven National Laboratory \\
Greg Landsberg	Brown University  \\
Katharine Leney	Southern Methodist University \\
Zoltan Ligeti	Lawrence Berkeley National Laboratory\\
Don Lincoln	Fermi National Accelerator Laboratory \\
Zhen Liu	University of Minnesota \\
Henry Lubatti University of Washington \\
Joseph Lykken	Fermi National Accelerator Laboratory \\
Yang Ma University of Pittsburgh \\
Christopher Madrid	Fermi National Accelerator Laboratory \\
Jeremy Mans	University of Minnesota  \\
Verena Ingrid Martinez Outschoorn University of Massachusetts \\
Simone Michele Mazza	University of California, Santa Cruz  \\
Petra Merkel	Fermi National Accelerator Laboratory \\
Corrinne Mills	University of Illinois at Chicago  \\
Rabindra Mohapatra	University of Maryland \\
Benjamin Nachman LBNL \\
Sergei Nagaitsev	Fermilab/U.Chicago \\
Steve Nahn	Fermi National Accelerator Laboratory \\
Christopher Neu	University of Virginia  \\
Mark Neubauer	Univ. Illinois at Urbana Champaign  \\
David Neuffer	Fermi National Accelerator Laboratory \\
Harvey Newman	California Institute of Technology \\
Jennifer Ngadiuba	Fermi National Accelerator Laboratory \\
Jason Nielsen	University of California,Santa Cruz  \\
Yuri Nosochkov SLAC National Accelerator Laboratory \\
Radek Novotny University of New Mexico \\
Isobel Ojalvo Princeton University \\
Yasar Onel	University of Iowa \\
Toyoko Orimoto	Northeastern University \\
Jennifer Ott	University of California,Santa Cruz  \\
Chris Palmer	University of Maryland  \\
Vaia Papadimitriou	Fermi National Accelerator Laboratory \\
John Parsons	Columbia University  \\
Christoph Paus	Massachusetts Inst. of Technology  \\
Michael Peskin	SLAC National Accelerator Laboratory\\
Marc-André Pleier	Brookhaven National Laboratory  \\
Mason Proffitt University of Washington \\
Jianming Qian	University of Michigan \\
Chris Quigg	Fermi National Accelerator Laboratory \\
Salvatore Rappoccio	The State University of New York SUNY \\
Tor Raubenheimer SLAC / Stanford University \\
Laura Reina Florida State University \\
Thomas Rizzo SLAC National Accelerator Laboratory \\
Jennifer Roloff	Brookhaven National Laboratory  \\
David Saltzberg	University of California, Los Angeles  \\
David Sanders	University of Mississippi \\
Deepak Sathyan University of Maryland \\
Todd Satogata	Jefferson Lab \\
Bruce Andrew Schumm	University of California,Santa Cruz  \\
Thomas Andrew Schartz University of Michigan \\
Ariel Gustavo Schwartzman	SLAC National Accelerator  Laboratory  \\
Reinhard Schwienhorst Michigan State University \\
Andrea Sciandra	University of California, Santa Cruz (US) \\
Sally Seidel	University of New Mexico (US) \\
Andrei Seryi	Jefferson Lab \\
Elizabeth Sexton-Kennedy	Fermi National Accelerator Laboratory \\
Vladimir Shiltsev	Fermi National Accelerator Laboratory \\
Mel Shochet	University of Chicago \\
Louise Skinnari Northeastern University \\
Maria Spiropulu	California Institute of Technology  \\
Giordon Holtsberg Stark	University of California,Santa Cruz  \\
Kevin Stenson	University of Colorado Boulder  \\
James Strait	Fermi National Accelerator Lab.  \\
Nadja Strobbe	University of Minnesota  \\
John Stupak	University of Oklahoma  \\
Shufang Su	University of Arizona \\
Indara Suarez	Boston University  \\
Raman Sundrum	University of Maryland, College Park \\
Anyes Taffard University of California Irvine \\
Rafael Teixeira De Lima	SLAC National Accelerator  Laboratory  \\
Evelyn Jean Thomson	University of Pennsylvania  \\
Alessandro Tricoli	Brookhaven National Laboratory  \\
Dmitri Tsybychev	Stony Brook University  \\
Carlos E.M. Wagner University of Chicago and Argonne National Laboratory \\
LianTao Wang University of Chicago \\
Bennie Ward	Baylor University \\
Gordon Watts University of Washington \\
James Wells	University of Michigan, Ann Arbor \\
Stephane Willocq	University of Massachusetts  \\
Darien Wood	Northeastern University  \\
Frank Wuerthwein	UC San Diego \\
Si Xie	California Institute of Technology  \\
Zijun Xu	SLAC National Accelerator  Laboratory \\
Hongtao Yang	Lawrence Berkeley National Laboratory \\
Charlie Young	SLAC National Accelerator  Laboratory  \\
Jinlong Zhang	Argonne National Laboratory  \\
Bing Zhou	University of Michigan, Ann Arbor  \\
Junjie Zhu	University of Michigan, Ann Arbor  \\

\end{document}